\documentclass{aa} 
\usepackage{graphicx}

\usepackage{subfigure}

\usepackage{txfonts}
\usepackage{lscape}
\newcommand{\kms}{km s$^{-1}$\xspace}
\newcommand{\HI}{{\rm H\,{\scriptsize I}}\xspace}

\usepackage{natbib}
\usepackage[colorlinks, citecolor={blue}]{hyperref}
\usepackage{threeparttable}

\usepackage{amsmath}    
\usepackage{tabularx}   
\usepackage{booktabs}   
\usepackage{siunitx}    
\usepackage{multirow}   
\usepackage{array}      
\usepackage{lscape}             
\usepackage{placeins}           
\usepackage{tabularx}       
\usepackage{tabularx}
\usepackage{threeparttable}
\usepackage{booktabs}
\usepackage{array}

\begin{document}

\title{\HI 21-cm Line Properties of the Nearby LIRG IRAS 04296+2923}

   \author{Guixiang Feng
          \inst{1}
          \and
          Zhongzu Wu\inst{1}\fnmsep\thanks{zzwu08@gmail.com}   
          \and
          Chuan-Peng Zhang\inst{2,3}
          \and
          Ming Zhu\inst{2,3}
        }

   \institute{College of Physics, Guizhou University, 550025 Guiyang, PR China \email{zzwu08@gmail.com}
   \and
         National Astronomical Observatories, Chinese Academy of Sciences, Beijing 100101, China
         \and
         Guizhou Radio Astronomical Observatory, Guizhou University, Guiyang 550000, China
         }
   \date{  }

\abstract
{We present an analysis of archival Very Large Array (VLA) and Five-hundred-meter Aperture
Spherical radio Telescope (FAST) \HI\ 21 cm data, together with archival multi-band radio continuum observations, of the nearby luminous infrared galaxy IRAS~04296+2923. The system, located behind the Taurus dark cloud at a distance of $\sim$29 Mpc, forms a small galaxy group consisting of five members as revealed by the \HI\ imaging. IRAS~04296+2923 has a close companion, HI~0432+2926, with a projected separation of $\sim$40 kpc, a small line-of-sight velocity difference of $\Delta$ v = 26 km s$^{-1}$, and comparable total \HI\ masses of order $10^{9}$~$M_{\odot}$. Both galaxies exhibit regular \HI\ velocity fields and characteristic double-horn profiles in the VLA and FAST data, accompanied by only subtle asymmetries and extended \HI\ structures, indicating rotation-dominated kinematics with early signs of weak tidal interaction. Radio continuum emission is detected only from IRAS~04296+2923 and is confined to its nuclear region, consistent with previous studies. Modeling of its multi-band radio spectrum reveals a significant contribution from free--free emission at high frequencies ($>$30 GHz) and a high FIR-to-radio flux ratio ($q_{8.4}\simeq3.2$), implying a young, dust-obscured nuclear starburst. Taken together, the regular \HI\ kinematics, the small velocity offset, and the group-scale environment favor an interpretation in which IRAS~04296+2923 and HI~0432+2926 form a gravitationally bound, orbiting galaxy pair embedded in a small group, rather than an advanced merger. In this context, the luminous infrared galaxy (LIRG) nature of IRAS~04296+2923 is more plausibly driven by internal processes, such as bar-induced gas inflow, possibly modulated by long-timescale, low-level tidal interactions with nearby group companions.
}

   \maketitle

\section{Introduction}
\label{sec:intro}

The {\it Infrared Astronomical Satellite} (IRAS; \citealt{1984ApJ...278L...1N}) revealed a population of luminous and ultraluminous infrared galaxies (LIRGs and ULIRGs) with $L_{\rm IR} = 10^{11}$--$10^{12}$\,$L_{\odot}$ and $>10^{12}$\,$L_{\odot}$, respectively. These systems are predominantly major mergers or strongly interacting pairs, with more than 90\% showing morphological or kinematic evidence of interaction \citep{2011AJ....141..100H}. Galaxy mergers drive dramatic evolution by transporting large amounts of gas into the central kiloparsec through gravitational torques, thereby triggering intense starbursts and, in some cases, feeding active galactic nuclei \citep{1996ARA&A..34..749S,2006ApJS..163....1H}. Enhanced star formation is observed in interacting pairs out to projected separations of $\sim$150\,kpc \citep{2012MNRAS.426..549S,2013MNRAS.433L..59P}, in agreement with hydrodynamical simulations that predict tidal perturbations can induce central inflows long before final coalescence \citep{2010A&A...518A..56M}.

Late-stage mergers typically exhibit strong tidal tails, bridges, and widespread star formation. However, an increasing number of LIRGs classified optically as isolated or “pre-merger” systems have been found, through \HI imaging, to possess extended or asymmetric neutral gas reservoirs \citep{2015AAS...22541101P,2014PhDT.......295P}. These observations indicate that intense nuclear star formation can be triggered well before the appearance of prominent large-scale tidal features. 
IRAS\,04296+2923 is a nearby LIRG ($L_{\rm IR} = 9.8 \times 10^{10}$\,$L_{\odot}$; \citealt{2010AJ....140.1294M}) located at a distance of $\sim$29\,Mpc behind the Taurus molecular cloud complex. Its nuclear starburst is extremely compact (confined within $\sim$150--250\,pc) and has been interpreted as driven by a stellar bar \citep{2010AJ....140.1294M,2014ApJ...795..107M,2004A&A...416..515D}. At the same time, Very Large Array (VLA)  \HI observations by \cite{2009AAS...21344503M} hinted that IRAS\,04296+2923 resides in a small group environment \citep{2009AAS...21344503M}, raising the possibility that weak tidal interactions may also contribute to fuelling its activity. This combination of a powerful but compact nuclear starburst and a potentially influential group environment makes IRAS\,04296+2923 an ideal laboratory for investigating the onset of merger-induced star formation in the very earliest interaction phases.

In this paper, we present \HI 21-cm imaging from the archival VLA and Five-hundred-meter Aperture
Spherical radio Telescope (FAST), combined with archival multi-frequency radio continuum data, to characterise the neutral gas environment, kinematics, and star-formation properties of IRAS\,04296+2923 and its companions. Our main goals are (i) to map the distribution and kinematics of the neutral gas in the group, (ii) to search for early signatures of tidal interaction, and (iii) to examine the link between the extended gas reservoir and the compact nuclear starburst. The data collection, reduction, and analysis methods are described in Sect.~\ref{data_collect}. Results are presented in Sect.~\ref{sec:results}, discussed in Sect.~\ref{sec:discussion}, and summarised in Sect.~\ref{sec:conclusions}. Throughout this work we adopt a flat $\Lambda$CDM cosmology with $H_0 = 67.8$\,km\,s$^{-1}$\,Mpc$^{-1}$, $\Omega_{\rm m} = 0.31$, and $\Omega_{\Lambda} = 0.69$, and the radio spectral index convention $S_\nu \propto \nu^\alpha$.

\section{Data collection, reduction and analysis}
\label{data_collect}
\subsection{The archival radio data}
\label{2.1}
We have collected the archival radio data including the \HI and radio continuum projects: The \HI observations including two VLA projects, one is AM960 (PI: D. Meier ) in VLA D array and the other is AM985 (PI: D. Meier)  in VLA C array  (see Table \ref{table1}).  The other is the \HI image cube of this source from the FAST ALL Sky  \HI Survey (FASHI) project (PI: Ming Zhu), the details about this project including observation, data reduction and generating the \HI image cube can be seen in \cite{2024SCPMA..6719511Z}.

The radio continuum project were mainly selected with beam size larger than 1 arcsec (see Table \ref{2923data})),  because the dominant radio continuum emission are believed to be compact at this scale \cite{2010AJ....140.1294M}. The radio continuum flux densities at 150 MHz 200 MHz and 325 MHz were from \cite{2021A&A...655A..17S}, which made the cross-identifications from three low-frequency radio surveys,  including the TIFR GMRT Sky Survey (TGSS), GaLactic and Extragalactic All-skyMWA (GLEAM) survey, the Westerbork Northern Sky Survey (WNSS). As well as the Arcminute Microkelvin Imager Galactic Plane Survey (AMIGPS) at 16 GHz \citep{2015MNRAS.453.1396P}.

\begin{table*}
\centering
\begin{threeparttable}
    
       \caption{Parameters of \HI line archive data of IRAS 04296+2923. \label{table1} }
   
  \centering
  \begin{tabular}{c c c c c c c c c }     
  \hline\hline
 Epoch  &   Array& Phase calibrator & Program& $\Delta_{V}$ &beam &PA  & rms\\ 
   &        &                  &        & (\kms)         &(")$\times$(")   & ($^\circ$)& (mJy $beam^{-1}$) \\
        \hline                                                                                                                         
2008Jul20   & VLA-D &0403+260 &AM960  & 10.3 &54.9$\times$49.4 & 49 &1.07\\         
2008Aug04   & VLA-D &0403+260 &AM960  & 10.3 &54.24$\times$49.1 & 71 &1.04\\   
2008Aug07   & VLA-D &0403+260 &AM960  & 10.3 &56.8$\times$47.3 & -82&1.49\\         
2008Aug08   & VLA-D &0403+260 &AM960  & 10.3 &51.4$\times$49.2 & 64 &0.81\\          
2009Aug11   & VLA-C &0403+260 &AM985  & 10.3 &16.9$\times$14.6 & -88&0.81\\ 
-          & FAST  & -       & FASHI    & 6.44  & 175$\times$175    & 0     &0.67\\
      \hline
       \end{tabular}
       \vskip 0.1 true cm \noindent Notes. Column (5) The velocity width of each channel image. Column (6) and (7) are the beam FWHM of these data. Column (8) the achieved 1 $\sigma$ noise level for channel images with channel width listed in Column (5).
         \end{threeparttable} 
   \end{table*}

     \begin{table*}[h]
         \centering
         \begin{threeparttable}
         \caption{Mult-band radio continuum emission of IRAS 04296+2923}
         \label{2923data}
         \begin{tabular}{c c c c c c c c c}
         \hline\hline
         Epoch&freq.& Prog.& Beam&PA & $F_{peak}$  & $F_{total}$  & rms & reference\\
            && &(")$\times$(")&($^\circ$)&(mJy $beam^{-1}$)&(mJy)&(mJy $beam^{-1}$) & \\
            \hline
            -    & 150 MHZ  & TGSS    &  -& -  & -&252 $\pm$50&-&a\\
            -    & 200 MHZ  & GLEAM    &  -& -   & -&385 $\pm$77&-&a\\
            -    & 325 MHz  & WNSS    &  -& -  & -&320 $\pm$64&-&a\\
        2008Jul20& 1.4 GHz & AM960 &41.8$\times$38.6&-85 & 120.3 &135$\pm$13&3.9& b \\
        2009Aug11& 1.4 GHz & AM985 & 13.2$\times$11.6&-89 & 111.7& 132$\pm$3&1.7& b \\
          -      & 3 GHz    &  VLASS  & 2.2$\times$2.3&-89&33.4& 64$\pm$3& 0.2& c \\
        1992Jul13& 5 GHz    & AC326& 14.2$\times$13.7&-84 & 52.1& 60$\pm$2&0.5& b \\
        --       & 16 GHZ   & --   &  180$\times$180&-               & -& 20  & 3 & d\\
        2005Jun24& 22.5 GHz& AT309& 0.9$\times$0.3&76&3.7&23$\pm$5&0.2 & e \\
        2008Jun30& 43.3 GHz& AT367&1.6$\times$1.0&-42&8.8&13$\pm$ 2&0.9& b \\
           -     & 111 GHz  & -    &    4.6$\times$3.8&8   &  9.9 &11$\pm$ 2&0.7& e\\
        \hline
         \end{tabular}
        \vskip 0.1 true cm \noindent Notes. Column (2) and (3) the observed radio frequency and program. TGSS GLEAM and WNSS are three radio surveys by GMRT, MWA and Westerbork array, respectively (see the text in section \ref{2.1}). Column(4) and (5) are the beam FWHM of the images. Column (6)-(7) are the peak and total flux densities. Column(8) are the achived noise level for these continuum images.  Column (9)  the reference of the total and peak flux. a: \cite{2021A&A...655A..17S,2020yCat.8104....0S}.   b: this work, c: VLASS survey d: \cite{2015MNRAS.453.1396P}   e: \cite{2010AJ....140.1294M}.
         \end{threeparttable} 
     \end{table*}

\subsection{Radio data reduction}
The Data calibration of the VLA archive data was performed using the Common Astronomy Software Applications (CASA) package developed by the National Radio Astronomy Observatory (NRAO). The main processing steps included inspecting and flagging the bad data, amplitude calibration, bandpass calibration, phase calibration, and applying the calibration solutions to the target source. The `tfcrop` command was used to remove some radio frequency interference (RFI). For the spectral line data, we have performed continuum subuction in CASA software, the poorly quality channels at the beginning and end were deleted, and `intphase` and `scanphase` phase interpolations were also performed. The steps have referenced the CASA User Manual for reducing VLA data. The calibrated data were imported into the Difmap package for makeing the radio continuum, \HI channel images, all the images were made with natural weighting. 

For the \HI channel images from VLA project, we have also generated a primary beam image and performed primary beam correction and generated a image cube using CASA software. Then we also used CASA to produce moment 0, 1, and 2 images which stands for images of total intensity, velocity centroid, and velocity dispersion, respectively. The moment maps have set a threshold for regions with pixal values greater than 3 $\sigma$. Meanwhile, we also used the SoFiA software \citep{2015MNRAS.448.1922S,2021MNRAS.506.3962W} to analysie the \HI image cube and generated similar momment maps. Following the official tutorial online \footnote{\href{https://gitlab.com/SoFiA-Admin/SoFiA-2/-/wikis/SoFiA-Tutorial}{SoFiA-Tutorial}}, the parameter settings are as follows: the spatial smoothing filter sizes (scfind.kernelsXY) were set based on the number of pixels per beam, kernelsXY=0,4,8; the spectral smoothing filter sizes (scfind.kernelsZ) were chosen according to the typical widths of the \HI emission and absorption line profiles, kernelsZ=3,7,15,31. Reliability checking was enabled, a scale factor for the size of the Gaussian kernel (reliability.scaleKernel) was set to 0.6 which is obtained from varing the the parameter and checking the diagnostic plots generated when the parameter reliability.plot = true as shown in the online tutorial.  While all other parameters remained at their default values. The source detection threshold (scfind.threshold) was set to 3.8, which lies within the recommended range of 3.5 to 4.5 and corresponds to approximately four times the noise level, typically yielding reliable results.

For each detected galaxy, the \HI parameters (see Table \ref{sources}) including the central velocity ($V_{\mathrm{HI}}$), 
the full width at half maximum ($W_{50}$), and the integrated flux density ($S_{21}$) 
were measured from the total \HI line profiles. 
The dynamical centers were determined using the SoFiA source-finding software. 
Optical counterparts were identified from the $I$-band Pan-STARRS image \citep{2016arXiv161205560C}. 
For IRAS~04296+2923 and HI~0432+2926, the line profiles cannot be well fitted with Gaussian components; 
therefore, $V_{\mathrm{HI}}$ was obtained from the intensity-weighted mean velocity and $S_{21}$ from direct profile integration. Following \citet{2004AJ....128...16K}, the uncertainties are calculated as
$\sigma(V_{\mathrm{HI}}) $ = $3\,(\mathrm{S/N})^{-1}\sqrt{0.5\,(W_{20}-W_{50})\,\Delta v}$,
where S/N is the signal-to-noise ratio, $W_{20}$ and $W_{50}$ represent the velocity widths at 20\% and 50\% of the peak line intensity, respectively, and $\Delta v$ is the velocity resolution. 
The uncertainty of $W_{50}$ is estimated as
$\sigma(W_{50})$ = $2\sigma(V_{\mathrm{HI}})$,
and the uncertainty of the integrated flux density, $S_{21}$, obtained by directly integrating over the H~\textsc{i} line profile, is given by
$\sigma(S_{21})$ = $4\,(\mathrm{S/N})^{-1}\sqrt{S_{\mathrm{peak}}\,S_{21}\,\Delta v}$.
For the other three sources, $V_{\mathrm{HI}}$, $W_{50}$, $S_{21}$, and their corresponding uncertainties are derived through Gaussian fitting in CASA.

We have also used the 3D-Based Analysis of Rotating Objects via Line Observations, also known as BBarolo \citep[][]{2015MNRAS.451.3021D} to analysis the \HI image cubes, including the construction of position–velocity (P–V) diagrams images and to perform tilted-ring modeling of the rotating disk of IRAS 04296+2923 and other detected \HI galaxies.  The software fits the observed \HI cube  directly in three dimensions in a automatic mode, minimizing the residuals between the data and the model, and allows us to recover the kinematic parameters such as rotation velocity, inclination, \HI surface density and systemic velocity.

\subsection{Photometry of the Pan-STARRS images}

For each \HI\ detection, we retrieved the corresponding Pan-STARRS images to identify the optical counterparts and to perform basic photometric measurements. The photometric analysis and stellar mass estimates were carried out following the procedures described in \citet{2024MNRAS.529.3469G} and \citet{2022MNRAS.510.1716R}. Source detection and masking of contaminating foreground and background objects were performed using the \textsc{Photutils} package. The photometry was conducted on the $g$- and $r$-band images, with segmentation maps generated via \textsc{Segmentation}. Isophotes were fitted using \textsc{Isophote}, and the total $r$-band flux was measured via aperture photometry within apertures defined by the reliable isophotes, yielding integrated magnitudes for each source. Assuming the Galactic dust extinction law described by \cite{1989ApJ...345..245C}, the attenuation can be written as $A_{V} = R_{V}E(B-V)$. The coefficients $R_{V}$ are taken to be 3.793 and 2.751 for the $g$ and $r$ bands, respectively \citep{2007ApJS..173..293W}. The Galactic reddening $E(B-V)$ for each galaxy was obtained from the IRSA Dust Extinction Service\footnote{\url{https://irsa.ipac.caltech.edu/applications/DUST/index.html}} at the sky position of each \HI\ detection. This service provides two estimates of the dust extinction from \cite{2011ApJ...737..103S} and \cite{1989ApJ...345..245C}; in this work we adopt the values from \cite{2011ApJ...737..103S}. The $g$- and $r$-band magnitudes were then corrected for Galactic extinction using $m_{\rm cor}$ = $m_{\rm obs}$ - $A_{V}$,
and the corrected magnitudes are listed in Table~\ref{sources}.
The photometric analysis in this work is based on Python scripts and an accompanying cookbook publicly released by the WALLABY collaboration\footnote{\url{https://github.com/tflowers15/wallaby-analysis-scripts}}.

Stellar masses were estimated using the empirical color--mass-to-light relation from \citet{2011MNRAS.418.1587T}:
\begin{align*}
\log\left(\frac{M}{M_\odot}\right) &= -0.840 + 1.654\,(g-r) + 0.4\,(D_{\mathrm{mod}} + M_{\odot,r} - m) \\
&\quad - \log(1+z) - 2\,\log\left(\frac{h}{0.7}\right),
\end{align*}
where the coefficients $-0.840$ and $1.654$ are empirically calibrated constants \citep{2009MNRAS.400.1181Z}. The $g-r$ color and the apparent magnitude $m$ are measured in the SDSS photometric system, $D_{\mathrm{mod}}$ is the distance modulus, $h$ is the Hubble constant in units of 100~km~s$^{-1}$~Mpc$^{-1}$, and $M_{\odot,r} = 4.64$ is the absolute magnitude of the Sun in the $r$ band \citep{2018ApJS..236...47W}. The typical uncertainty in the derived stellar masses is $\sim$0.16~dex, dominated by uncertainties in the photometry and the empirical mass-to-light relation \citep[see][for details]{2024MNRAS.529.3469G,2022MNRAS.510.1716R}.

\section{Results}
\label{sec:results}
 \subsection{Detection of \HI emission in IRAS 04296+2923 and its Companion Galaxies}\label{3.1}
\begin{figure*}[h]
       \centering
        \includegraphics[width=9cm]{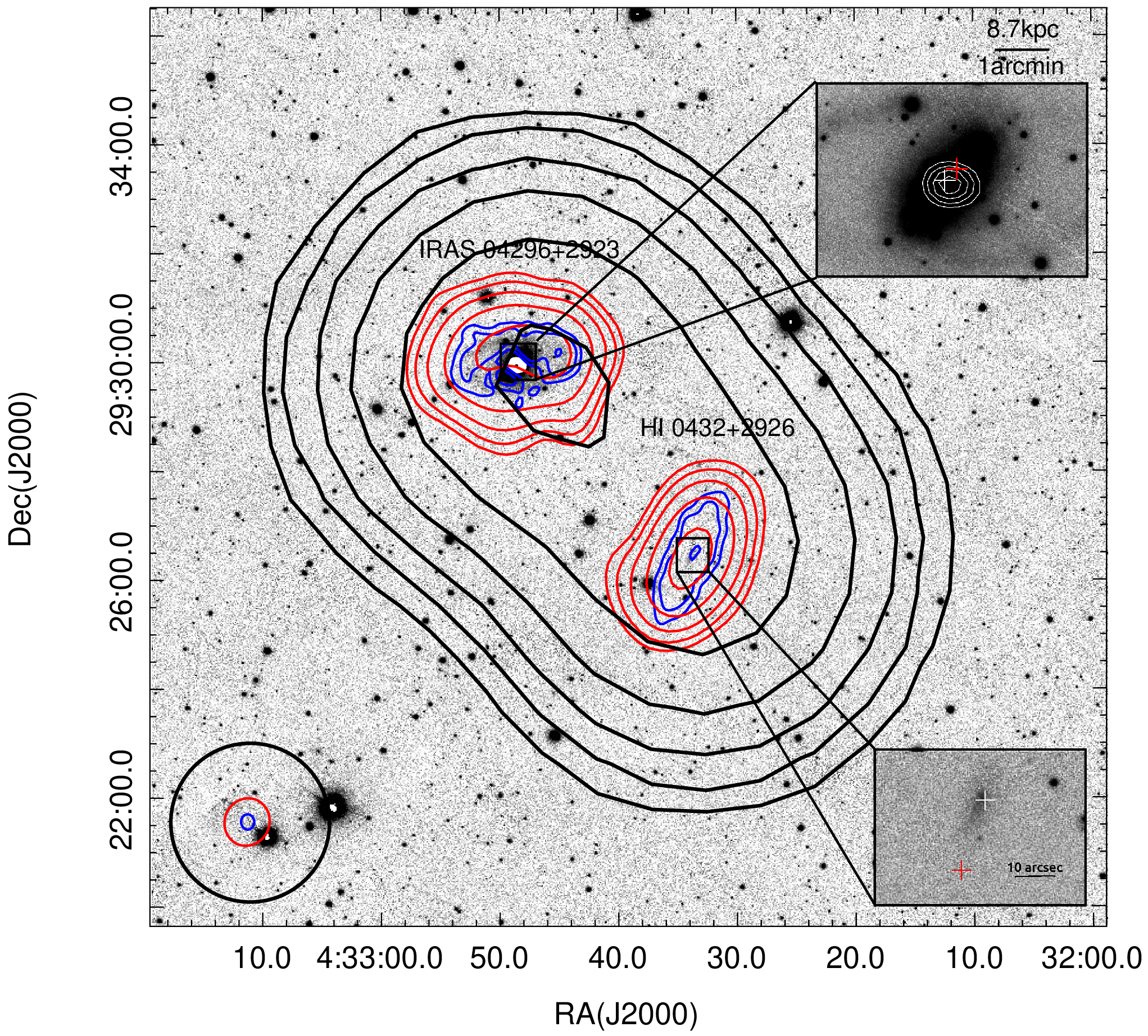}
        \includegraphics[width=9cm,height=8.1cm]{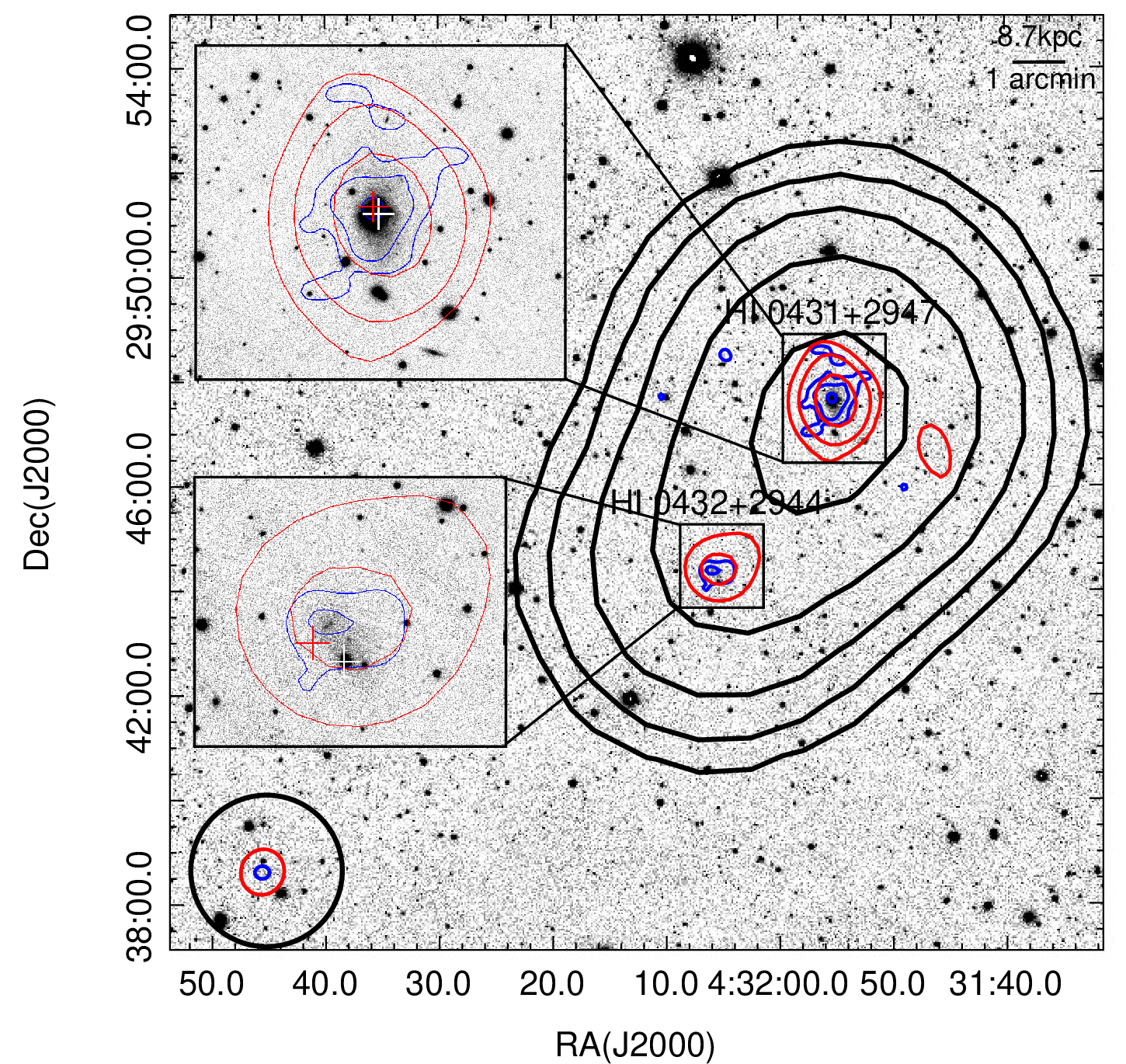}
        \includegraphics[width=10.5cm,height=8.1cm]{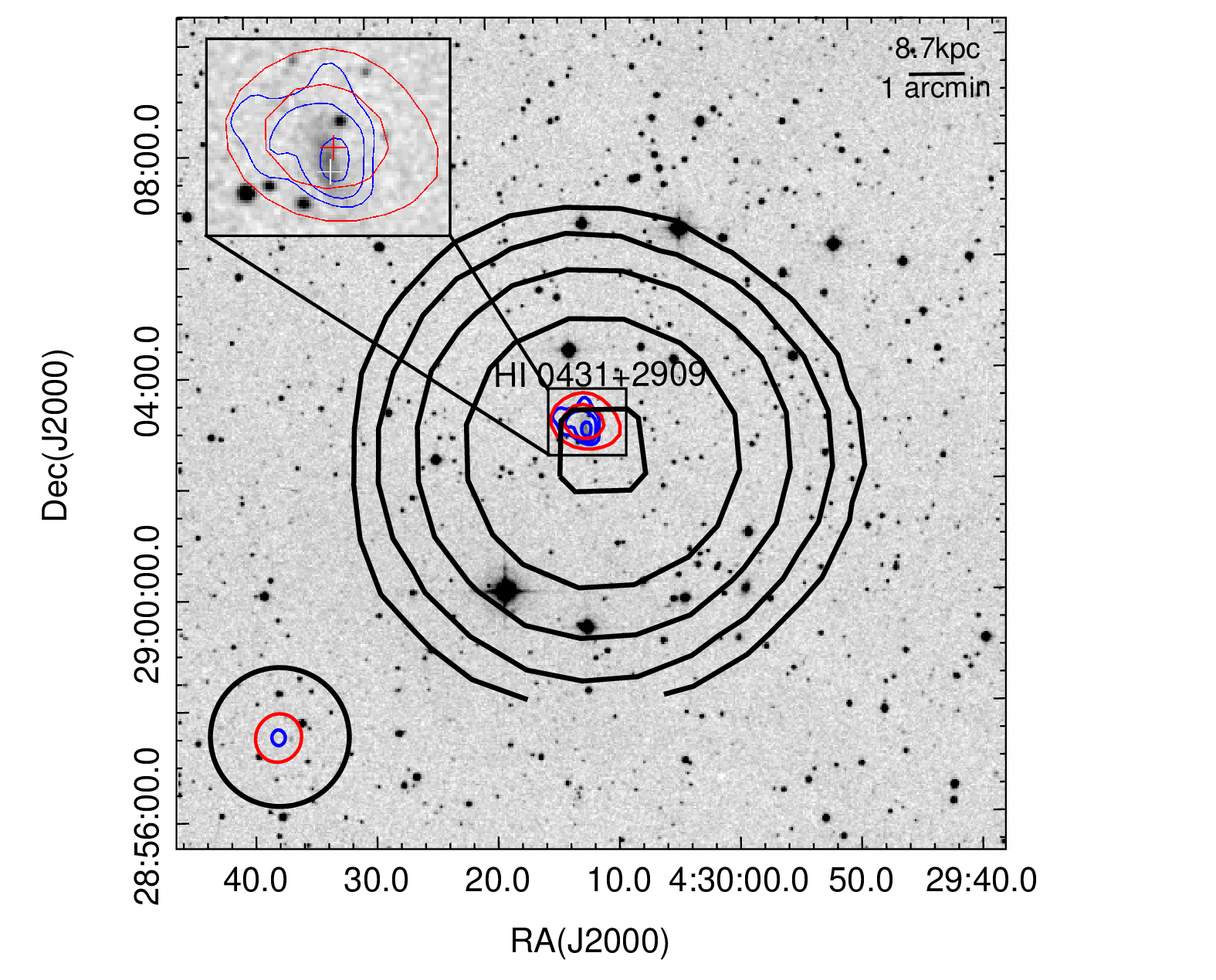}
        \caption{\HI\ emission contours overlaid on the optical Pan-STARRS images of IRAS~04296+2923 and the other members of its galaxy group.
Red, blue, and black contours represent the combined \HI\ total intensity (moment\,0) maps from the VLA D-array, VLA C-array, and FAST observations, respectively.
White contours in the upper-left panel indicate central \HI\ absorption toward IRAS~04296+2923 as detected in the VLA C-array data.
For the upper-left panel, the contour levels are:
red: 0.03, 0.06, 0.12, 0.24, 0.48, 0.96, 1.30, and 1.54~Jy\,beam$^{-1}$\,km\,s$^{-1}$;
blue: 0.09, 0.18, and 0.36~Jy\,beam$^{-1}$\,km\,s$^{-1}$;
black: 0.10, 0.20, 0.40, 0.80, 1.60, and 3.20~Jy\,beam$^{-1}$\,km\,s$^{-1}$;
white (absorption): $-$0.30, $-$0.40, $-$0.50, and $-$0.60~Jy\,beam$^{-1}$\,km\,s$^{-1}$.
For the remaining panels, the contour levels are:
red: 0.40, 0.80, and 1.60~Jy\,beam$^{-1}$\,km\,s$^{-1}$;
blue: 0.18, 0.36, and 0.72~Jy\,beam$^{-1}$\,km\,s$^{-1}$;
black: 0.10, 0.20, 0.40, 0.80, 1.60, and 3.20~Jy\,beam$^{-1}$\,km\,s$^{-1}$.
In all panels, the lowest contour corresponds to the $3\sigma$ level.
White and red crosses mark the optical and \HI\ centroid positions of the detected sources listed in Table~\ref{sources}.
Insets show zoomed-in optical views of the individual galaxies.
The lower-left inset indicates the synthesized beam sizes for each dataset (see Table~\ref{table1} for beam parameters).
}
        \label{fig.1:HIpos}
     \end{figure*}

Using both VLA-D and VLA-C array data, we imaged the \HI\ emission in IRAS 04296+2923 within a region of approximately 30 arcminutes, corresponding to the VLA primary beam at 1.4 GHz. We also analyzed the \HI\ image cube of the same source from the FASHI project, which covers a comparable field of view.
The integrated total intensity (moment-0) maps of the detected \HI\ emission are presented in Figs. \ref{fig.1:HIpos},  and \ref{fig:fast}. From these maps, we identified five \HI\ galaxies, including IRAS 04296+2923 and its companion HI 0432+2926, separated by about 4.7 arcmin, as well as three additional galaxies located roughly 17–22 arcmin from IRAS 04296+2923. The basic image parameters of these galaxies are listed in Table \ref{sources}. 

The 3$\sigma$ contour image from the VLA-D data shows that the \HI\ emission of IRAS 04296+2923 extends over a box-shaped region of approximately $3.5 \times 3$ arcmin, which is about 26 times larger than the synthesized beam. This extent is also larger than that seen in the near-infrared $J$-band image (Fig. \ref{fig.1:HIpos}), where the ring-like spiral arms appear weak and span about $2.5 \times 1.6$ arcmin. The 3$\sigma$ \HI\ contour image from the FAST data shows that the \HI\ emission is distributed over a much larger area (about 5.6 $\times$ 5.6 arcmin), encompassing both IRAS 04296+2923 and HI 0432+2926.

\begin{figure*}[h]
       \centering
         \includegraphics[width=8cm,height=7cm]{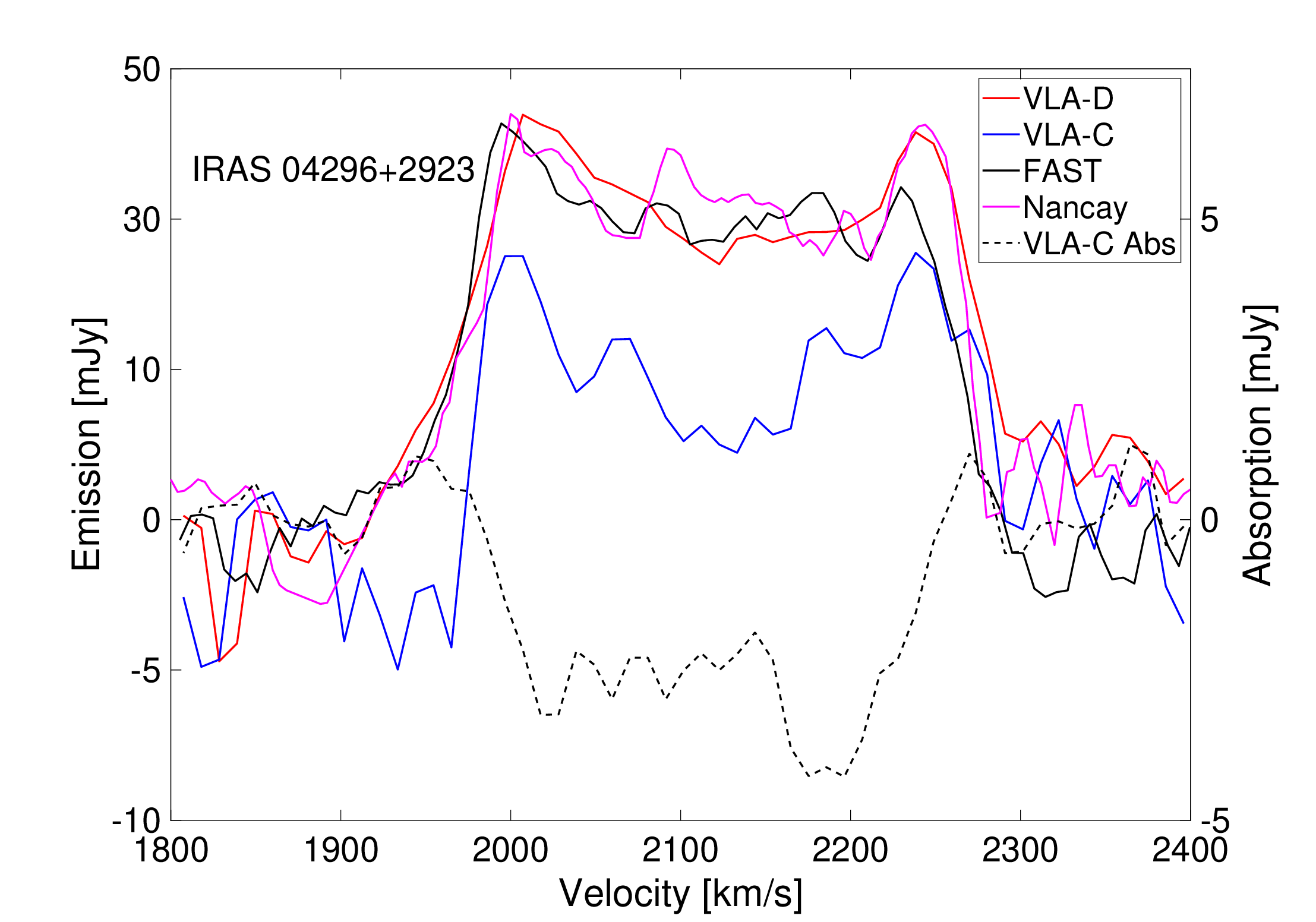}
         \includegraphics[width=8cm,height=7cm]{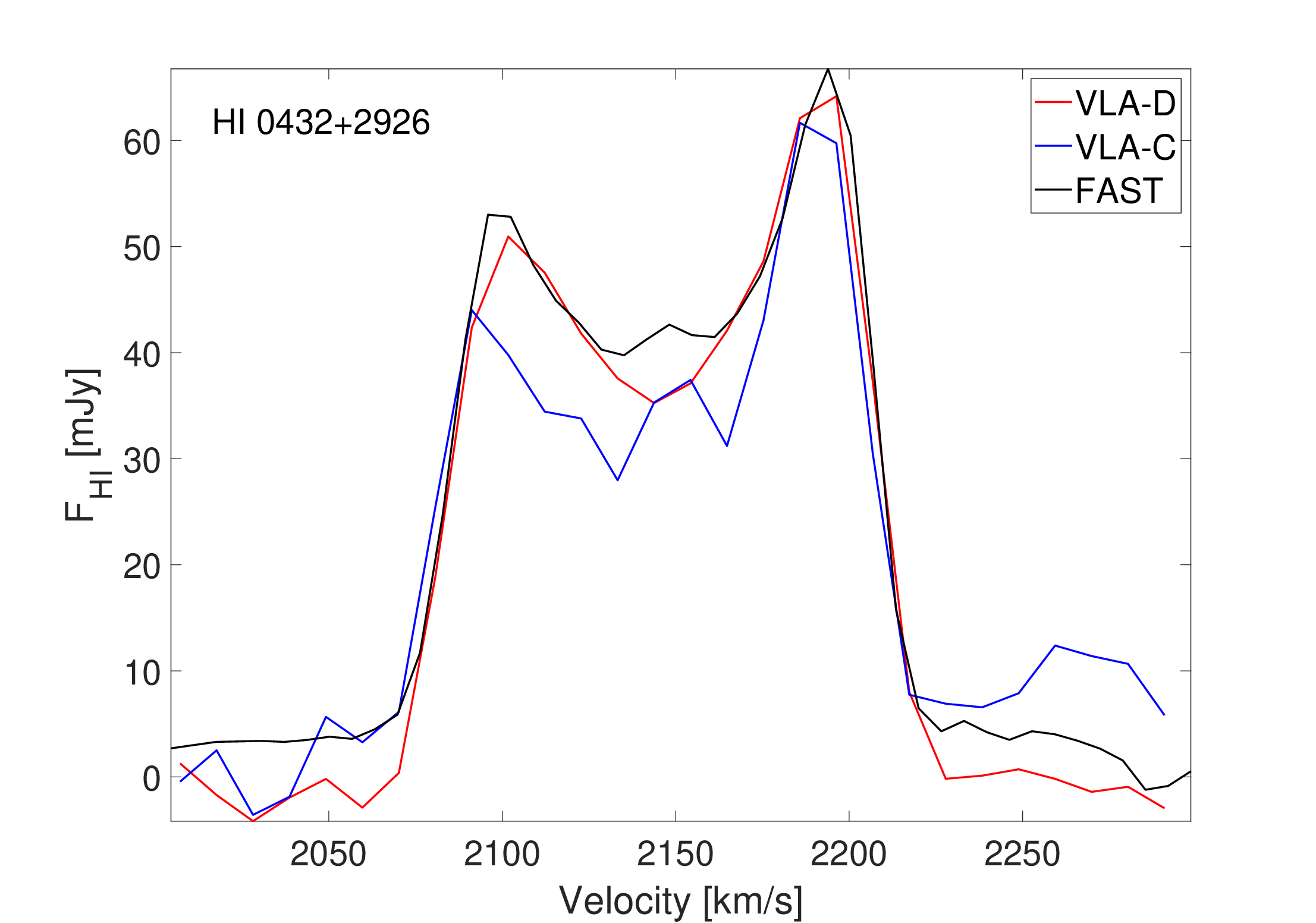}
        \caption{     \HI\ line profiles of IRAS~04296+2923 (\textit{left}) and HI~0432+2926 (\textit{right}) obtained from the VLA-C, VLA-D, and FAST observations. 
        For IRAS~04296+2923, the \HI\ emission lines were extracted from a circular region with a diameter of 3.4\arcmin, 
        centered at $\mathrm{RA}=04^{\mathrm{h}}32^{\mathrm{m}}48.9^{\mathrm{s}}$, 
        $\mathrm{Dec}=+29^{\circ}29^{\prime}53.1^{\prime\prime}$. 
        The \HI\ absorption spectrum was obtained from a box region centered at 
        $\mathrm{RA}=04^{\mathrm{h}}32^{\mathrm{m}}48.6^{\mathrm{s}}$, 
        $\mathrm{Dec}=+29^{\circ}29^{\prime}57.4^{\prime\prime}$, 
        with a size of about 20\arcsec~$\times$~20\arcsec. 
        The black line represents the Nançay \HI\ data from \citet{1995A&A...299..347C}. 
        For HI~0432+2926, the \HI\ emission lines were extracted from a box region centered at 
$\mathrm{RA}=04^{\mathrm{h}}32^{\mathrm{m}}33.7^{\mathrm{s}}$,    $\mathrm{Dec}=+29^{\circ}26^{\prime}26.5^{\prime\prime}$, 
        with a size of about 2.5\arcmin~$\times$~4.1\arcmin. 
        Because the angular resolution of the FAST telescope is relatively low, the two galaxies cannot be fully separated in the FAST \HI\ image. 
        Therefore, the FAST \HI\ line profiles were extracted from two box regions, as indicated in Fig.~\ref{fig:fast}. 
        In the left panel, the y-axis of the \HI\ absorption line is shown on the right, 
        while those of the emission lines are on the left.} 
        \label{fig.1:HIline} 
     \end{figure*}

The extracted \HI\ spectra of these five galaxies are shown in Fig. \ref{fig.1:HIline} and \ref{fig:another sources}, and their fitted line parameters are summarized in Table \ref{sources}. The \HI\ spectrum of IRAS 04296+2923 from the VLA-C data is about half the flux of that from the other projects, suggesting that part of the extended \HI\ emission may have been resolved out by the higher-resolution C-array. The \HI\ fluxes derived from the other projects are consistent with each other. Among the five galaxies, IRAS 04296+2923 has the highest \HI\ mass and the broadest linewidth. In addition, the VLA-C data reveal \HI\ absorption in the central region of IRAS 04296+2923 (see Fig. \ref{fig.1:HIpos}). The absorption profiles are broad and cover the same velocity range as the \HI\ emission lines (see Fig. \ref{fig.1:HIline}), indicating the presence of a dense column of cold atomic gas in front of the nucleus.

The \HI\ line profiles of IRAS 04296+2923 and HI 0432+2926 display double peaks, corresponding to the receding and approaching sides of rotating disks. In contrast, the other three galaxies show single-peaked profiles with FWHM values ranging from 50 to 90 \kms (see Table \ref{sources}). The systemic velocities of IRAS 04296+2923 and HI 0432+2926 are similar, with the latter being about 20–30 \kms\ higher. HI 0433+2909 has the highest systemic velocity, about 2200 \kms, while HI 0432+2944 and HI 0431+2947 have comparable velocities of approximately 1984 and 1973 \kms, respectively.

\begin{table*}
\caption{Properties of the detected \HI\ galaxies}
\label{sources}

\begin{threeparttable}
\footnotesize
\centering

\begin{tabular}{lccccccccccc}
\toprule
Name & Prog. & \HI\ Coords & Opt. Coords & $V_{\HI}$ & $W_{50}$ & $S_{21}$ & $D_{\rm sep}$ & $\log M_{\HI}$ & $\log M_{*}$ & $r$ & $g$ \\
 &  & (J2000) & (J2000) & (km\,s$^{-1}$) & (km\,s$^{-1}$) & (Jy\,km\,s$^{-1}$) & (arcmin/kpc) & ($M_{\odot}$) & ($M_{\odot}$) & (mag) & (mag) \\
\midrule

04296+2923 & D & 043248.39 & 043248.7 & 2128$\pm$3 & 283$\pm$6 & 10.6$\pm$0.6 & -- & 9.36 & 10.11 & 11.48 & 11.91 \\
 & C & +293002.0 & +292958.3 & 2135$\pm$7 & 283$\pm$15 & 4.0$\pm$0.6 & -- & 8.95 &  &  &  \\
 & FAST &  &  & 2113.5$\pm$0.5 & 267$\pm$1 & 9.2$\pm$0.1 & -- & 9.26 &  &  &  \\

\addlinespace

\mbox{HI 0432+2926} & D & 043234.36 & 043233.9 & 2153.7$\pm$0.8 & 115$\pm$2 & 6.0$\pm$0.3 & 4.7/40 & 9.12 & 7.07 & 17.21 & 17.19 \\
 & C & +292606.8 & +292624.7 & 2147$\pm$2 & 105$\pm$4 & 5.5$\pm$0.5 &  & 9.08 &  &  &  \\
 & FAST &  &  & 2145.5$\pm$0.3 & 117.6$\pm$0.5 & 9.2$\pm$0.1 &  & 9.16 &  &  &  \\

\addlinespace

\mbox{HI 0433+2909} & D & 043320.4 & 043320.5 & 2207$\pm$3 & 53$\pm$7 & 1.5$\pm$0.2 & 22.2/187 & 8.55 & 7.48 & 16.89 & 17.04 \\
 & C & +290919.8 & +290911.3 & 2211$\pm$6 & 47$\pm$13 & 1.8$\pm$0.6 &  & 8.63 &  &  &  \\
 & FAST &  &  & 2197.8$\pm$0.5 & 73$\pm$1 & 4.0$\pm$0.1 &  & 8.96 &  &  &  \\

\addlinespace

\mbox{HI 0432+2944} & D & 043206.3 & 043205.5 & 1984$\pm$2 & 78$\pm$5 & 3.7$\pm$0.2 & 17.2/144 & 8.84 & 8.01 & 16.29 & 16.61 \\
 & C & +294415.9 & +294409.8 & 1990$\pm$5 & 68$\pm$13 & 3.6$\pm$0.3 &  & 8.83 &  &  &  \\
 & FAST &  &  & 1980.4$\pm$0.2 & 59.9$\pm$0.5 & 3.79$\pm$0.03 &  & 8.85 &  &  &  \\

\addlinespace

\mbox{HI 0431+2947} & D & 043155.3 & 043155.1 & 1970$\pm$2 & 58$\pm$4 & 4.7$\pm$0.3 & 21/177 & 8.94 & 8.30 & 14.63 & 14.73 \\
 & C & +293002.0 & +292958.3 & 1978$\pm$5 & 90$\pm$11 & 5.4$\pm$0.6 &  & 9.00 &  &  &  \\
 & FAST &  &  & 1971.0$\pm$0.2 & 62.1$\pm$0.5 & 5.54$\pm$0.05 & -- & 9.01 &  &  &  \\

\bottomrule
\end{tabular}

\vskip 0.1cm
\textit{Notes.}
Column (2): Configuration of the VLA array and FAST survey data.
       "D" and "C" correspond to the VLA D and C configurations (projects AM960 and AM985 in Table~\ref{table1}, respectively). 
        Column (3): Dynamical center of the \HI image from the VLA-D project.
        Column (4): Coordinates of the possible optical counterpart from the $r$-band Pan-STARRS image \citep{2016arXiv161205560C}. 
        Columns (5)--(7): Central velocity, full width at half maximum (FWHM), and integrated flux density of the \HI line profile. 
        Column (8): Projected separation between each detected source and IRAS~04296+2923. 
        Column (9): \HI mass derived from the integrated \HI flux listed in Column~(7).
         Column (10): Stellar mass. 
Columns (11) and (12): Extinction-corrected apparent $r$- and $g$-band magnitudes measured from the Pan-STARRS images.
    \end{threeparttable}
\end{table*}

\subsection{Velocity structure analysis of IRAS 04296+2923 and HI 0432+2926}
\label{result.2}
\begin{figure*}[h]
         \centering
         \includegraphics[width=8cm]{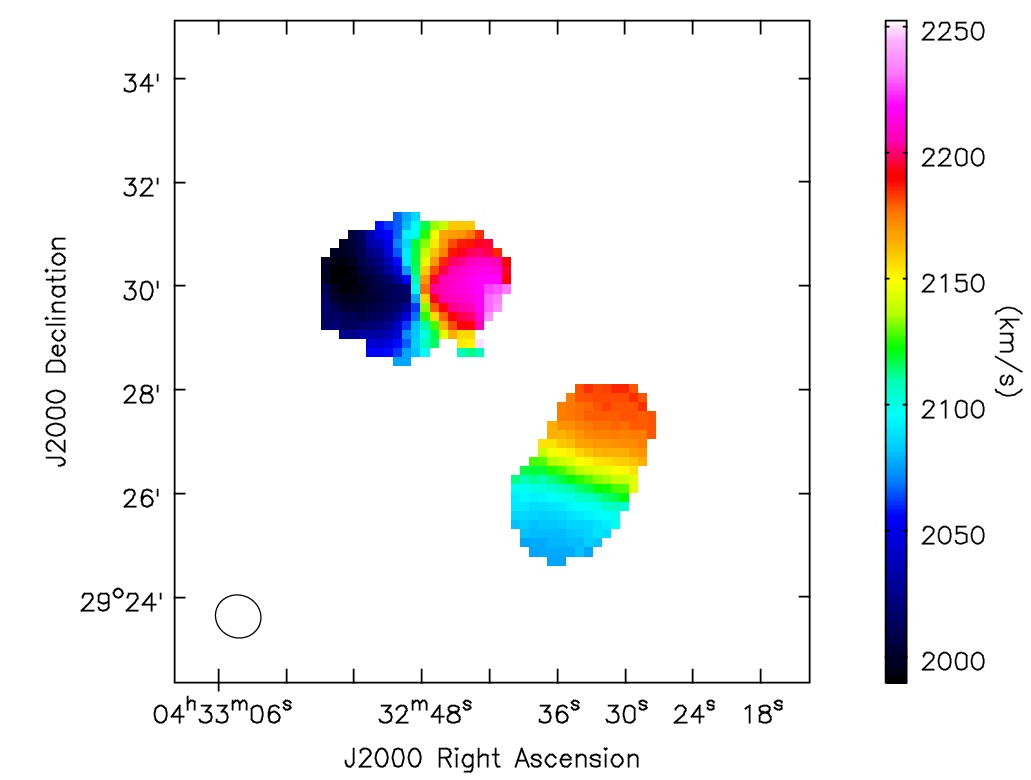}
         \hspace{1cm}
         \includegraphics[width=8cm]{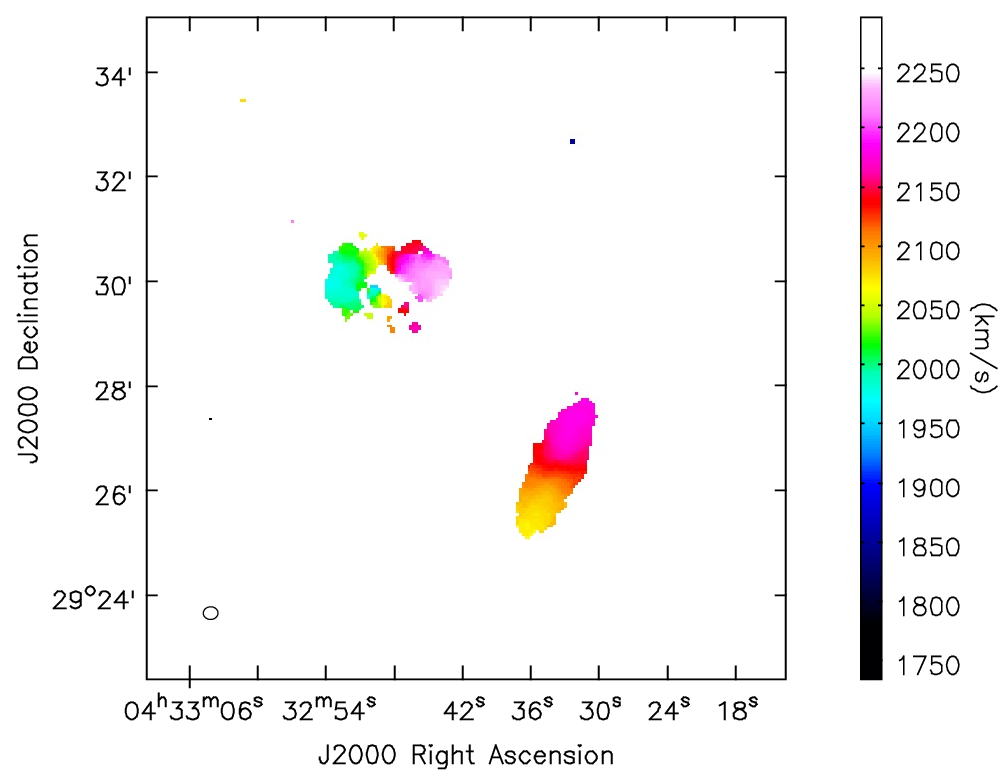}
         \includegraphics[width=9cm]{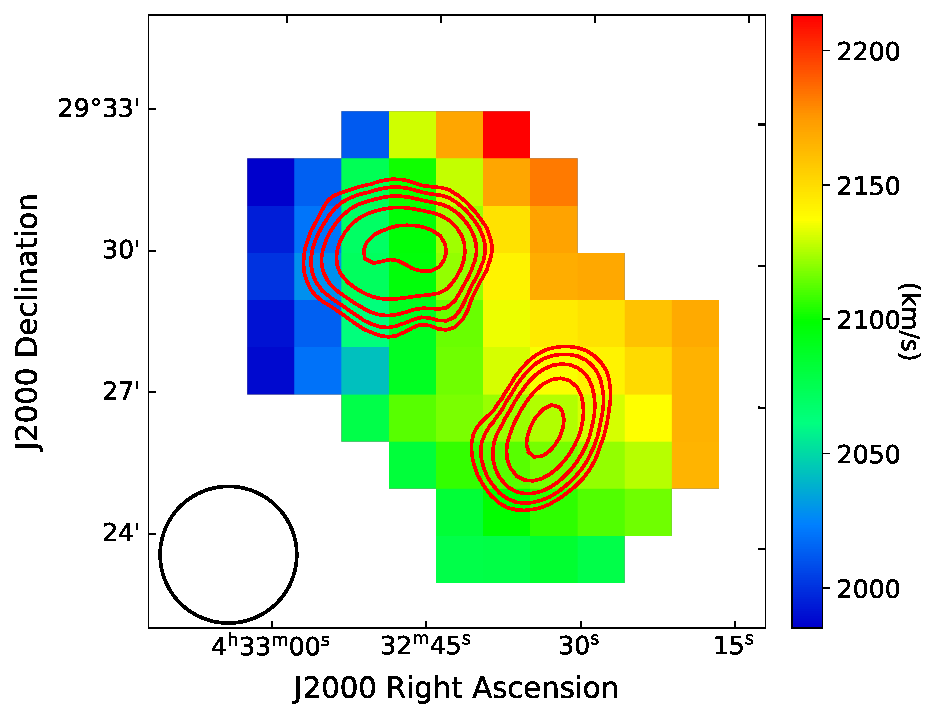}
         \caption{Velocity (moment-1) maps of IRAS~04296+2923 and HI~0432+2926. 
    The upper left and right panels show the velocity fields derived from the VLA-D and VLA-C array data, respectively. 
    The bottom panel presents the velocity map from the FAST \HI\ survey data, with red contours representing the \HI\ emission from the VLA-D project, as shown in Fig.~\ref{fig.1:HIpos}. 
    The synthesized beam of each observation is shown in the lower left corner of each panel.  }
         \label{fig2:v}
     \end{figure*}

\begin{figure*}[h]
         \centering
         \includegraphics[width=8cm]{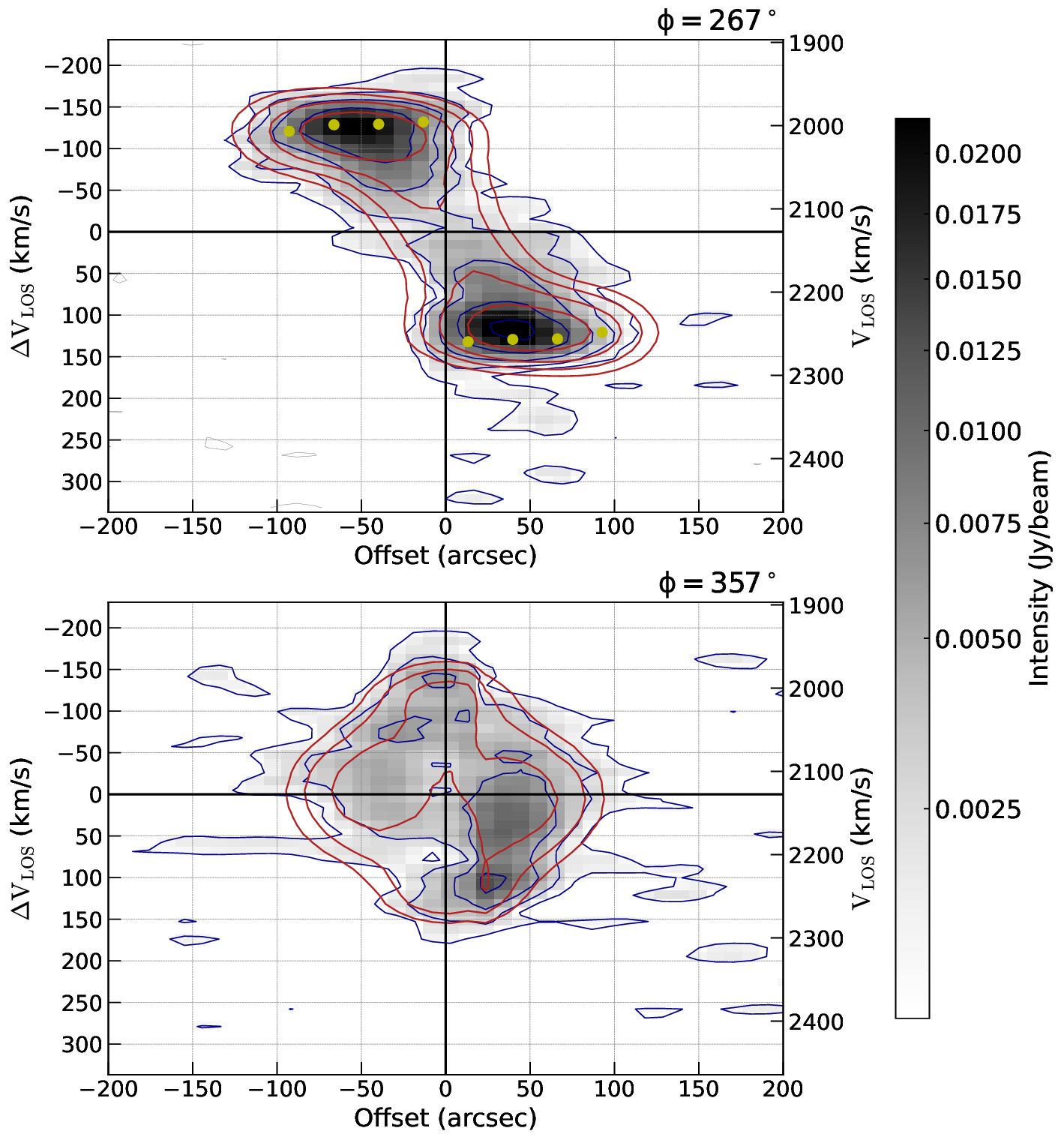} 
         \hspace{1cm}
         \includegraphics[width=8cm]{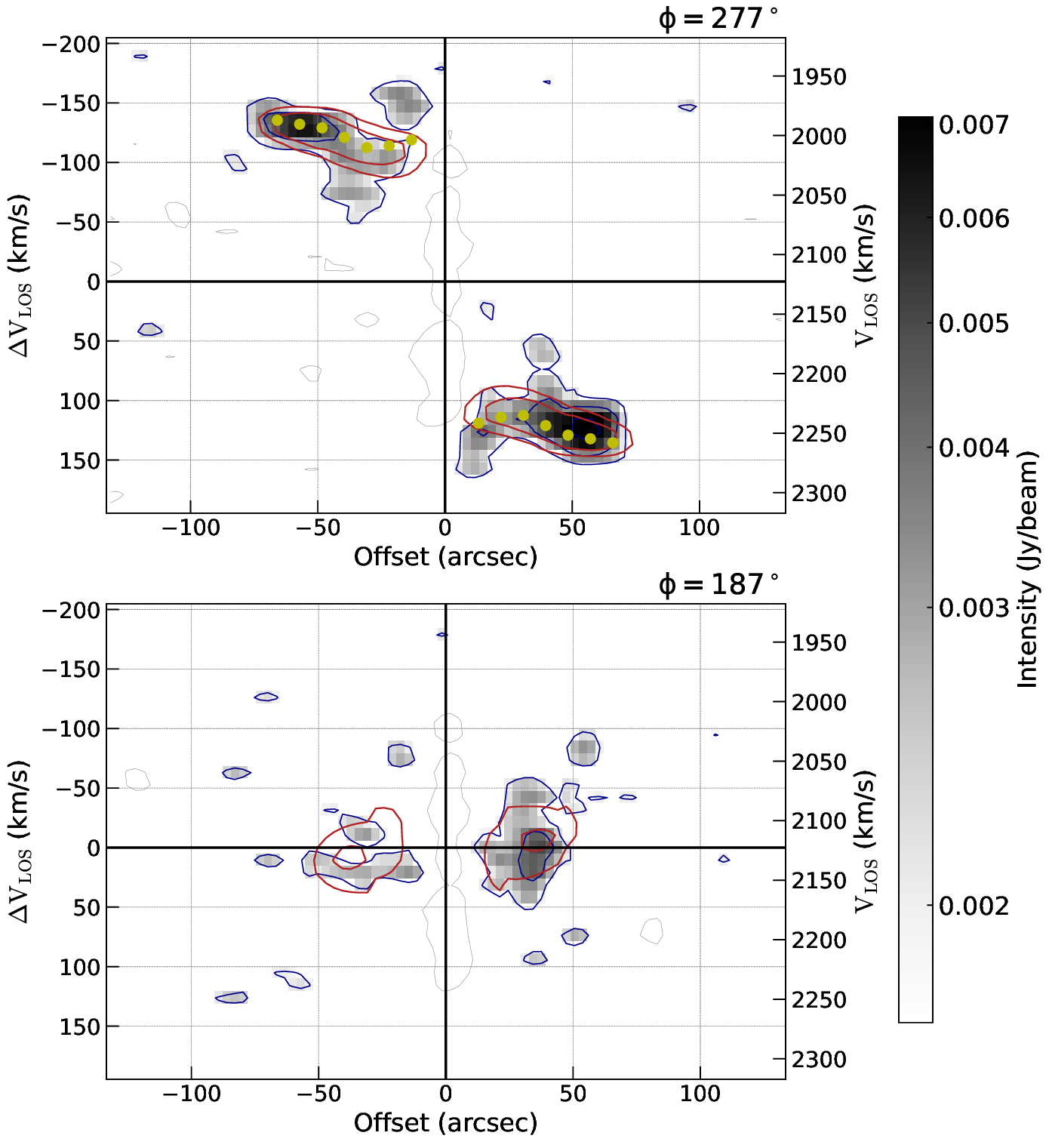}
         \caption{PV diagrams of IRAS 04296+2923 generated from the \HI cube images observed with the VLA-D (left) and VLA-C (right) configurations, using the 3D-BAROLO fitting software.
The blue contours show the observed data, while the red contours indicate the best-fit model results. The yellow dots mark the fitted rotation curve derived from the model.
The angles ($\phi$) in the upper-right corners denote the position angles obtained from the kinematic fitting.
The grey contour at the center of the right panels highlights the region where strong \HI absorption is detected in the C-configuration data.}
    \label{pv2923}
     \end{figure*}

We found that IRAS 04296+2923 and HI 0432+2926 are separated by about 40 kpc. Both galaxies exhibit strong \HI\ emission, and their \HI\ line profiles display double peaks corresponding to the receding and approaching sides of rotating disks. Fig.~\ref{fig2:v} shows the velocity maps generated using CASA from the VLA-C and VLA-D array data. The position–velocity (PV) diagrams of the two galaxies are presented in Figs~\ref{pv2923} and \ref{fig:PV-wise}, respectively. These maps clearly reveal ordered velocity fields consistent with disk rotation in both systems.

The velocity map derived from the FAST \HI\ cube exhibits a velocity gradient of about 200~\kms, running from the southeast to the northwest across both galaxies (see Fig.~\ref{fig2:v}). This is consistent with the VLA-C and VLA-D velocity maps, which show that the western side of IRAS~04296+2923 and the nothern side of HI~0432+2926 have the highest line-of-sight velocities. The velocity dispersion map (Fig~\ref{fig3:dispersed}) indicates a slightly enhanced velocity dispersion in the central region of IRAS~04296+2923, which may be caused by gas turbulence or overlapping motions associated with the stellar bar. This is consistent with the broad \HI\ absorption features seen in the central region from the VLA-C data (Fig~\ref{fig.1:HIline}).

\begin{figure*}
    \centering
    \includegraphics[width=15.0cm]{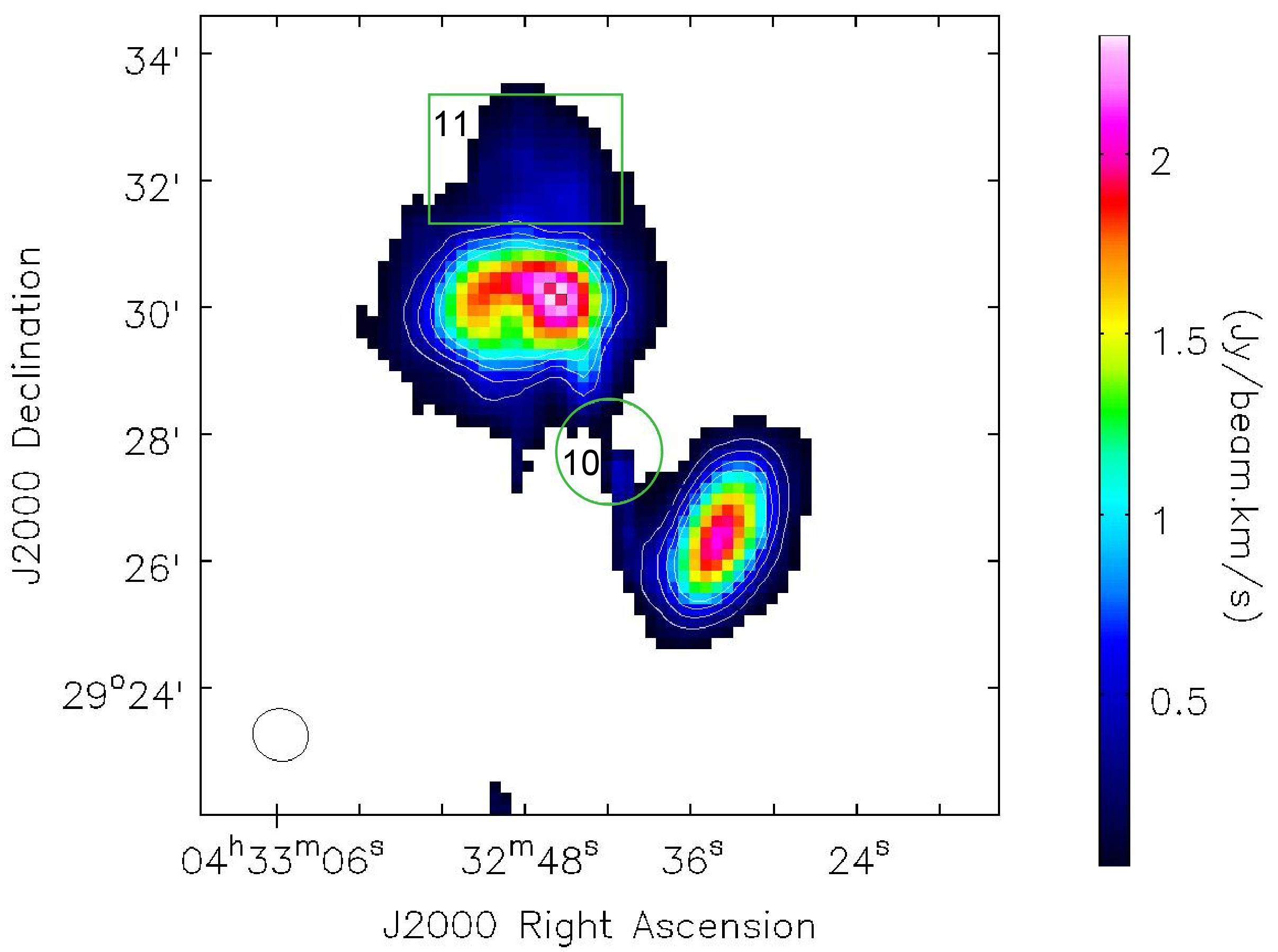}
\caption{
Integrated intensity (moment~0) map of IRAS~04296+2923 (upper left) and \HI~0432+2926 (lower right), derived from the VLA–D array \HI\ data using \textsc{SoFiA}. The white contours show the 3$\sigma$ \HI\ emission from the same VLA–D data, identical to those displayed in Fig.~\ref{fig.1:HIpos}. Region~11 (green box) covers an area of 3\arcmin\ $\times$ 2\arcmin, while Region~10 (green circle) covers 100\arcsec\ $\times$ 100\arcsec. The corresponding \HI\ spectral profiles extracted from these regions are presented in Fig.~\ref{fig:regionline}, and the fitted parameters are summarized in Table~\ref{region}.
\label{Fig:sofia_D1}
}
\end{figure*}

The moment maps generated using SoFiA for the \HI data are presented in Fig.~\ref{Fig:sofia_D1} and \ref{Fig:sofia_D}. These maps incorporate both spatial and velocity smoothing procedures. The identified \HI\ emission region (approximately $4\times5$~arcmin) is comparable to, but slightly larger than, the 3$\sigma$ \HI\ contour region (about $3.5\times3$~arcmin) shown in Fig.~\ref{fig.1:HIpos}. The SoFiA-derived \HI\ emission extends toward the northern part of the system and possibly forms a connecting bridge between the two galaxies. Since the \HI\ emission region identified by SoFiA—due to its smoothing in both spatial and velocity dimensions—differs slightly from the conventional 3$\sigma$ \HI\ region defined per pixel, we further extracted \HI\ line profiles from several specific regions as shown in Fig~\ref{Fig:jiugongge}. The corresponding spectra are displayed in Fig.~\ref{fig:regionline}. Regions~1–9 correspond to IRAS~04296+2923, while Regions~10 and~11 are selected from the SoFiA moment map. Region~10 lies between IRAS~04296+2923 and HI~0432+2926, representing the bridge region that shows weak, redshifted \HI\ emission. Region~11 corresponds to the extended northern region, which exhibits the most redshifted velocities in the velocity map. The fitted parameters of these \HI\ line profiles are summarized in Table~\ref{region}.

The extracted spectra are consistent with the velocity map: the eastern regions (1, 4, and 7) show the lowest velocities ($\sim$2030~\kms), the western regions (3, 6, and 9) exhibit the highest velocities ($\sim$2220~\kms), and the central regions (2, 5, and 8) have intermediate velocities ($\sim$2120~\kms) (see Fig~\ref{fig:regionline} and Table~\ref{region}). Both Regions~10 and~11, identified in the SoFiA map, show weak \HI\ emission. The spectrum of Region~10 in the VLA-D data displays a weak, redshifted line with a central velocity of $\sim$2450~\kms, whereas Region~11 shows a broad, weak \HI\ profile ranging from 1800–2500~\kms, which may indicate turbulent gas motions in this area.

\subsection{The Multi-band Radio Continuum Emission of IRAS 04296+2923}

\citet{2010AJ....140.1294M} suggested that the starburst activity in IRAS~04296+2923 is confined within a compact region of 1\arcsec–2\arcsec\ (150–250 pc) around the nucleus, based on the spatial extent of its radio continuum emission. Radio images with spatial resolutions comparable to that of the VLA A-configuration at L band ($\sim$2\arcsec) therefore provide a reliable measure of the total radio emission from the system.

\begin{figure}[h]
    \centering
    \includegraphics[width=\linewidth]{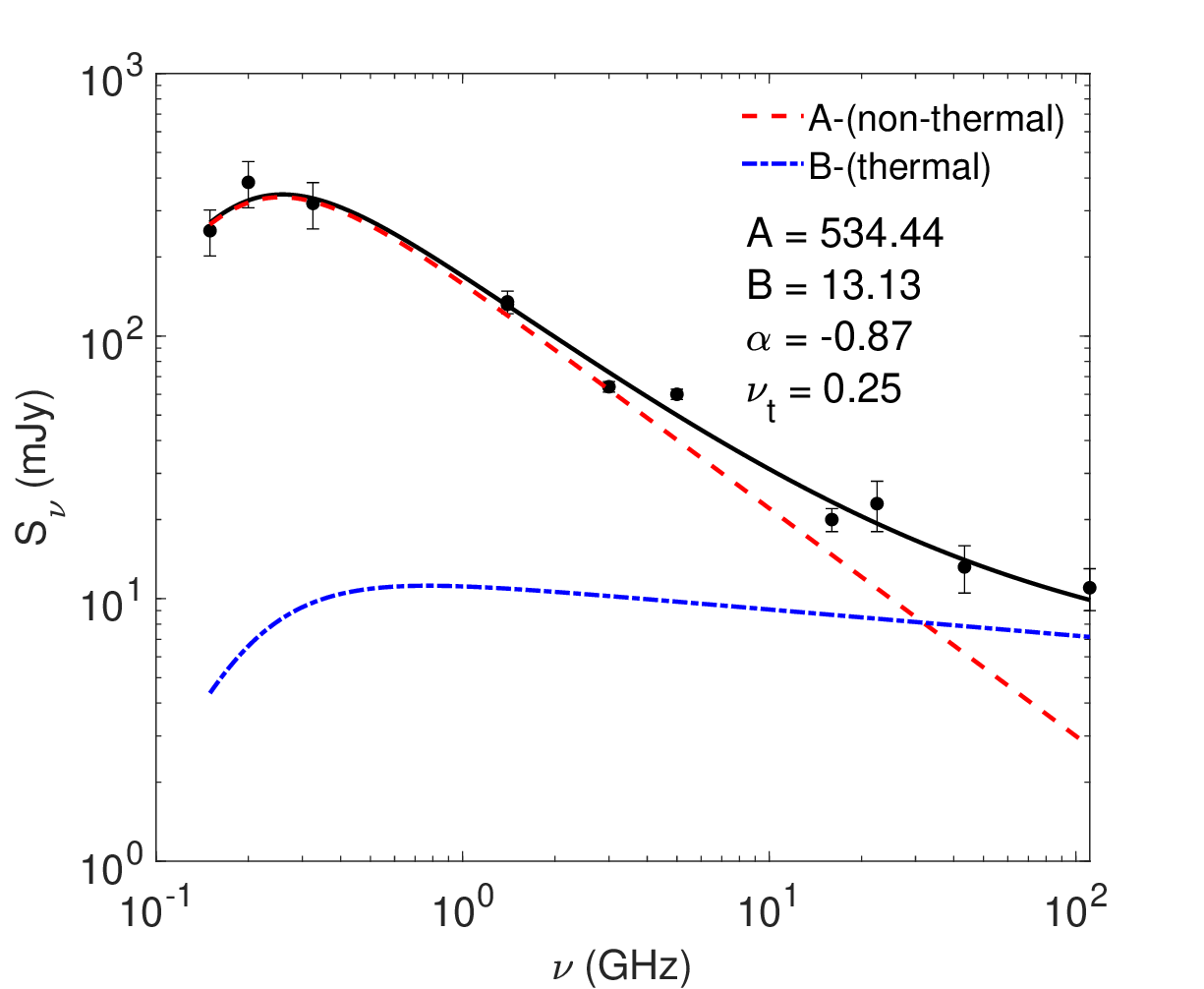}
    \caption{Multi-band integrated flux densities of IRAS 04296+2923 (listed in Table \ref{2923data}). The solid line represents the best-fitting model described by Equation \ref{sedfiting}. The dashed and dot-dashed lines indicate the thermal (free-free) and non-thermal (synchrotron) components, respectively.
}
    \label{fig:spix}
\end{figure}

We present the radio continuum maps extracted from the line-free channels of the VLA C-configuration data. To complement these data, we also compiled archival observations with synthesized beams larger than 2\arcsec\ (see Table~\ref{2923data}) and generated the corresponding continuum images shown in Fig.~\ref{fig:duoboduan}. From these images, we measured both the integrated and peak flux densities for each dataset. The resulting spectral index ($\alpha \sim -0.87$; Fig.~\ref{fig:spix}) is consistent with values commonly found in luminous infrared galaxies (LIRGs), indicative of non-thermal synchrotron emission associated with intense star formation \citep[e.g.][]{2011ApJ...739L..25L}.

\section{Discussion}
\label{sec:discussion}
\subsection{Properties of the IRAS 04296 galaxy group}
\citet{2009AAS...21344503M} suggested that IRAS~04296+2923 may reside in a small galaxy group. To examine this possibility, we searched for nearby galaxies using \HI\ images from multiple VLA projects and the FAST survey. Five \HI-detected galaxies are identified as likely members of the same system.

\subsubsection{Characterizing the galaxies in this group}

We identify optical counterparts for all five galaxies in the IRAS~04296+2923 group. Among them, IRAS~04296+2923 has the largest \HI\ mass as well as the highest stellar mass (see Table~\ref{sources}). In contrast, HI~0432+2926 possesses a relatively high \HI\ mass but the lowest stellar mass in the group, and correspondingly exhibits the faintest apparent magnitudes in both the \textit{g}- and \textit{r}-bands. For most group members, the derived \HI\ and stellar masses broadly follow the empirical $M_{\HI}$--$M_{\ast}$ relation reported in the literature \citep[e.g.,][]{2024MNRAS.529.3469G,2015MNRAS.447.1610M}. A notable exception is HI~0432+2926, which deviates significantly from this relation. Based on its \HI\ mass and extinction-corrected optical luminosity inferred from the \textit{g}- and \textit{r}-band magnitudes, we estimate an unusually high \HI\ mass-to-light ratio of $M_{\HI}/L_{B} \sim 9$. Such an extreme value places HI~0432+2926 within the class of so-called ``almost dark'' galaxies, which are characterized by inefficient or suppressed star formation \citep{2015AJ....149...72C}. The spatial coincidence between the \HI\ emission and the faint optical counterpart indicates that HI~0432+2926 is not a truly ``dark'' galaxy, in which neutral gas fails to reach the critical surface density required to initiate star formation \citep{2026arXiv260112513S}. Nevertheless, deeper optical spectroscopy observations will be required to further constrain the nature of its stellar component and star formation activity.

All five galaxies lie at similar Galactic latitudes ($b \simeq -12^{\circ}$) and have systemic velocities of $\sim$2000~km~s$^{-1}$ derived from their \HI\ profiles, making this system well suited for comparison with nearby galaxy groups in the Local Supercluster. \citet{2009AstBu..64...24M} showed that well-populated groups with more than four members are typically characterized by a velocity dispersion of $\sim$74~km~s$^{-1}$ and a harmonic radius of $\sim$204~kpc. Comparable values have also been reported for galaxy groups and clouds in the local Universe ($z \sim 0.01$) by \citet{2011MNRAS.412.2498M}. Using the systemic velocities listed in Table~\ref{sources}, we estimate a velocity dispersion of $\sim$100~km~s$^{-1}$ for the IRAS~04296+2923 group. The corresponding harmonic radius is $\sim$214~kpc. Both values are consistent with those of typical nearby galaxy groups reported in the literature \citep[e.g.][]{2009AstBu..64...24M,2005AJ....129..178K}.



\subsubsection{\HI view of the large scale environment}
\label{section:4.1.2}
The projected extent of the group is approximately 50~arcmin (corresponding to $\sim$435~kpc; see Fig.~\ref{fig:fast}), fully covered within the VLA field of view. Although the FAST \HI\ data span a much larger area ($\sim$200~arcmin across), no additional \HI\ sources are detected near the group. The nearest other \HI\ galaxy \text{HI 0430+2846}, located at (RA = 04:30:10.5, Dec = 28:46:29.1), with a system velocity $V_{c}$=2086$\pm$19 \kms (see Fig. \ref{fig:another sources}) lies 106~arcmin away ($\sim$922~kpc from IRAS~04296+2923). Given this large separation, it is unlikely to be a bound member of the group. 

As shown in Fig.~\ref{fig:fast} and Table~\ref{sources}, the two northwestern members (HI~0431+2947 and HI~0432+2944) have the lowest systemic velocities ($\sim$1970–1990~km~s$^{-1}$), while the southeastern galaxy (HI~0433+2909) exhibits the highest value ($\sim$2210~km~s$^{-1}$). The two central galaxies (IRAS~04296+2923 and HI~0432+2926) lie between these extremes and possess the largest \HI\ masses. This spatial–kinematic distribution suggests that the central pair may constitute the gravitational core of the group, with the three outer galaxies orbiting around them. Assuming a simple virial relation, $M = RV^{2}/G$, where $R$ is the projected separation and $V$ is the corresponding velocity offset, we estimate the total mass of the central pair using the positions and velocities of the three outer members (Table~\ref{sources}). The inferred masses are $2.7\times10^{11}M_\odot$ (from HI~0433+2909), $7\times10^{11}M_\odot$ (from HI~0432+2944), and $1\times10^{12}M_\odot$ (from HI~0431+2947). Because the orbital inclinations of these satellites are unknown, these estimates represent lower limits, implying a group mass exceeding $1\times10^{12}M_\odot$. This is roughly 300 times the combined \HI\ mass of the two central galaxies ($3.6\times10^{9}M_\odot$; Table~\ref{sources}).

Using the rotation curves fitted with 3D-Barolo (Fig.~\ref{fig.nihe}), we obtain dynamical masses of $8\times10^{10}M_\odot$ within 18~kpc for IRAS~04296+2923 and $2.9\times10^{10}M_\odot$ within 14~kpc for HI~0432+2926, yielding a combined mass of $\sim$$1.1\times10^{11}M_\odot$. This accounts for only $\sim$11\% of the total mass derived from the outer members, implying that at least 89\% of the mass in this system must be in the form of an extended dark matter halo far beyond the observed \HI\ disks. This is broadly consistent with the commonly accepted picture that dark matter contributes 80–90\% of the total mass in galaxies and dominates at large radii beyond the stellar and gaseous components \citep{2020ApJ...905...28H,1986RSPTA.320..447V,2018ARA&A..56..435W}. Nevertheless, the possibility of undetected massive group members not visible in \HI\ cannot be ruled out.

\subsection{\HI\ Kinematic Structure and Modeling}
\label{4.2}

The \HI\ velocity structure of IRAS~04296+2923 and its companion \HI~0432+2926 were investigated using VLA D- and C-array observations, supplemented by the FAST \HI\ survey (see Section \ref{result.2}). To quantify the gas dynamics, we further modeled the three-dimensional \HI\ data cubes with \textsc{3D-Barolo} (Fig.~\ref{fig.model}). For both galaxies, the models reproduce the main kinematic structures, as reflected by the small residuals in intensity and velocity. In IRAS~04296+2923, the residual maps show localized deviations that may arise from non-circular motions, turbulent gas, or beam-smearing effects. The intensity residuals reveal a systematic north–south asymmetry, with enhanced \HI\ emission on the northern side of the disk. Given that the \HI\ 21~cm line is optically thin, this asymmetry does not reflect radiative transfer or obscuration effects, but instead indicates a genuine asymmetry in the \HI\ column density distribution. The northern side of the galaxy corresponds to the near side of the disk \citep{2010AJ....140.1294M}, and the observed enhancement is plausibly associated with bar-driven gas dynamics. In particular, gas compression along the downstream sides of bar-induced streaming flows can locally increase the \HI\ surface density, while the opposite side may be characterized by more diffuse gas or mild kinematic dilution in the upstream regions. In contrast, \HI~0432+2926 shows a more symmetric gas distribution and regular velocity field, with low velocity dispersions ($\sigma \lesssim 30$ km s$^{-1}$) and weak residuals, consistent with a dynamically settled rotating disk.

The fitted rotation curves and systemic velocities are shown in Fig.~\ref{fig.nihe}. For IRAS~04296+2923, the rotation curve flattens at $V_{\rm rot}\sim 180$–$200$ km s$^{-1}$ beyond $\sim 20''$, consistent with a massive, rotation-supported disk. The residual velocity field shows small-scale asymmetries suggestive of mild non-circular motions related to gravitational interaction. In contrast, \HI~0432+2926 shows a gradually rising rotation curve that levels off at $V_{\rm rot}\sim70$–$100$ km s$^{-1}$, typical of a lower-mass, less evolved system. The systemic velocities remain stable across rings in both galaxies, confirming well-constrained kinematic centers. The CO(1–0) position–velocity (PV) diagram of \citet{2010AJ....140.1294M} (their Fig.~9) shows a nearly solid-body rise in the inner 15", reaching $\gtrsim 100$ km s$^{-1}$ within 5–10". This steep gradient, well reproduced by their bar model, indicates strong bar-driven non-circular motions in the molecular gas. PV slices extracted along the kinematic major and minor axes using \textsc{3D-Barolo} are shown in Fig.~\ref{pv2923} for IRAS~04296+2923 and Fig.~\ref{fig:PV-wise} for \HI~0432+2926. The \HI\ PV slices of IRAS~04296+2923 show a shallower velocity gradient and only mild S-shaped distortions within $\pm20$–$30''$, consistent with beam smearing and the intrinsic absence of strong non-circular motions in the atomic gas. By contrast, \HI~0432+2926 shows a narrow velocity width and no clear non-circular signatures, emphasizing its dynamically regular nature. Together, these differences highlight that IRAS~04296+2923 is currently experiencing the earliest kinematic effects of tidal interaction within the system. A comparison of the HI and CO PV diagrams further reveals that the HI disk extends to offsets of $\sim150''$, whereas the CO emission is confined to the central $\sim20''$. This indicates that a compact molecular core is embedded within a much more extended atomic disk. The \HI kinematics, together with the steep central CO gradient, are consistent with bar-driven inflows that may be feeding nuclear star formation.


Large-scale, diffuse \HI\ emission is detected in both galaxies, especially in the FAST and VLA-D data (see Figs.~\ref{fig.1:HIpos}, \ref{fig2:v}, and \ref{fig:fast}). The \HI\ disks extend several times beyond the optical extents, indicating that both systems are extremely gas rich. The diffuse \HI\ emission in regions 10 and 11 (Fig.~\ref{Fig:sofia_D1}, ~\ref{fig:regionline}) may trace gas stirred by the ongoing tidal interaction. The FAST image further suggests that the westernmost \HI\ of IRAS~04296+2923 and the northernmost \HI\ of \HI~0432+2926 may already be in contact. Although this region is not detected in the VLA-D map, deeper observations are required to confirm the physical connection. Overall, the combination of high-resolution VLA imaging and wide-field FAST mapping shows that IRAS~04296+2923 is a massive, rotation-supported disk undergoing mild dynamical disturbance, while \HI~0432+2926 is less massive and more kinematically regular.


\subsection{Comparsion with other LIRGs at similar merging stages}
\subsubsection{comparsion with ngc 253}
\cite{2014ApJ...795..107M} investigated the properties of dense molecular gas in the inner disk and found a strong morphological similarity between NGC 253 and IRAS 04296+2923. Both are strongly barred spiral galaxies hosting nuclear starbursts that are likely triggered by bar-driven gas inflows. The gas properties in IRAS 04296+2923 are comparable to those in the well-studied starburst galaxy NGC 253.

\begin{figure}
    \centering

    \includegraphics[width=9cm]{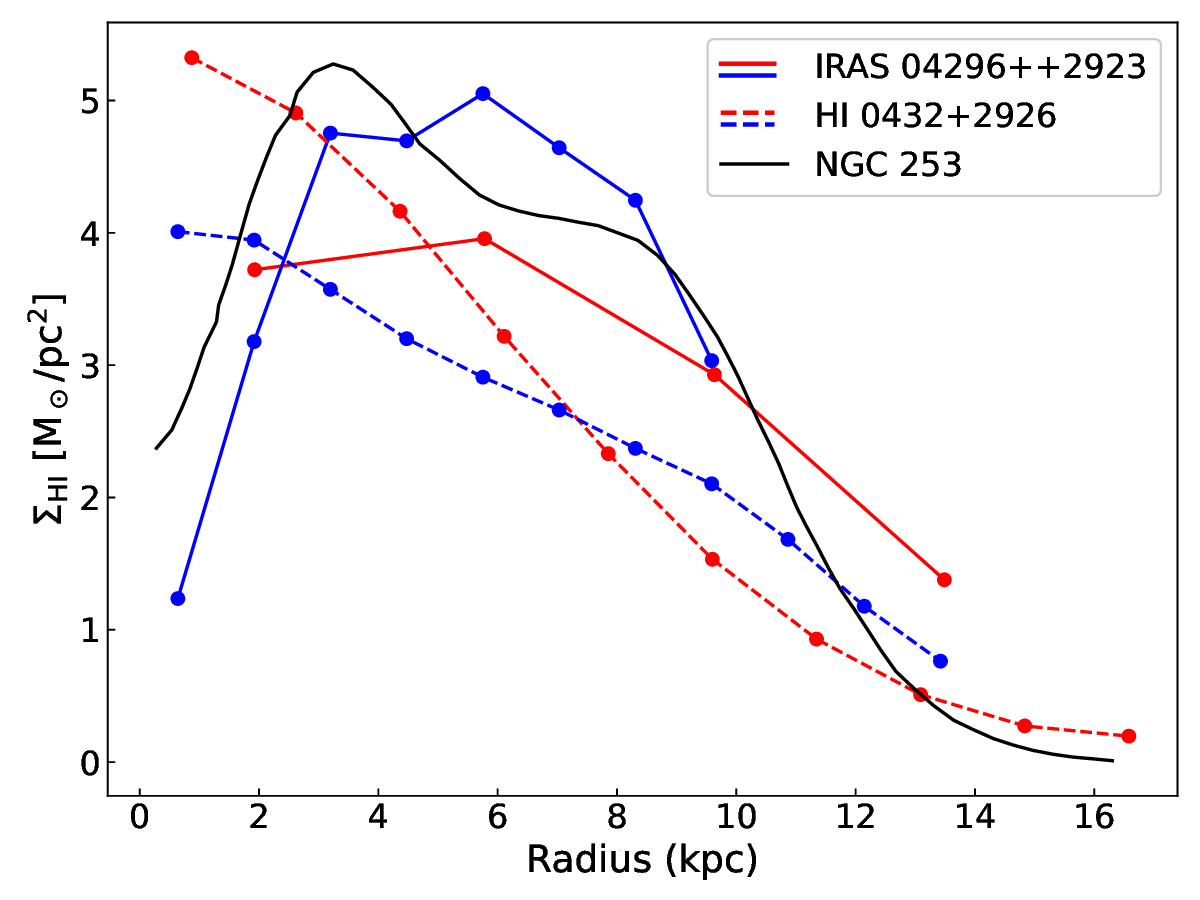}
    \caption{
Radial \HI\ surface density profiles of IRAS~04296+2923 and \HI~0432+2926. 
For each galaxy, the profiles derived from the VLA–D and VLA–C configurations are shown in red and blue, respectively, with solid lines representing IRAS~04296+2923 and dashed lines representing \HI~0432+2926. 
For comparison, the \HI\ radial profile of NGC~253 from the KAT-7 observations \citep{2015MNRAS.450.3935L} is plotted as the black solid line.
}
    \label{fig:radiprofile}
\end{figure}

\cite{2015MNRAS.450.3935L} presented \HI\ observations of NGC 253 obtained with the SKA precursor KAT-7. We also compared the \HI\ properties of IRAS 04296+2923 and NGC 253. The total \HI\ mass derived from the VLA-D data for IRAS 04296+2923 is about $2.3\times10^9M_{\odot}$(see Table \ref{sources}), which is similar to that of NGC 253 \citep[$\sim2.1\times10^9M_{\odot}$, see][]{2015MNRAS.450.3935L}. Using the inclination-corrected \HI\ surface densities from the 3D-BAROLO modeling, the radial profile of IRAS~04296+2923 (see Fig.~\ref{fig:radiprofile}) shows a clear mid-disk peak similar to that observed in the starburst galaxy NGC~253. Such a feature is typical of massive, actively star-forming spiral disks and is consistent with the \HI\ distributions reported for normal spirals \citep[e.g.][]{2016A&A...585A..99M}. In contrast, \HI~0432+2926 displays a smoothly declining \HI\ profile without a mid-disk enhancement, a form characteristic of low-mass late-type galaxies that generally lack the ring-like \HI\ structure seen in more massive systems \citep{2002A&A...390..829S}. This contrast suggests different gas distributions and possibly different evolutionary or mass regimes between the two galaxies.

\citet{2005A&A...431...65B} reported that the \HI\ disk of NGC 253 is relatively compact compared to its deep optical extent. In contrast, IRAS 04296+2923 exhibits a comparable \HI\ size but a more extended disk relative to its optical radius. This suggests that IRAS 04296+2923 retains a larger neutral gas reservoir in its outer disk. Combined with the unusual CO isotopic ratios and other molecular gas diagnostics in the inner disk \citep{2014ApJ...795..107M}, these similarities in \HI\ structure and total gas content support the interpretation that IRAS 04296+2923 and NGC 253 represent highly similar systems, with the starburst in IRAS 04296+2923 occurring at an earlier evolutionary phase, before significant gas depletion or feedback has altered the disk properties as extensively as in NGC 253.

\subsubsection{Comparison of HI line profile concentration with M1 and M2 merger-stage LIRGs}
\cite{2010AJ....139.2066F} showed that a large fraction of IRAS galaxies exhibit distorted \HI\ spectral features, suggesting that many of them are interacting or merging systems. \cite{2020ApJ...898..102Y} defined the concentration parameter of the \HI\ line profile as $C_{V} = V_{85}/V_{25}$, where $V_{85}$ and $V_{25}$ represent the velocity widths enclosing 85\% and 25\% of the total integrated flux, respectively. \cite{2022ApJ...929...15Z} further demonstrated that this parameter effectively distinguishes mergers from non-mergers based on their \HI\ line profiles, whereas neither the line width nor the asymmetry parameters alone provide clear merger diagnostics.

Following the method of \cite{2022ApJ...929...15Z}, we calculated the $C_{V}$ values for the M1 and M2 merger-stage samples defined in \cite{2016ApJ...825..128L}. As shown in Fig.~\ref{fig:Cv}, the $C_{V}$ values of sources in the M1 merging phase are significantly smaller than those in the M2 phase. The mean $C_{V}$ is 3.98 for M1 systems and 5.35 for M2 systems, indicating that the \HI\ line profiles of M2 galaxies are more concentrated (tending toward a single-peaked shape) compared to those in the M1 phase.

For comparison, the $C_{V}$ values of IRAS 04296+2923 and \HI\ 0432+2926 are 2.88 and 3.35, respectively. These values are even smaller than those typically found in M1 systems and are instead comparable to the $C_{V}$ values of non-merging galaxies reported by \cite{2022ApJ...929...15Z}. This suggests that the interaction between IRAS 04296+2923 and \HI\ 0432+2926 has not significantly affected the \HI\ kinematics or overall line profiles of the two galaxies. This interpretation is consistent with our VLA imaging results presented in Section~\ref{4.2}, which show no strong morphological or kinematic disturbances in their \HI\ distributions.

\subsection{Implications for Star Formation Activity in IRAS 04296+2923}

Multi-band radio continuum observations detect emission only from IRAS~04296+2923, while HI~0432+2926 and other nearby group members remain undetected (see Fig. \ref{fig:duoboduan} and \ref{fig:vlassandvlac}). The radio continuum spectrum of IRAS~04296+2923 (Fig.~\ref{fig:spix}) is well described by a steep non-thermal synchrotron component ($\alpha \approx -0.87$) plus a thermal free–free contribution that becomes increasingly important above $\sim$30 GHz. The SED fitting follows the model
\begin{equation}
\label{sedfiting}
S_\nu = \left( 1 - e^{-\tau(\nu)} \right)
\Bigg[ B + A \left( \frac{\nu}{\nu_t} \right)^{0.1 + \alpha} \Bigg]
\left( \frac{\nu}{\nu_t} \right)^2,
\end{equation}
where $A$ and $B$ are the synchrotron and free–free normalization constants, and the optical depth is parameterized as $\tau(\nu) = (\nu/\nu_t)^{-2.1}$ \citep{2018MNRAS.474..779G}. The derived turnover frequency, $\nu_t \approx 0.25$ GHz, indicates strong free–free absorption in dense ionized gas, consistent with a compact, obscured nuclear starburst.

The radio SED of IRAS~04296+2923 is notably similar to that of the central starburst region in NGC~253 \citep{2017ApJ...838...68K}, which also shows a turnover near 0.23,GHz. The physical scales of the central starburst regions in both galaxies are likewise comparable, each confined within $\lesssim$500 pc \citep{2017ApJ...838...68K,2010AJ....140.1294M}. However, after accounting for distance, the 1.4 GHz radio luminosity of IRAS~04296+2923 is approximately five times higher than that of the NGC~253 nuclear starburst (based on 140 mJy and 2 Jy flux densities, respectively; \citealt{2017ApJ...838...68K}). IRAS~04296+2923 also exhibits a higher IR luminosity \citep[log $L_{\rm IR}\approx10.97L_\odot$,][]{2003AJ....126.1607S} compared to NGC~253 \citep[log $L_{\rm IR}\approx10.44L_\odot$,][]{2003AJ....126.1607S}. The elevated radio and IR luminosities of IRAS~04296+2923 may reflect a more gas-rich central environment, even though several physical conditions of their nuclear regions appear broadly similar \citep[see][]{2014ApJ...795..107M}.

\cite{2002A&A...392..377B} and \cite{2008A&A...477...95C} showed that very young starbursts often display an excess of FIR emission relative to radio emission, because thermal free–free processes dominate before the onset of strong supernova-driven synchrotron radiation. To examine this effect, we compute the FIR–radio flux ratio \citep{1985ApJ...298L...7H}:
\begin{equation}
q = \log_{10} \left( \frac{\mathrm{FIR}}{3.75 \times 10^{12}} \cdot \frac{1}{S_\nu} \right),
\end{equation}
where $S_{\nu}$ is the radio flux density and the FIR flux is estimated as:
\begin{equation}
\mathrm{FIR} = 1.26 \times 10^{-14} \left( 2.58, S_{60\mu{\rm m}} + S_{100\mu{\rm m}} \right)\ {\rm Wm^{-2}}.
\end{equation}
IRAS~04296+2923 has IRAS flux densities of 42.13 Jy at 60 $\mu$m and 48.27 Jy at 100 $\mu$m \citep{2003AJ....126.1607S}. Using the 8.4 GHz flux density extracted from the SED (34.92 mJy), we obtain $q_{8.4} = 3.2$. This value is significantly higher than the mean $q_{8.4}$ found for (U)LIRGs \citep[e.g.][]{2008A&A...477...95C,2006A&A...449..559B}, indicating that the radio emission is relatively weak for its FIR luminosity. Such a high $q$ implies a substantial thermal fraction and is likely consistent with IRAS~04296+2923 hosting a young nuclear starburst.


Neutral hydrogen (\HI) is a sensitive tracer of galaxy--galaxy interactions. Disturbed \HI\ morphologies, as well as asymmetric or broadened line profiles, are commonly associated with tidal interactions or external gas accretion \citep{1994Natur.372..530Y,2005MNRAS.358..202K}, whereas an ordered \HI\ velocity field accompanied by symmetric double-horn profiles typically indicates a dynamically settled rotating disk. The galaxy pair IRAS~04296+2923 and HI~0432+2926 exhibits a small line-of-sight velocity difference of $\Delta v = 26$~km~s$^{-1}$ and a projected separation of 40~kpc. Dynamical masses derived from 3D-Barolo rotation curve fits are $8 \times 10^{10}$~M$_\odot$ within 18~kpc for IRAS~04296+2923 and $2.9 \times 10^{10}$~M$_\odot$ within 14~kpc for HI~0432+2926 (see Section~\ref{section:4.1.2}). The low relative velocity, when considered together with the inferred dynamical masses and separation, is consistent with the system being gravitationally bound. Within the interaction-stage classification framework of \citet{2016ApJ...825..128L}, the combination of projected separation and velocity offset places this system in a regime broadly consistent with an early interaction stage (M1), corresponding to the initial approach prior to first pericentric passage. However, high-resolution infrared, optical, and radio imaging \citep{2010AJ....140.1294M,2014ApJ...795..107M}, together with the \HI\ data presented in this work, reveal no evidence for prominent tidal tails, significant velocity-field distortions, or other morphological signatures indicative of a strong interaction. The rotation curves of both galaxies remain regular and well fitted, consistent with largely undisturbed disk kinematics. Taken together, these kinematic and morphological properties favor the interpretation that IRAS~04296+2923 and HI~0432+2926 constitute a bound, orbiting galaxy pair rather than an actively merging system. A rough, order-of-magnitude estimate of the orbital timescale, based on Kepler’s third law and assuming a bound two-body configuration, yields a timescale of $\sim$1--2~Gyr, implying that tidal perturbations are currently weak. The luminous infrared galaxy (LIRG) nature of IRAS~04296+2923, with its elevated star formation rate, is therefore more plausibly driven by internal processes, such as bar-induced gas inflow \citep[e.g.,][]{2010AJ....140.1294M}, possibly modulated by long-timescale, low-level tidal forcing from the companion, rather than by an advanced merger phase, a scenario commonly observed in nearby galaxy groups \citep{2007ggnu.conf..123C}.

While a very early pre-merger configuration cannot be entirely excluded, the observed properties impose significant constraints on such a scenario. At the current projected separation of 40~kpc, the combination of the inferred galaxy masses and the small line-of-sight velocity difference ($\Delta v = 26$~km~s$^{-1}$) would require either a nearly face-on orbital geometry, in which the orbital plane lies close to the plane of the sky, or a highly eccentric orbit in which the galaxies spend most of their time near apocenter, where relative velocities are naturally low. In the former case, projection effects can significantly reduce the apparent surface-brightness contrast and morphological distinctiveness of tidal features, making them more difficult to detect observationally \citep[e.g.,][]{2013LNP...861..327D}. In the latter case, strong tidal perturbations would occur only briefly during pericentric passage, further limiting the visibility of tidal signatures at most orbital phases. An additional important constraint arises from the group environment: IRAS~04296+2923 and HI~0432+2926 reside in a small galaxy group that includes at least three lower-mass members, as revealed by our \HI\ observations. The presence of multiple apparently stable satellite galaxies suggests a dynamically mature, low-velocity-dispersion group potential, in which long-lived bound orbits are common and merger timescales are substantially prolonged.
Nevertheless, given the absence of detectable tidal disturbances in both the \HI\ morphology and kinematics presented in this work, together with the group-scale dynamical context, we favor an interpretation in which IRAS~04296+2923 and HI~0432+2926 form a long-period, weakly interacting bound pair.


\section{summary}
\label{sec:conclusions}
We have analyzed archival VLA and FAST \HI 21 cm data, together with archival multi-band radio continuum observations, to investigate the neutral gas distribution and star-forming activity in the nearby luminous infrared galaxy IRAS 04296+2923. The \HI maps reveal that IRAS 04296+2923 and its companion \HI 0432+2926 are members of a small group of five galaxies. The two main galaxies have comparable total \HI masses and a projected separation of $\sim$40 kpc. Both systems exhibit regular velocity fields and symmetric double-horn line profiles, typical of rotation-dominated disks. Apart from minor outer-disk asymmetries, the \HI morphology and kinematics show no convincing evidence for significant tidal disturbance, suggesting that the system remains dynamically regular.

Radio continuum emission is detected only from IRAS 04296+2923 and is confined to the nuclear region. The compact morphology and spatial extent of the emission are consistent with the 150–250 pc starburst region reported by \citet{2010AJ....140.1294M} and \citet{2014ApJ...795..107M}, and our analysis supports their interpretation of a compact, centrally concentrated starburst. The broadband radio spectrum exhibits a significant free–free contribution at high frequencies and a high FIR-to-radio ratio ($q_{8.4}$$\sim$3.2 ), both indicative of a young, dust-enshrouded nuclear starburst.

The coexistence of a dynamically regular large-scale \HI disk and a compact nuclear starburst, together with the presence of a stellar bar, strongly suggests that bar-driven inflow is the dominant mechanism funneling gas into the nucleus and triggering the current episode of star formation. Although the nearby gas-rich companion may exert a weak tidal influence, the absence of prominent large-scale tidal features or kinematic distortions implies that external interaction plays only a minor role at this stage. Based on the regular \HI\ morphology and kinematics, we conclude that IRAS~04296+2923 and HI~0432+2926 are best described as a long-period, weakly interacting bound pair. The embedding of this pair in a small galaxy group with additional low-mass members provides a natural dynamical context for such a configuration.

\begin{acknowledgements}
We thank the anonymous referee for his/her useful comments and suggestions on the manuscript.
This work is supported by the grants of NSFC (Grant No.12363001) and Guizhou Provincial Major Scientific and Technological Program XKBF (Grant No. (2025)010 and (2025)011.) This work makes use of archival data from the Very Large Array (VLA) and the the FAST All Sky HI survey (FASHI) project. The National Radio Astronomy Observatory is a facility of the National Science Foundation operated under cooperative agreement by Associated Universities, Inc. FAST is a Chinese national mega-science facility, op-
erated by the National Astronomical Observatories of Chinese Academy of Sciences (NAOC).
\end{acknowledgements}

\begin{appendix}

\section{Online materials}
\twocolumn

\begin{figure*}[h]
    \centering
     \includegraphics[width=6cm]{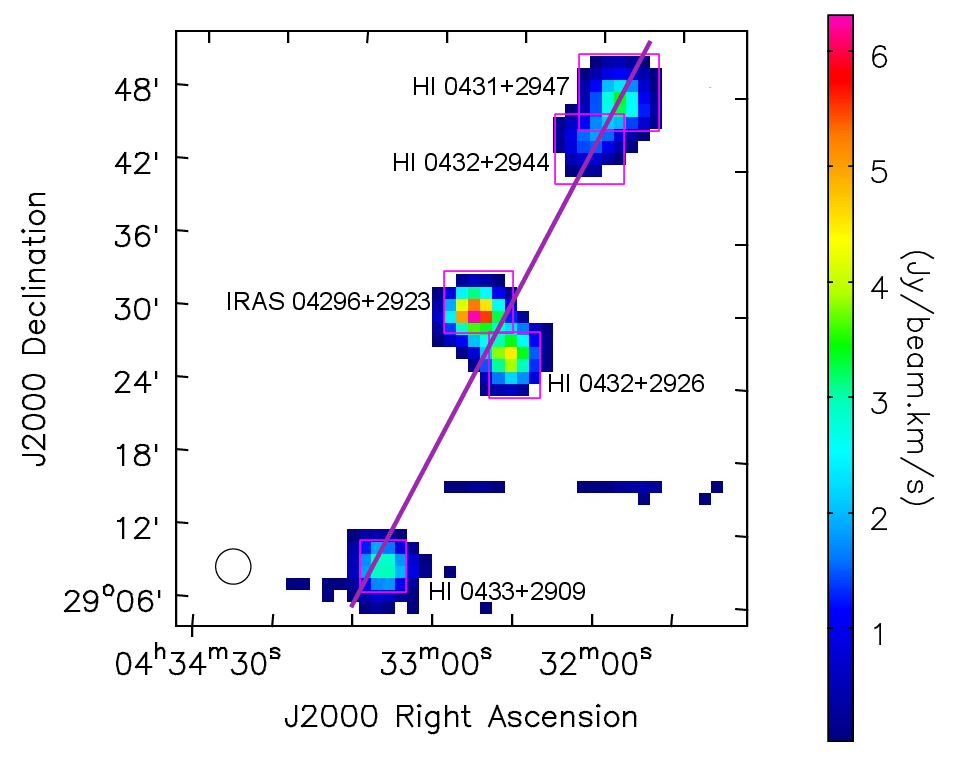}
     \includegraphics[width=6cm]{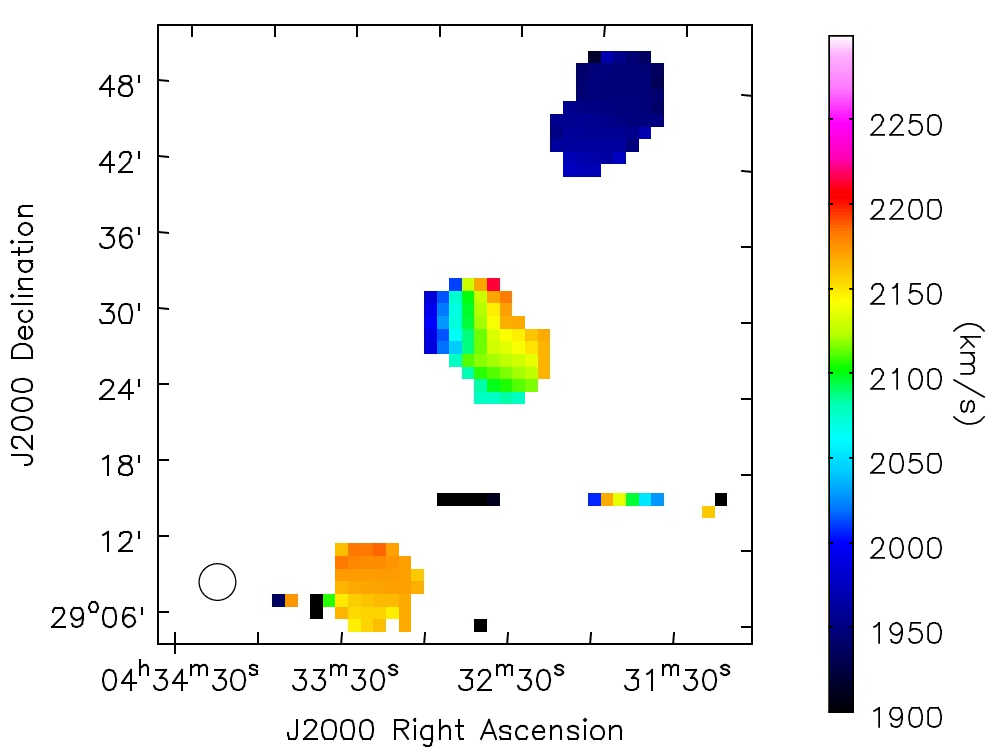}
     \includegraphics[width=6cm,height=7cm]{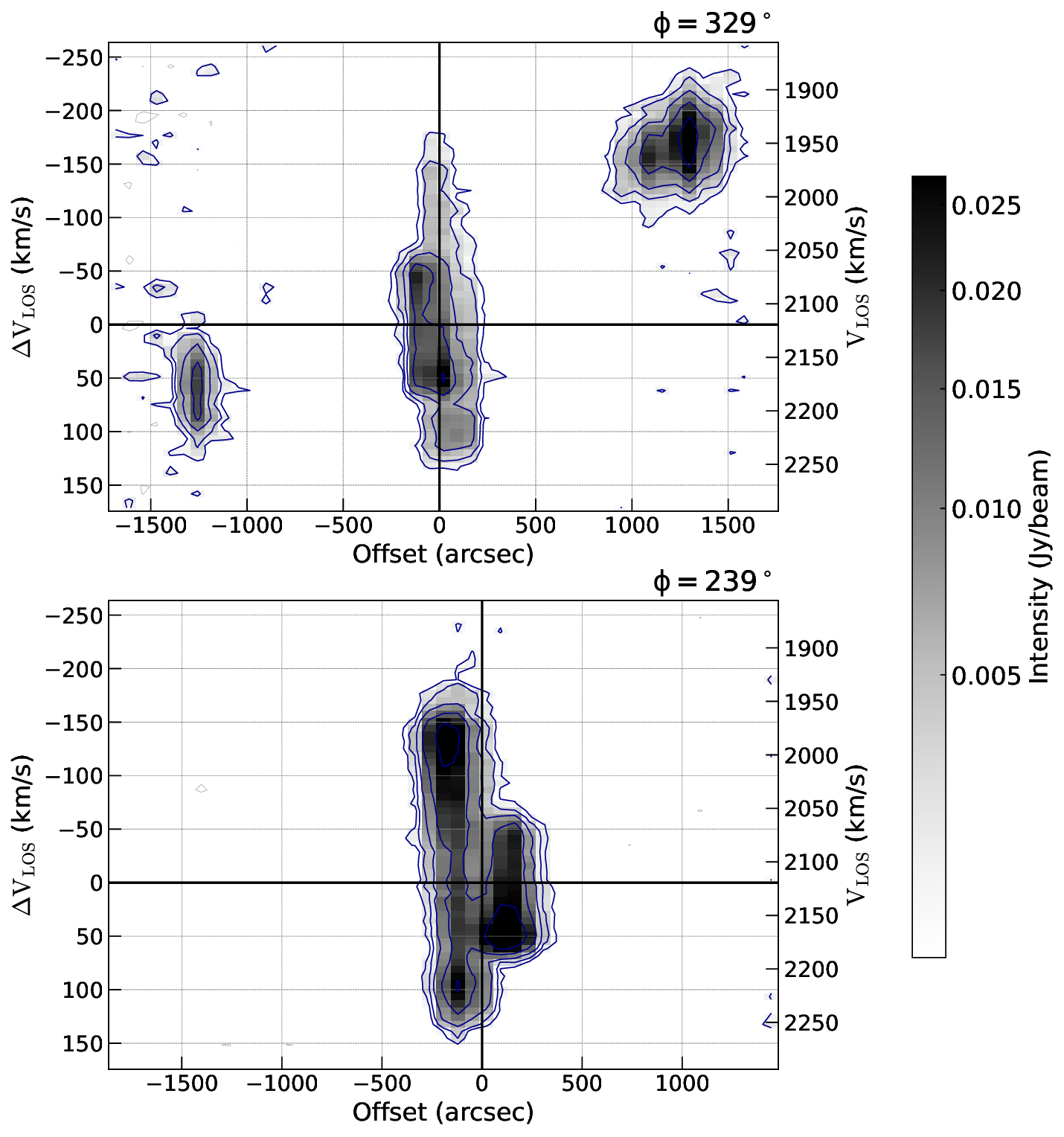}    
\caption{\HI\ moment~0 (integrated intensity) and moment~1 (velocity field) maps, together with the position--velocity (PV) diagrams derived from the FAST \HI\ cube. 
The purple regions in the left panels mark the selected areas used to extract the \HI\ line profiles of each source: 
HI~0431+2947, HI~0432+2944, IRAS~04296+2923, HI~0432+2926, and HI~0433+2909. 
Due to the limited spatial resolution of FAST, HI~0431+2947 and HI~0432+2944 partially overlap. 
The PV diagrams on the right were obtained using \textsc{BBarolo} along the purple solid line (center: 04$^{\mathrm h}$32$^{\mathrm m}$41.3$^{\mathrm s}$, +29$^\circ$27$'$55.2$''$, 
PA = 328$^\circ$, length = 50~arcmin).}
    \label{fig:fast}
\end{figure*}

\begin{figure*}
    \centering
    \includegraphics[width=8cm]{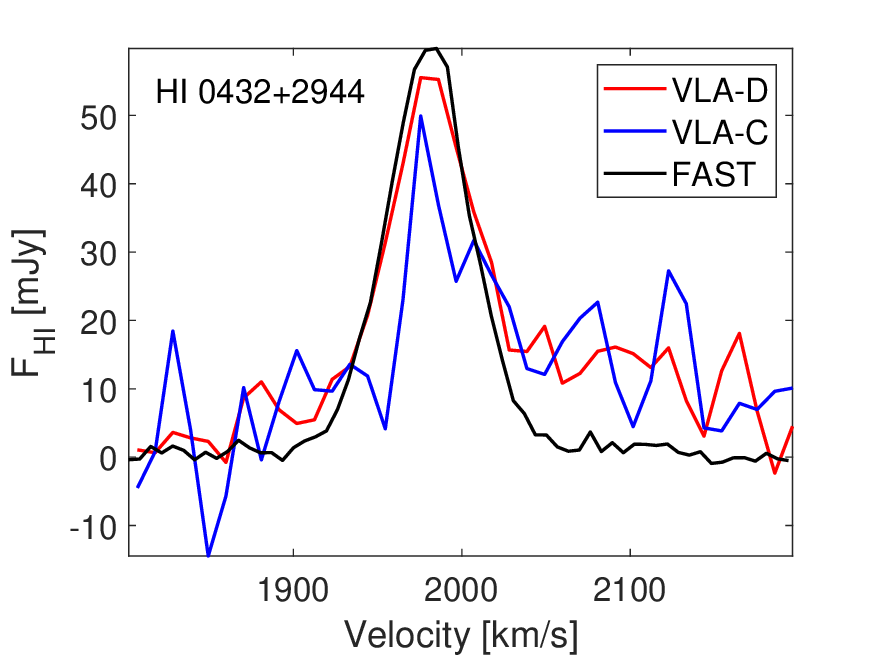}
    \includegraphics[width=8cm]{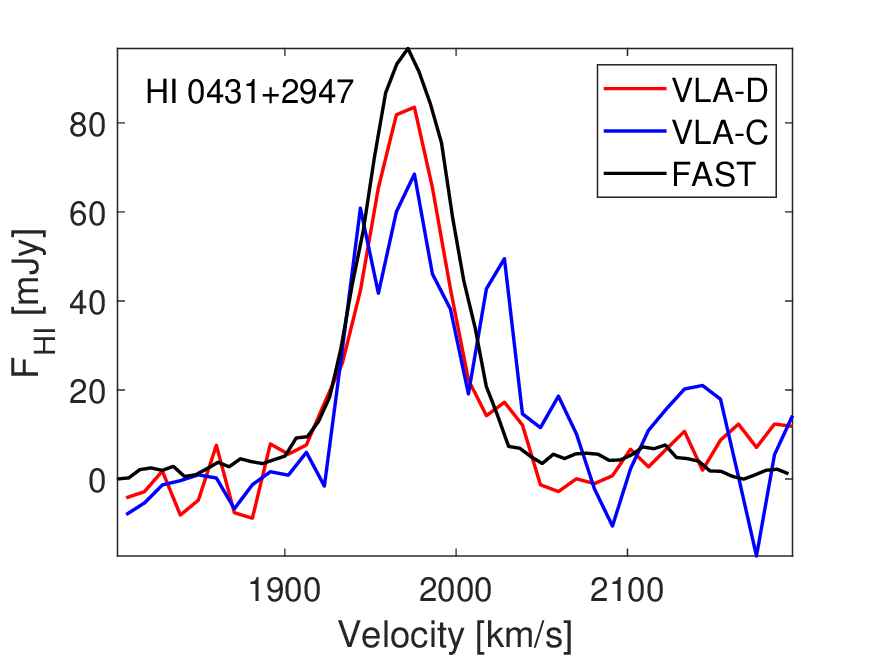}
    \includegraphics[width=8cm]{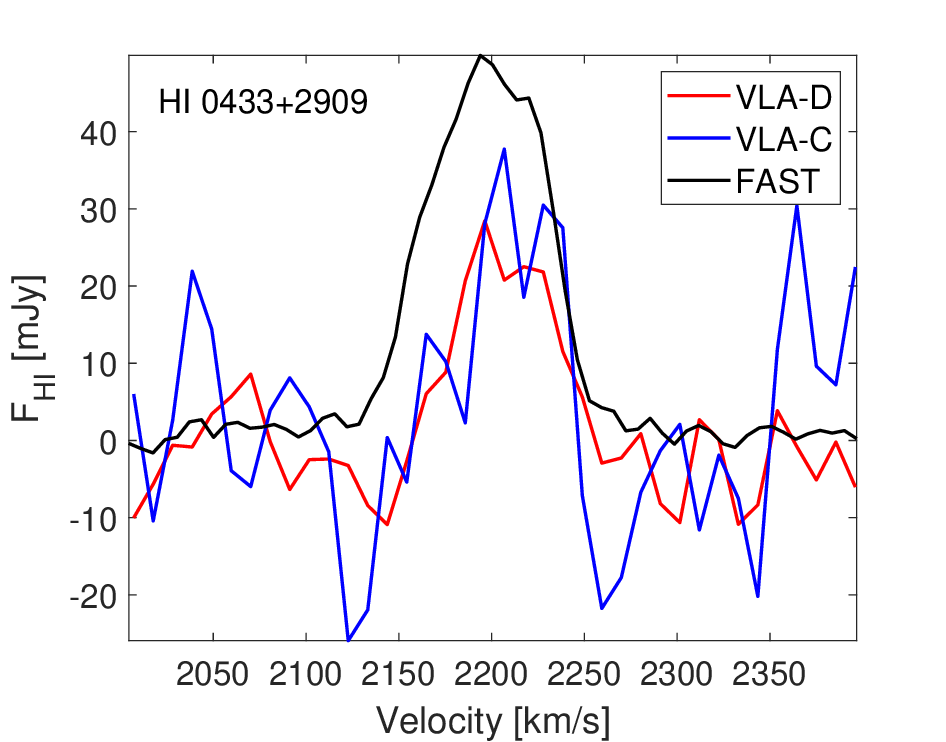}
     \includegraphics[width=8cm]{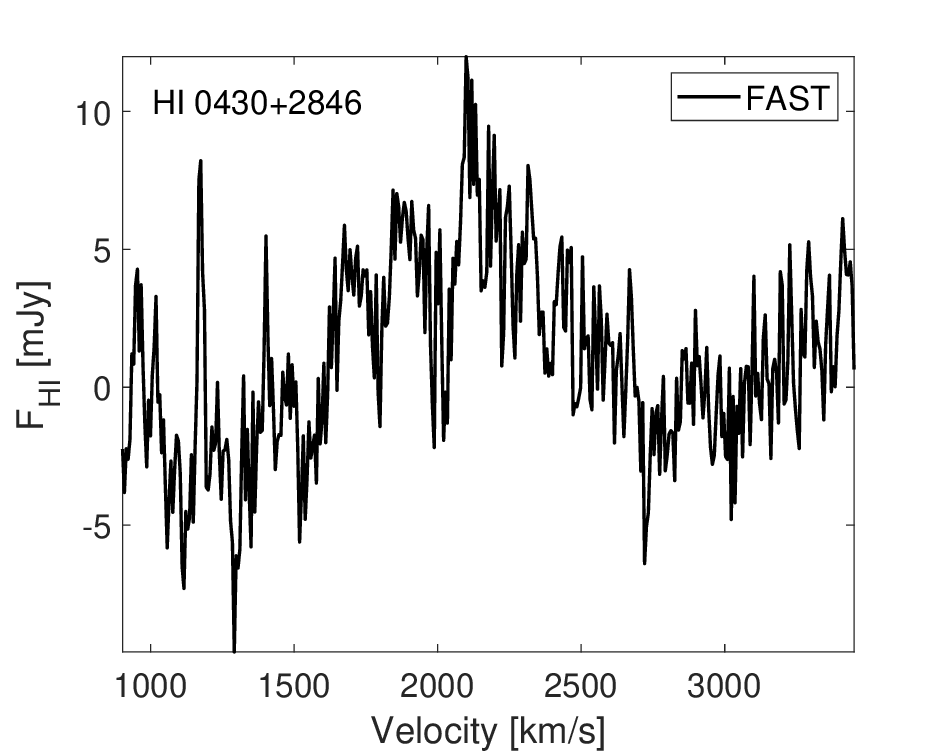}
    \caption{  \HI line profiles of \HI 0431+2947, \HI 0432+2944, \HI 0433+2909 and \HI 0430+2846, obtained from the VLA-C, VLA-D, and FAST observations.
    }
    \label{fig:another sources}
\end{figure*}


    \begin{figure*}[h]
         \centering
         \includegraphics[width=9cm]{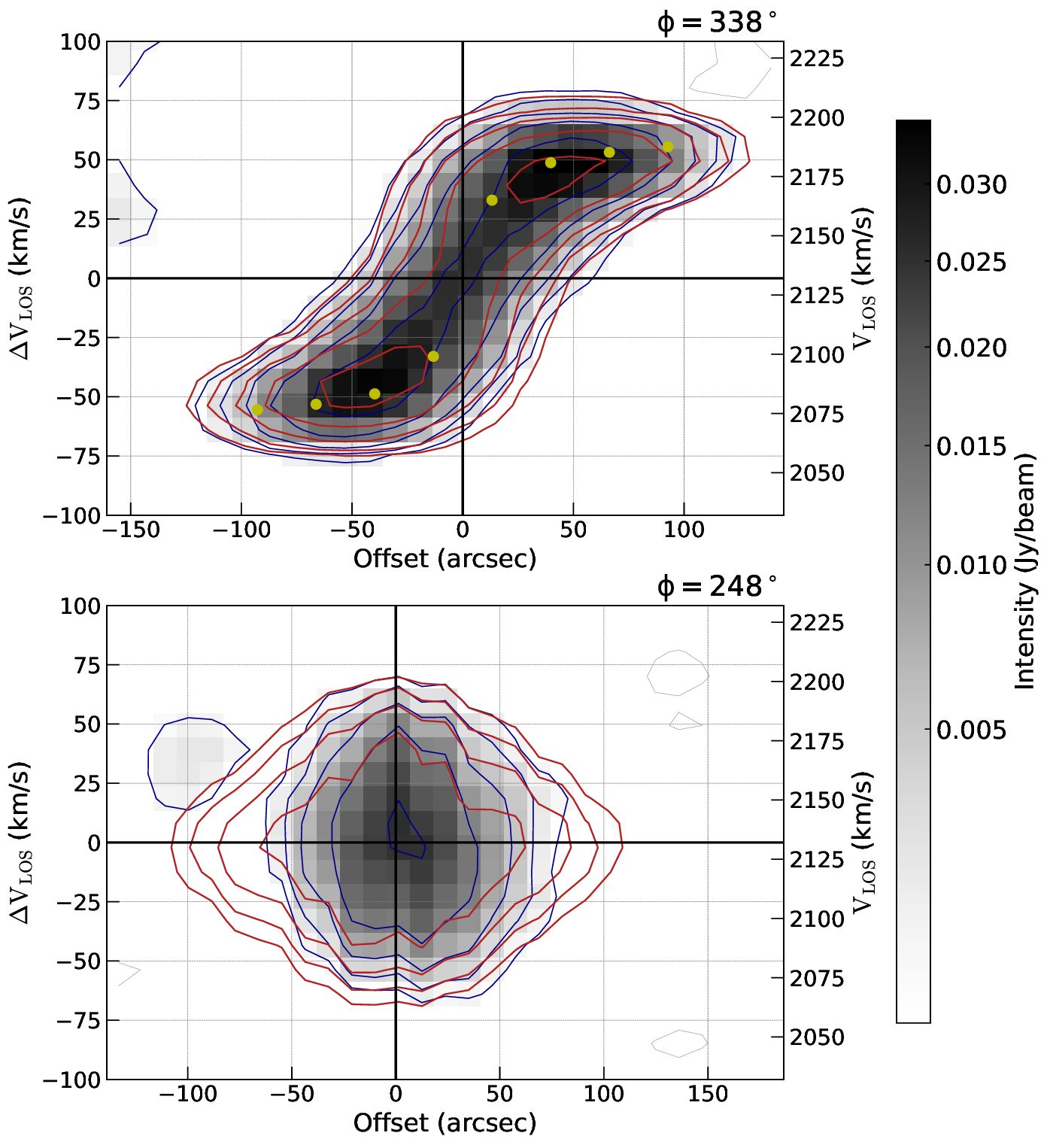}
         \includegraphics[width=9cm]{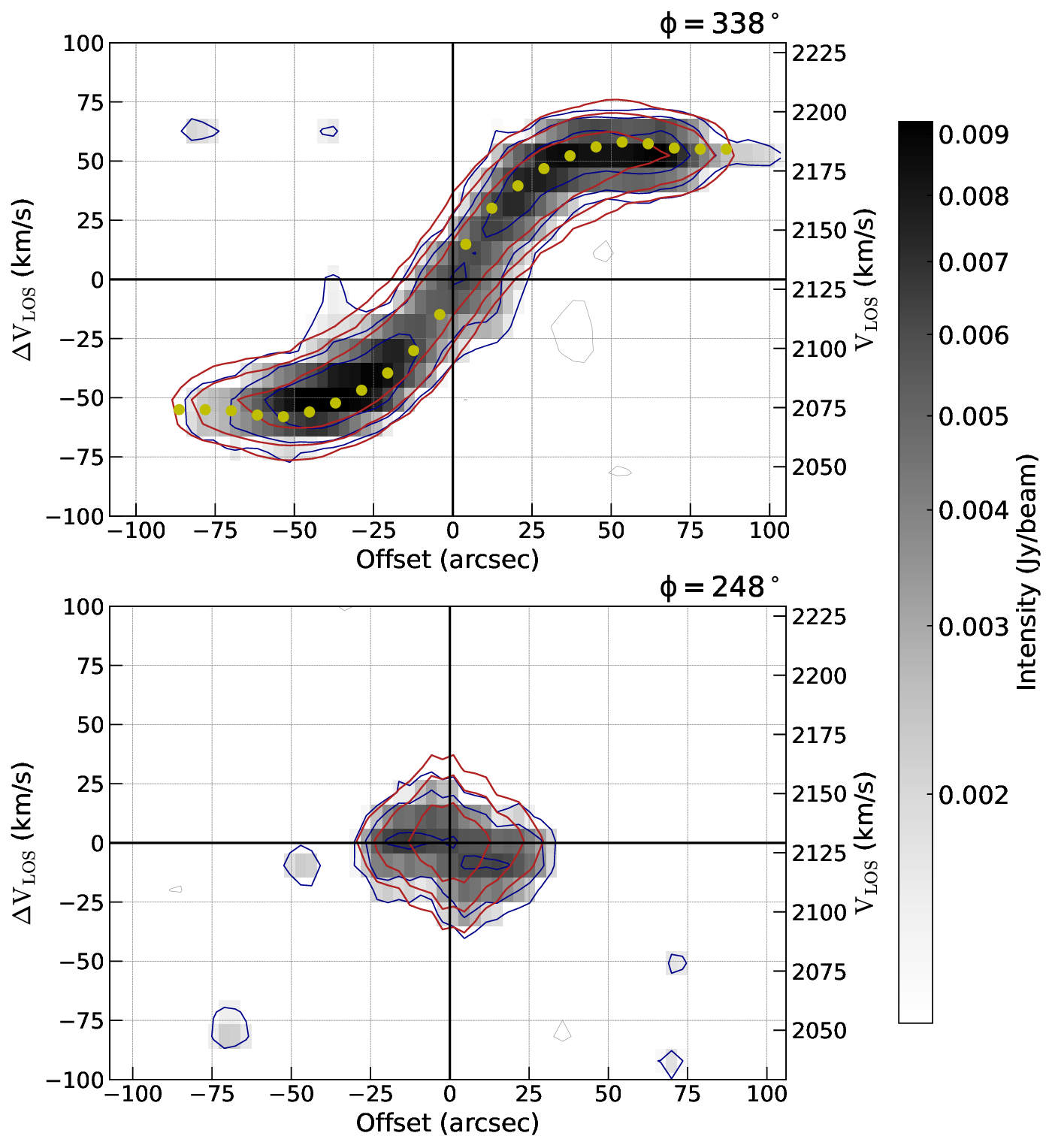}
         \caption{PV diagrams of HI 0432+2926 generated from the \HI cube images observed with the VLA-D (left) and VLA-C (right) configurations,
using the 3D-BAROLO fitting software. 
The blue contours show the observed data, while the red contours indicate the best-fit model results. The
yellow dots mark the fitted rotation curve derived from the model. The angles ($\phi$) in the upper-right corners denote the position angles obtained
from the kinematic fitting.}
         \label{fig:PV-wise}
     \end{figure*}

  \begin{figure*}
       \centering
       \includegraphics[width=8cm]{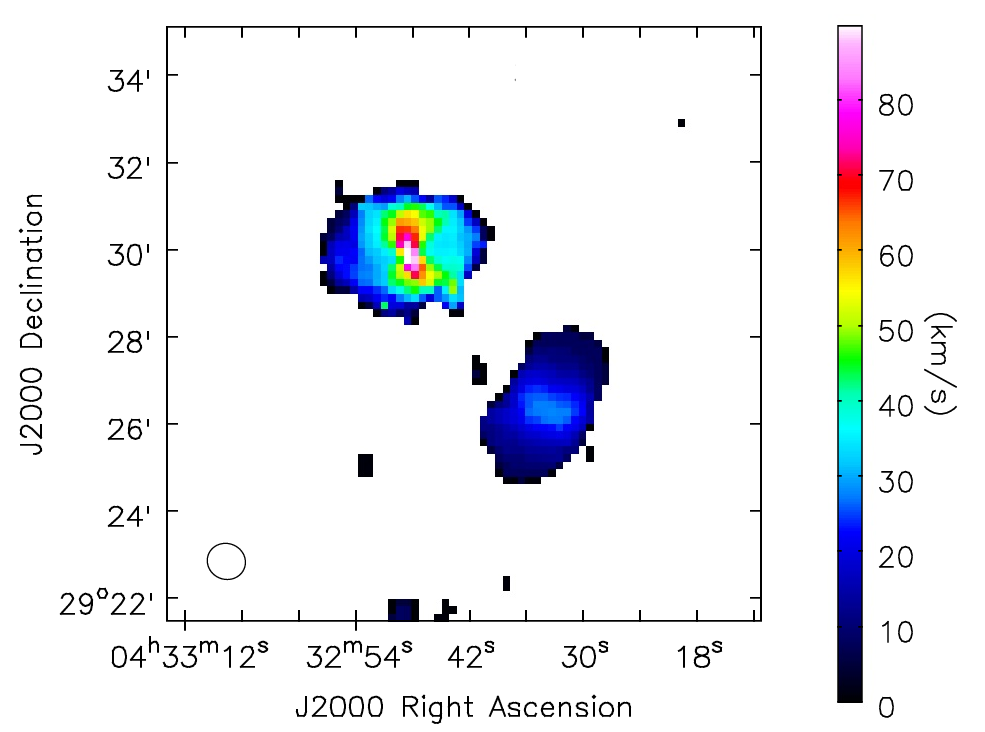}
       \hspace{1cm}
       \includegraphics[width=8cm]{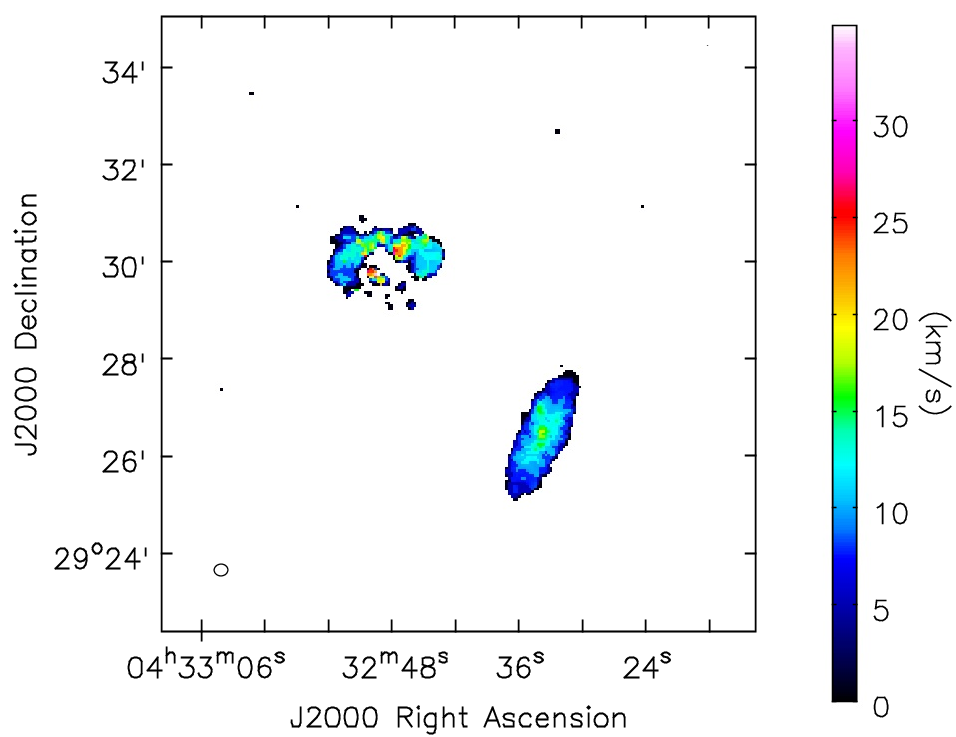}
       \caption{Velocity dispersion (moment~2) maps of IRAS~04296+2923 and HI~0432+2926. 
The left panel shows the map derived from the VLA-D configuration data, while the right panel shows that from the VLA-C configuration data.}
       \label{fig3:dispersed}
   \end{figure*}

\begin{figure*}
    \centering
    \includegraphics[width=8cm]{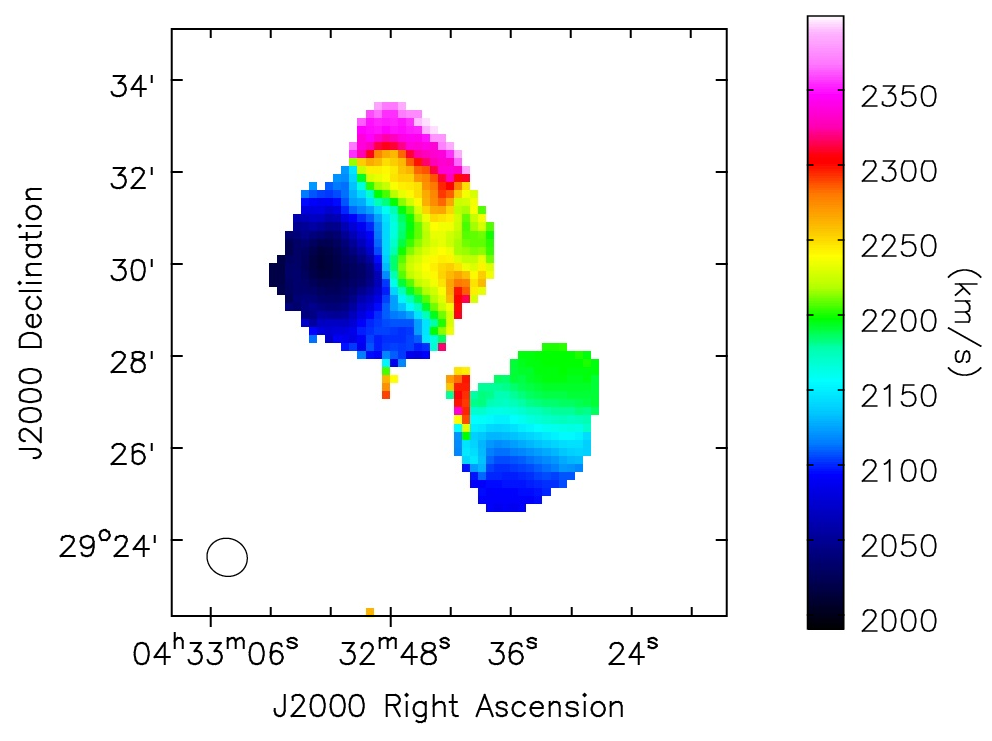}
    \includegraphics[width=8cm]{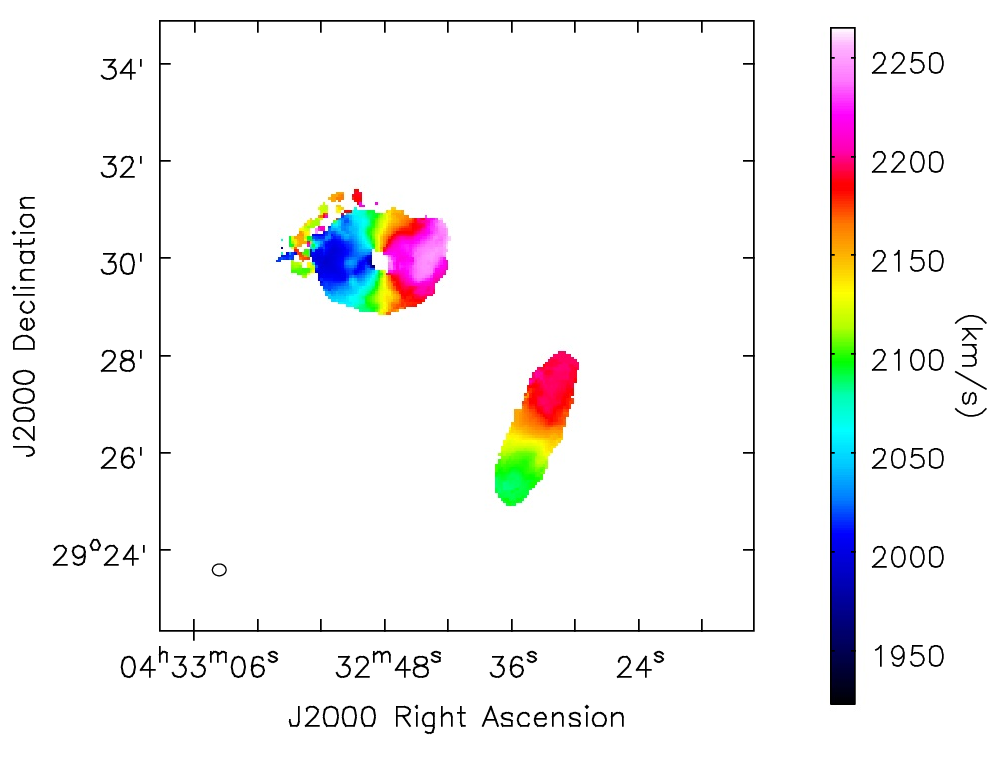}
    \includegraphics[width=7.8cm]{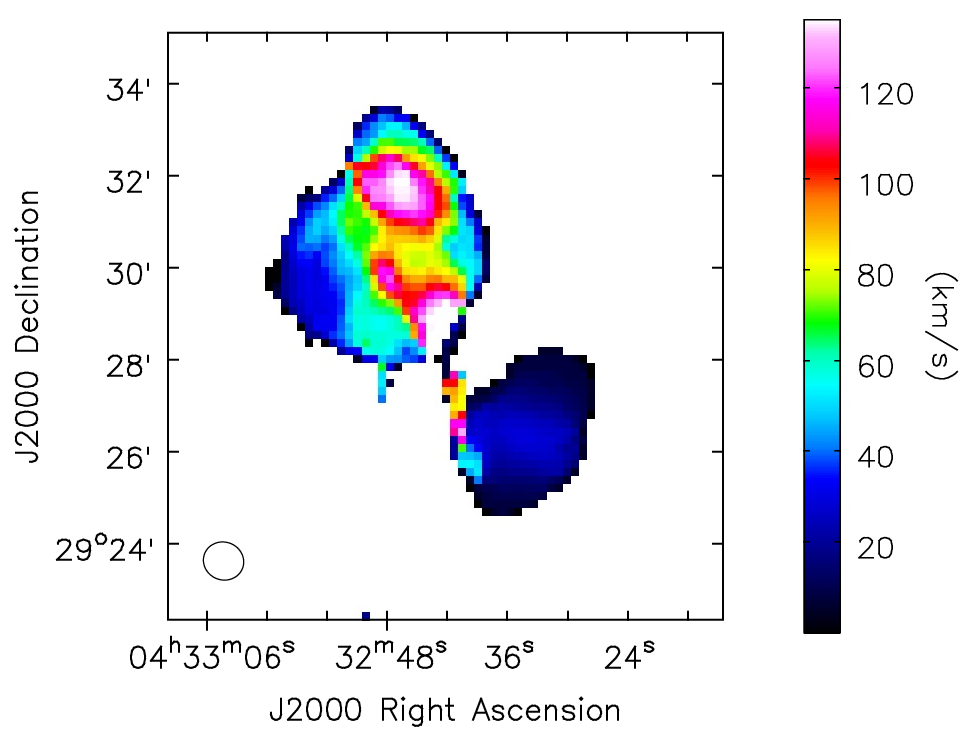}
      \includegraphics[width=7.8cm]{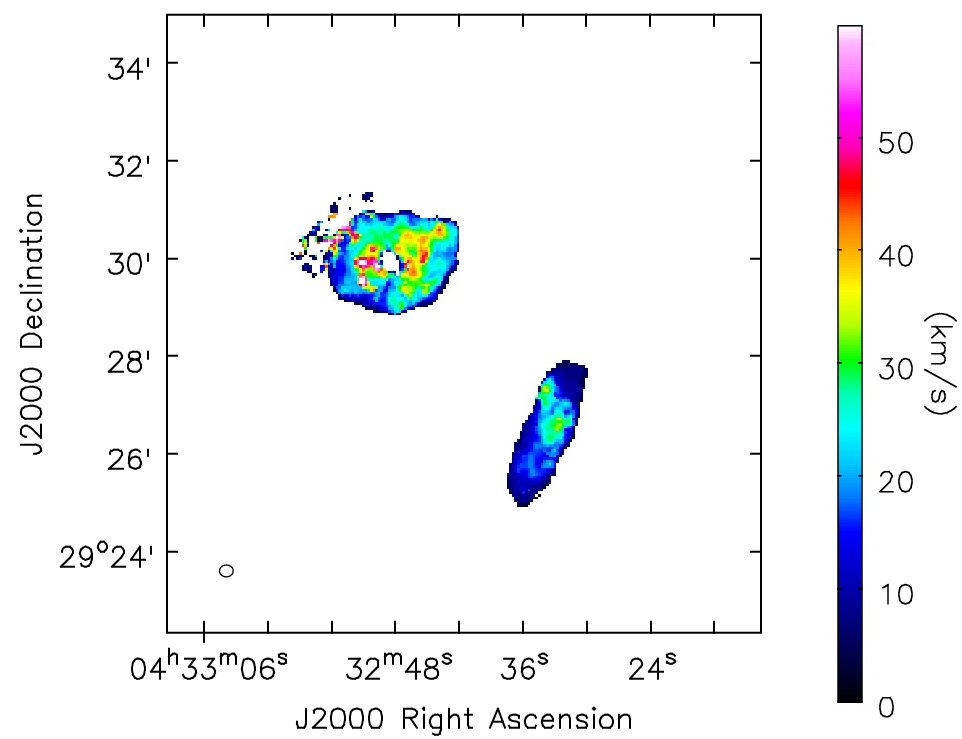}
\caption{Velocity field (moment~1; top), and velocity dispersion (moment~2; bottom) maps of IRAS~04296+2923 and HI~0432+2926. The maps are derived from VLA-D (left column) and VLA-C (right column) configuration \HI\ data using \textsc{SoFiA}.}

    \label{Fig:sofia_D}
\end{figure*}

\begin{figure*}
    \centering
    \includegraphics[width=9cm]{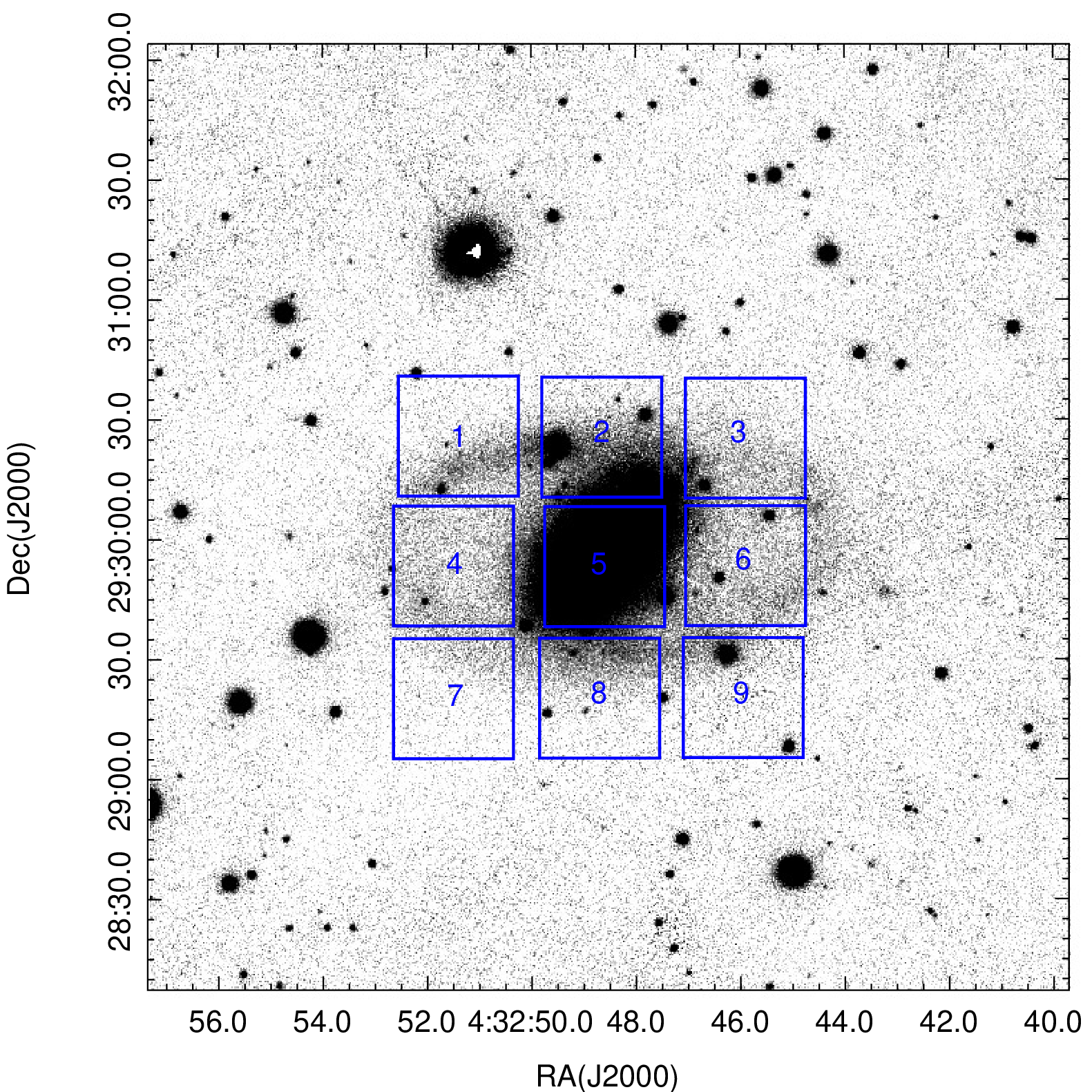}
    \caption{I-band optical image of IRAS 0432+2926 from the Pan-STARRS survey. The blue square boxes mark the regions selected for extracting \HI\ line profiles. Each region measures 30\arcsec\ $\times$\ 30\arcsec, with central coordinates listed in Table~\ref{region}. The corresponding \HI\ spectra are shown in Figure~\ref{fig:regionline}, and the fitted parameters are presented in Table~\ref{region}.}

    \label{Fig:jiugongge}
\end{figure*}

\begin{figure*}
    \centering
     \includegraphics[width=6cm]{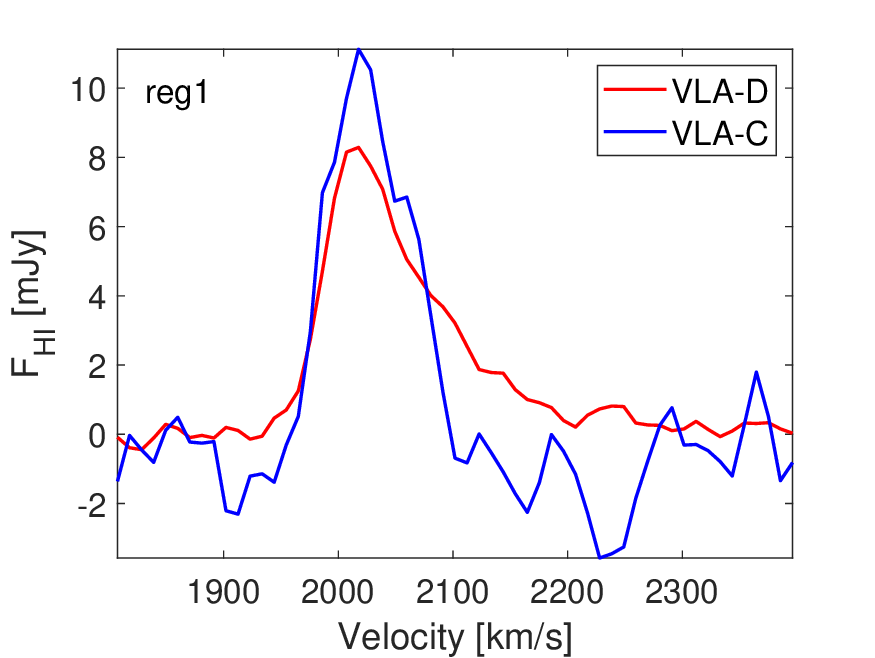}
     \includegraphics[width=6cm]{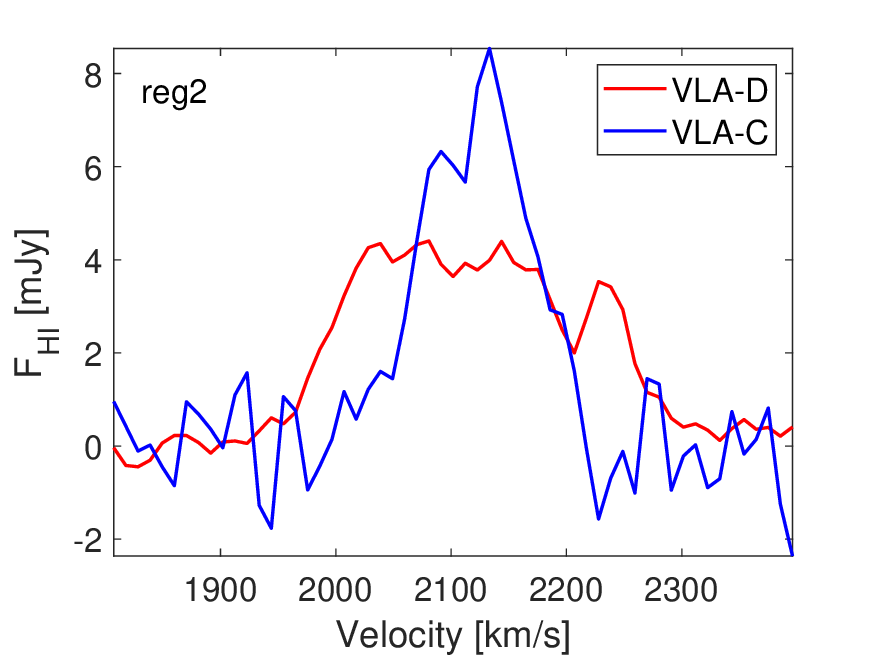}
     \includegraphics[width=6cm]{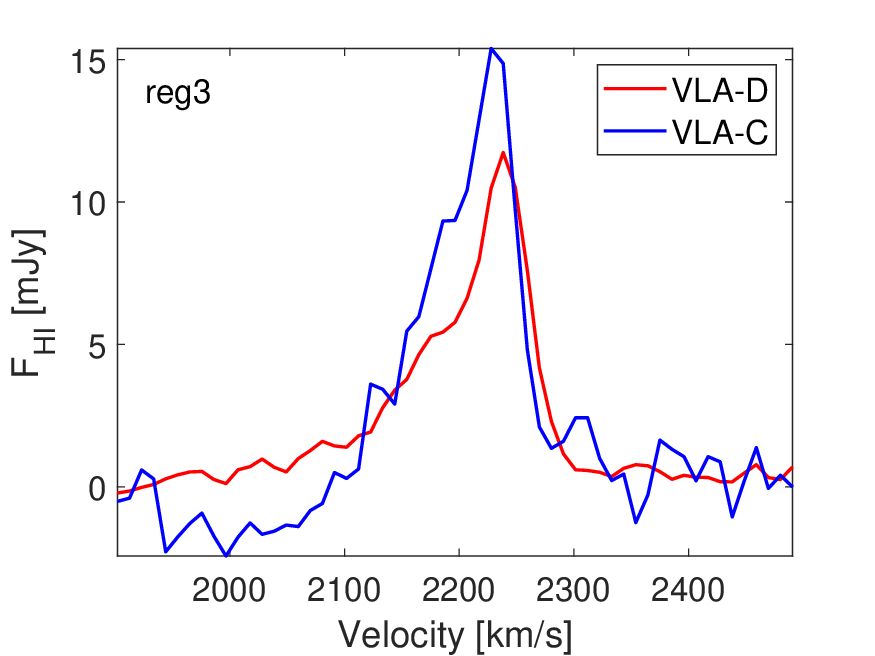}
     \includegraphics[width=6cm]{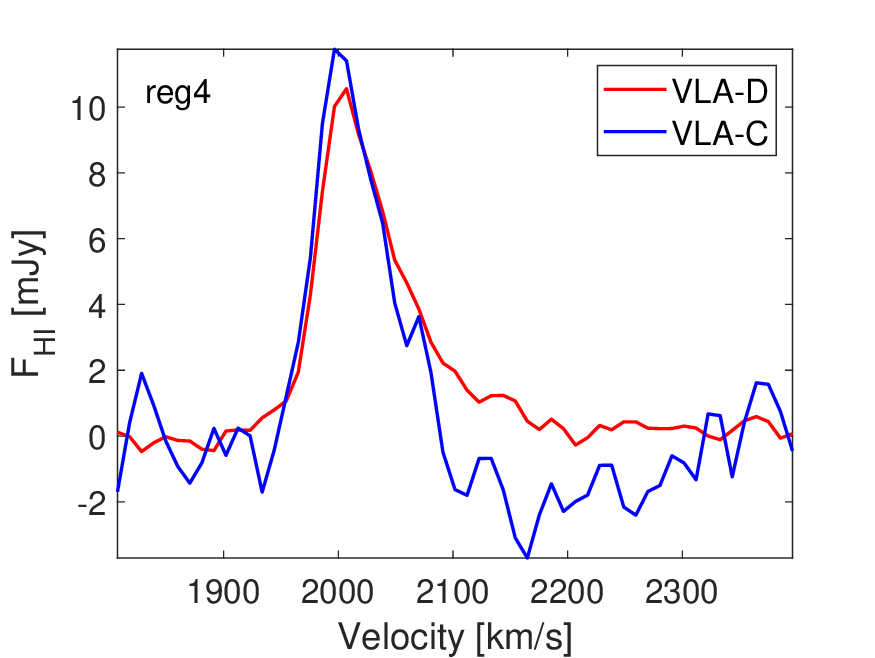}
     \includegraphics[width=6cm]{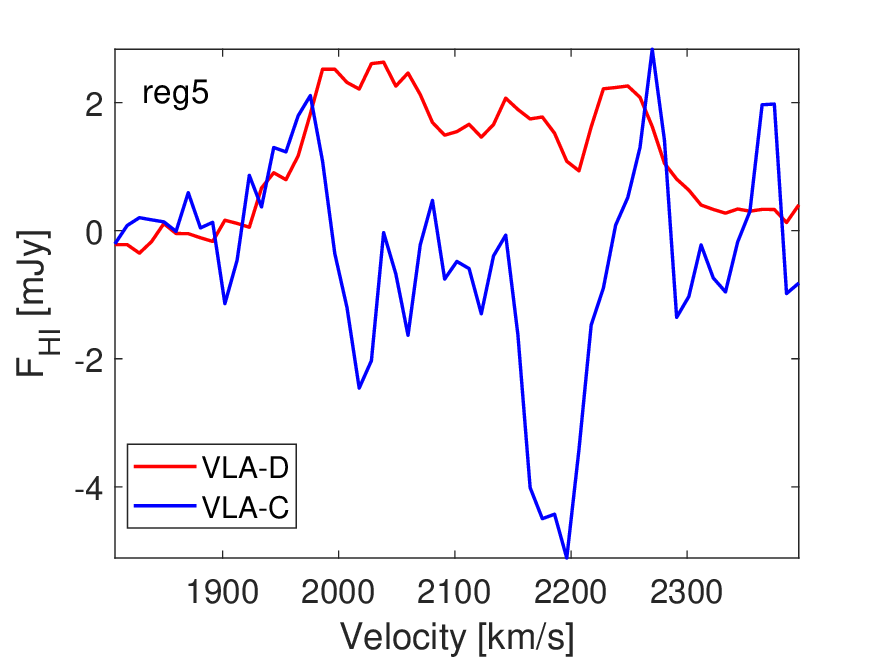}
     \includegraphics[width=6cm]{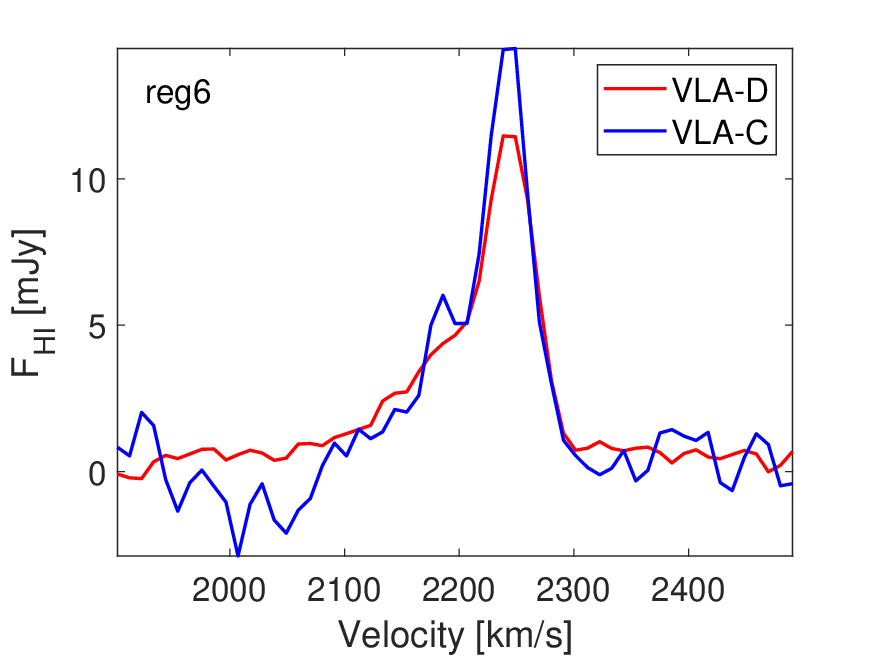}
     \includegraphics[width=6cm]{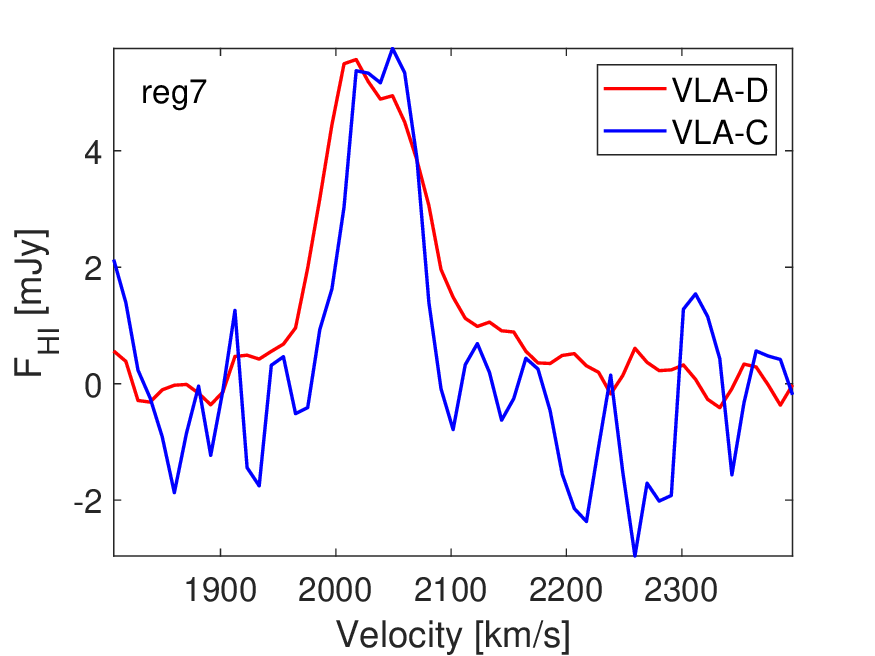}
     \includegraphics[width=6cm]{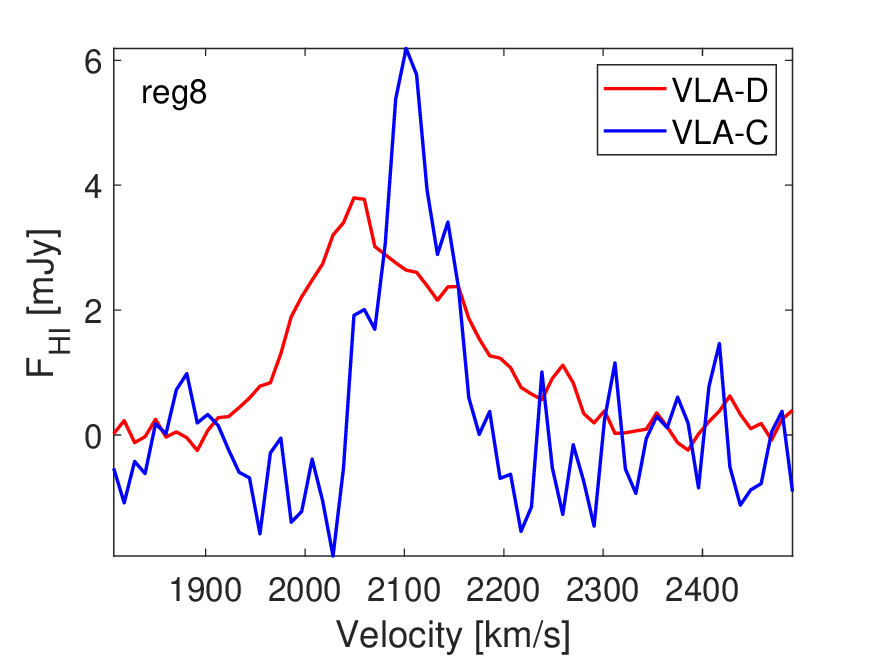}
     \includegraphics[width=6cm]{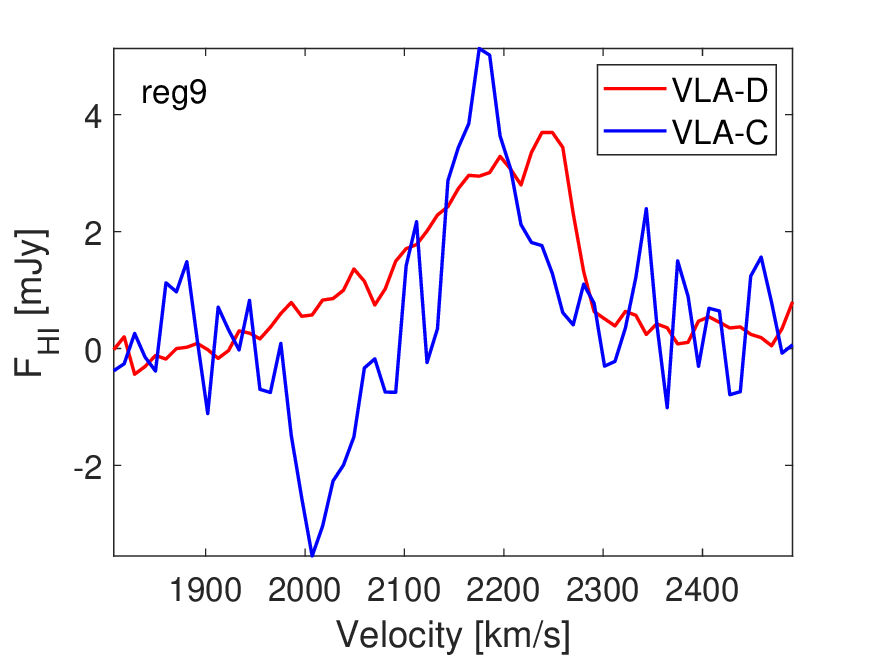}
     \includegraphics[width=6cm]{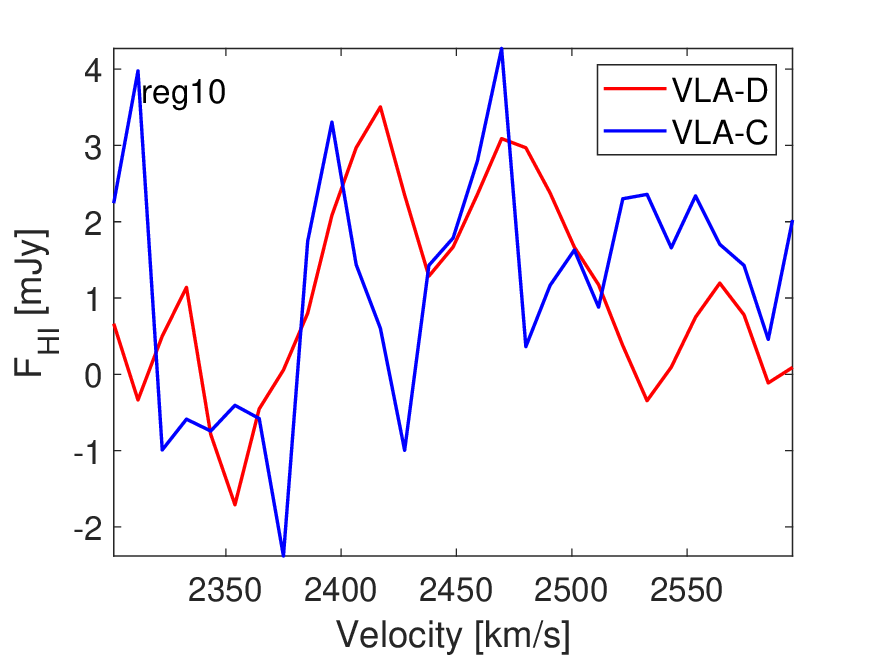}
     \includegraphics[width=6cm]{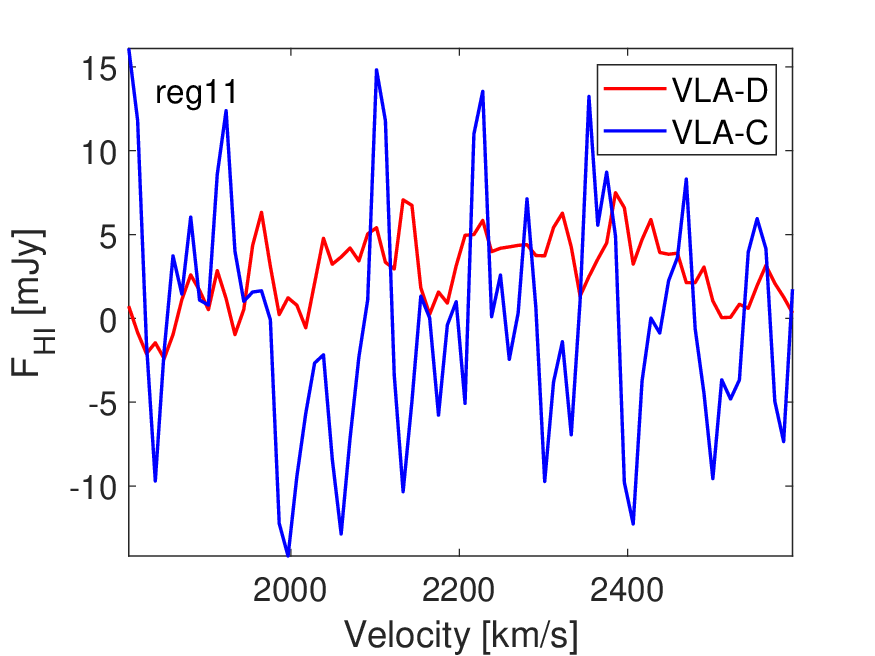}
     \caption{\HI\ line profiles extracted from regions 1--11, as marked in Figs.~\ref{Fig:sofia_D1} and \ref{Fig:jiugongge}. The fitting parameters are listed in Table~\ref{region}.}
    \label{fig:regionline}
\end{figure*}

\begin{figure*}
        \centering
        \includegraphics[width=9cm]{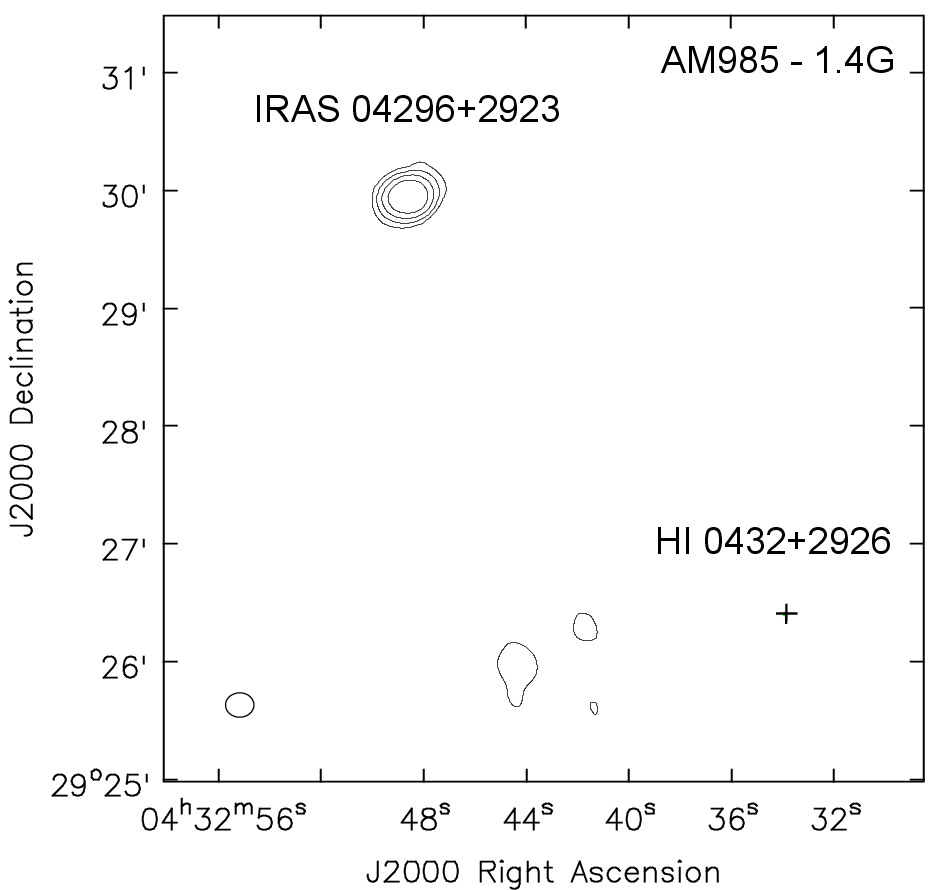}
        \includegraphics[width=9cm]{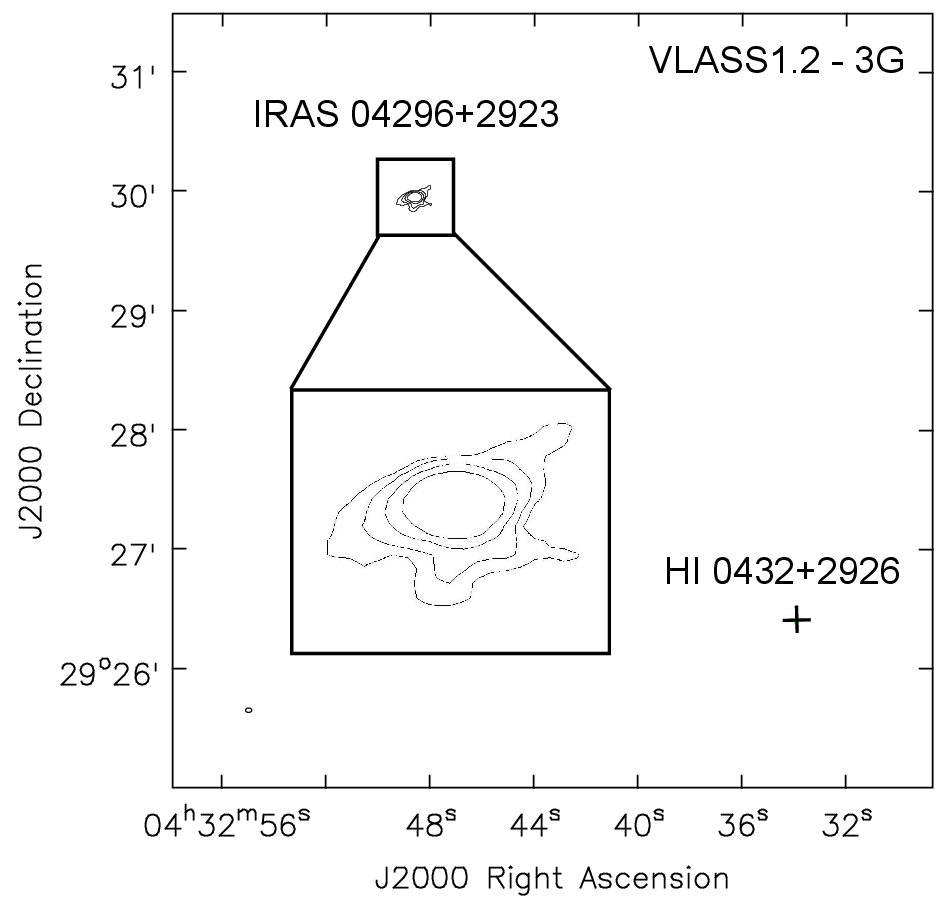}
        \includegraphics[width=9cm]{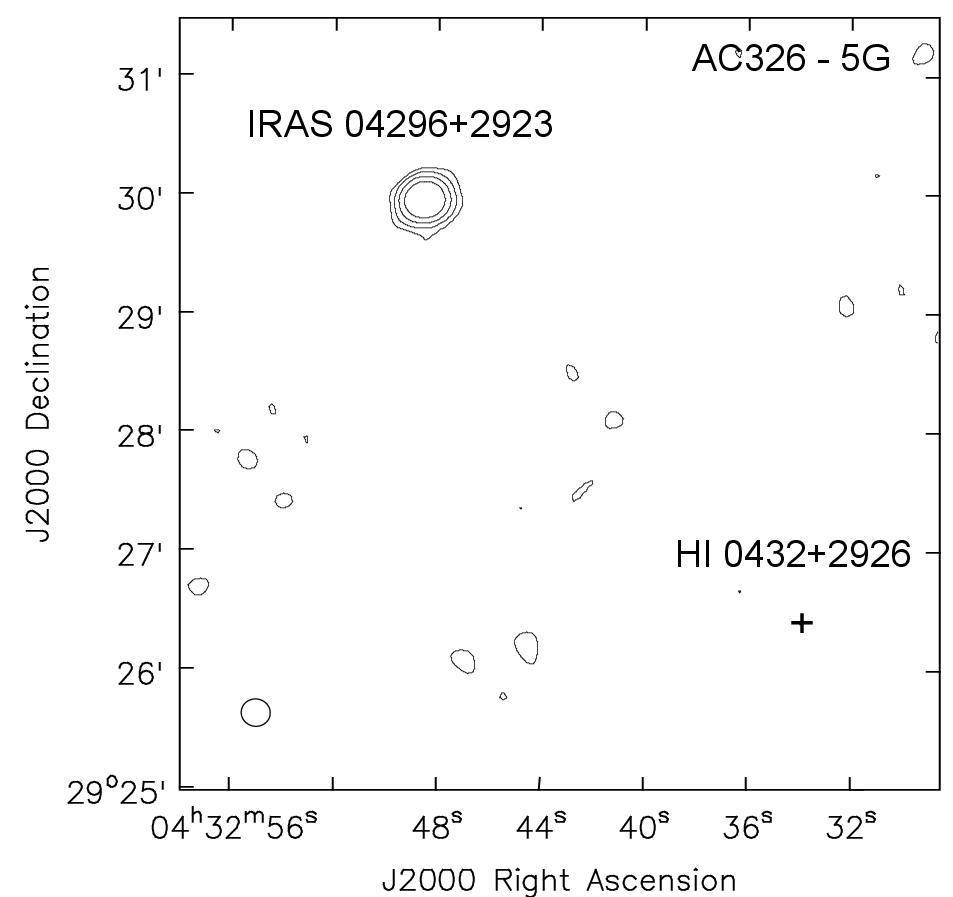}
        \includegraphics[width=9cm,height=9cm]{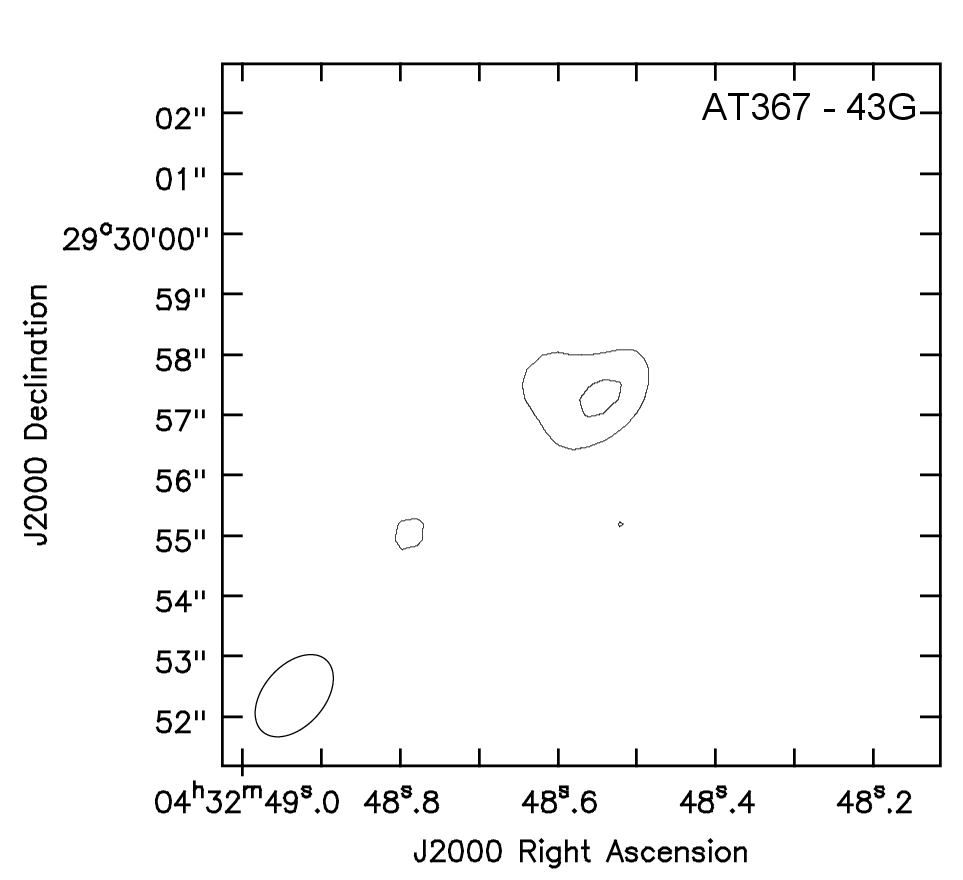}
        \caption{
Radio continuum contour maps of IRAS~04296+2923 at 1.4, 3, 5, and 43\,GHz. The project name and observing frequency are indicated in the upper-right corner of each panel, and additional information on these datasets is provided in Table~\ref{2923data}. The contour levels are:
1.4\,GHz: 1.5$\times$(1, 2, 4, 8)\,mJy\,beam$^{-1}$;
3\,GHz: 0.6$\times$(1, 2, 4, 8)\,mJy\,beam$^{-1}$;
5\,GHz: 1.5$\times$(1, 2, 4, 8)\,mJy\,beam$^{-1}$;
43\,GHz: 2.7$\times$(1, 2, 4, 8)\,mJy\,beam$^{-1}$.
The lowest contour corresponds to the 3$\sigma$ level in each map.
}
        \label{fig:duoboduan}
\end{figure*}

      \begin{figure*}
        \centering
         \subfigure[IRAS 04296+2923]
        {\includegraphics[width=8cm,height=11cm]{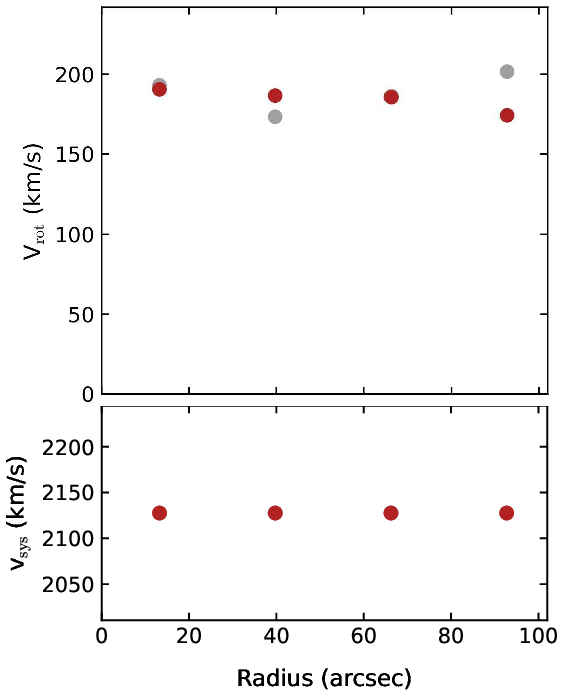}
        \hspace{1cm}
        \includegraphics[width=8cm,height=11cm]{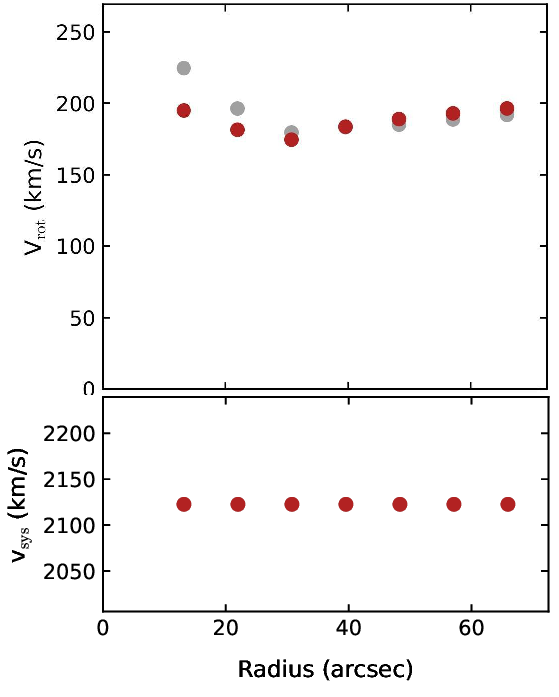}}
        \subfigure[\HI 0432+2926]
     {\includegraphics[width=8cm,height=11cm]{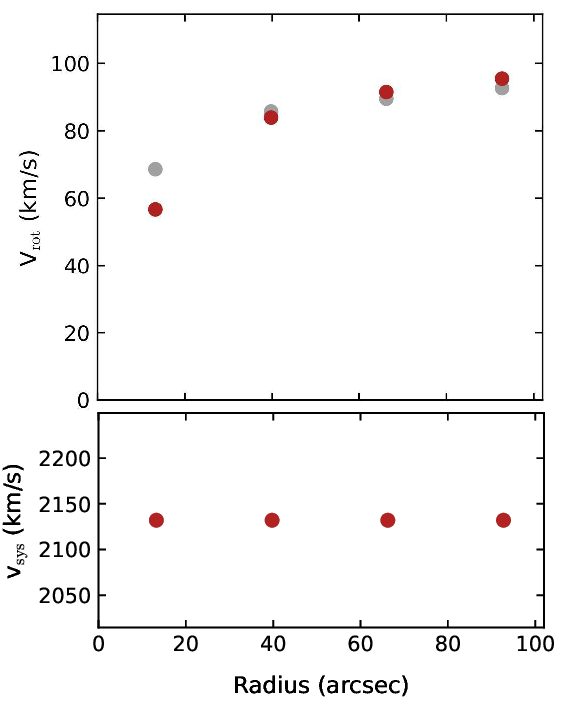}
        \hspace{1cm}
        \includegraphics[width=8cm,height=11cm]{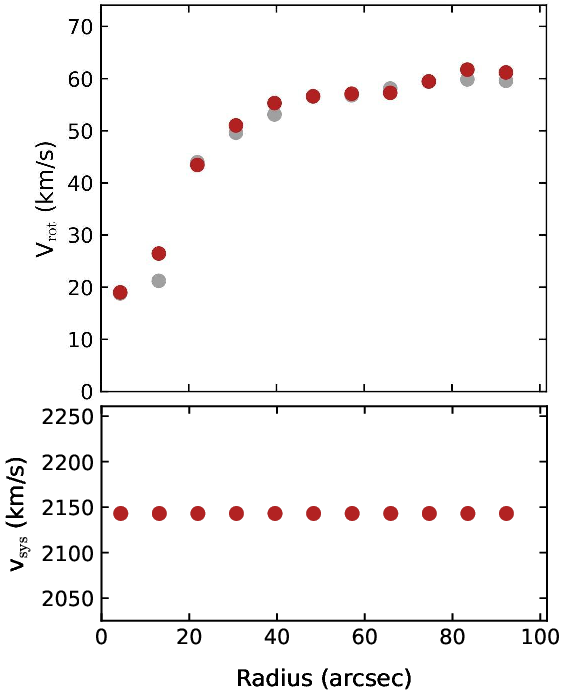}}        
        \caption{Radial profiles of rotation velocity ($V_{\rm rot}$, top) and systemic velocity ($V_{\rm sys}$, bottom) for IRAS 04296+2923 (top row) and HI 0432+2926 (bottom row), derived from VLA-D (left) and VLA-C (right) configuration \HI\ data using \textsc{3D-Barolo}. Red points represent the best-fit model values, while gray points show the observed data.  
        \label{fig.nihe} 
        }  
    \end{figure*}


\begin{figure*}
         \centering
           \subfigure[IRAS 04296+2923 (VLA-D)]
        { \includegraphics[width=8cm]{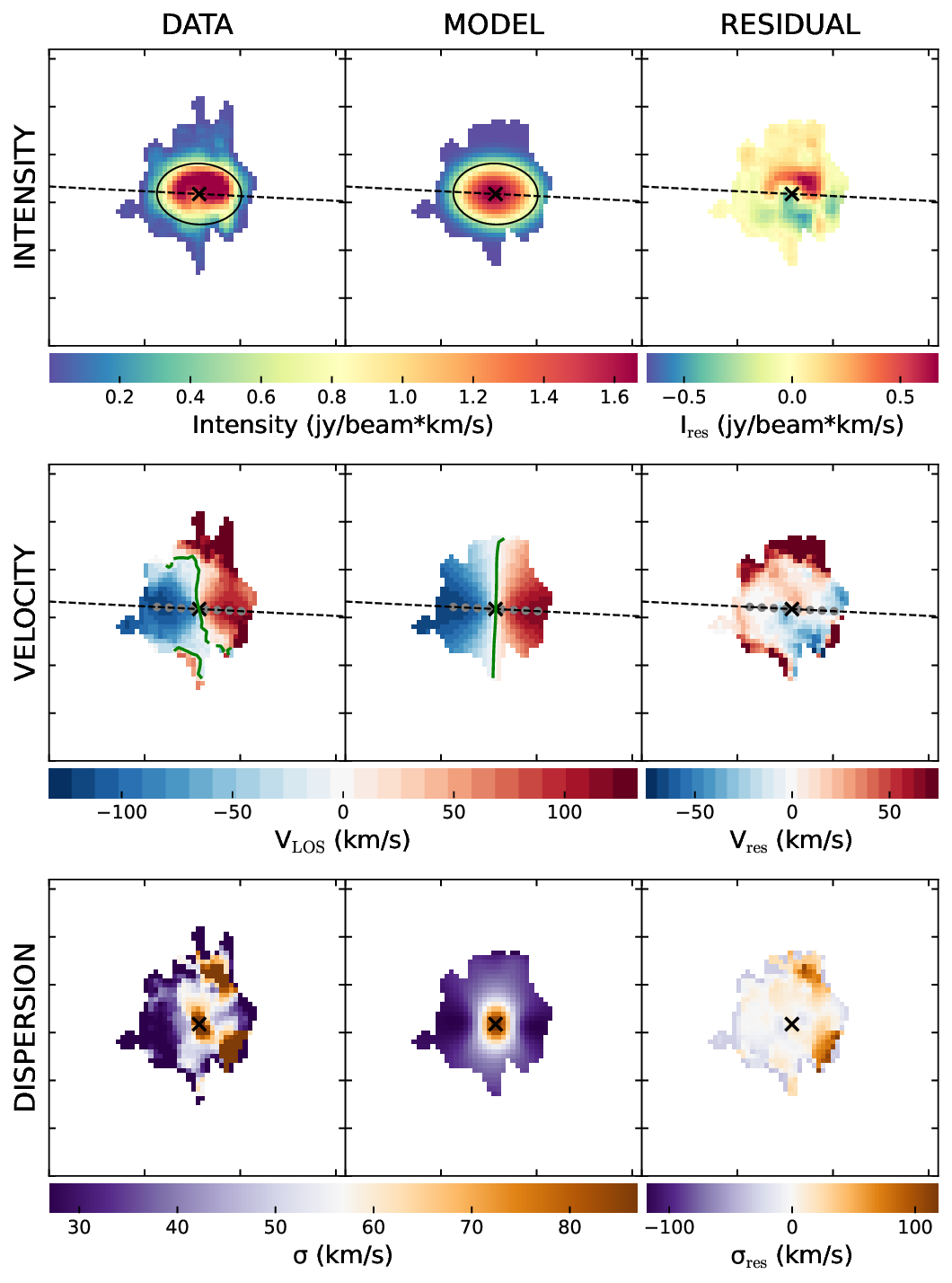}} 
         \hspace{1cm}
         \subfigure[IRAS 04296+2923 (VLA-C)]
        { \includegraphics[width=8cm]{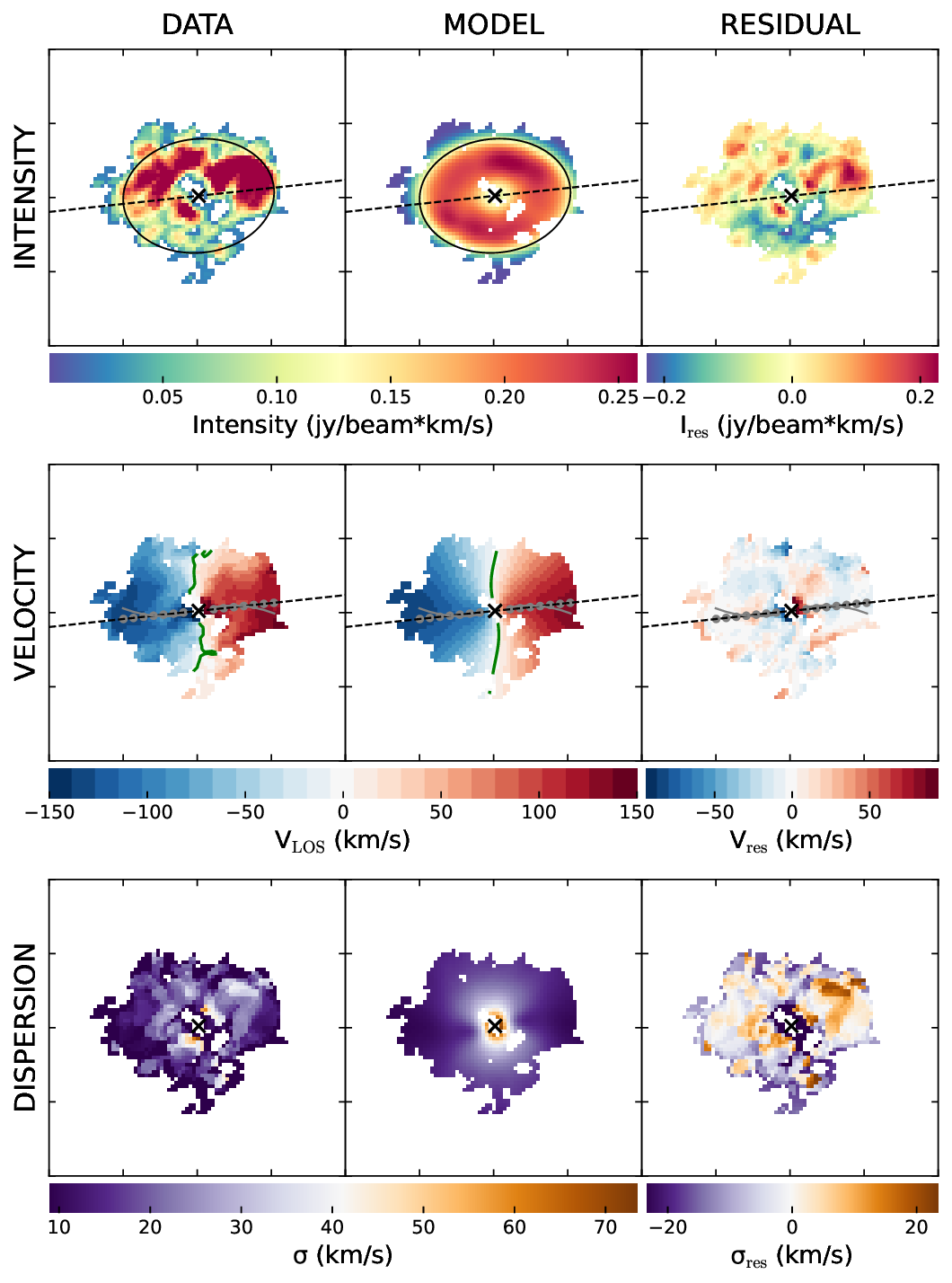}}
        
          \subfigure[HI 0432+2926 (VLA-D)]
        { \includegraphics[width=8cm]{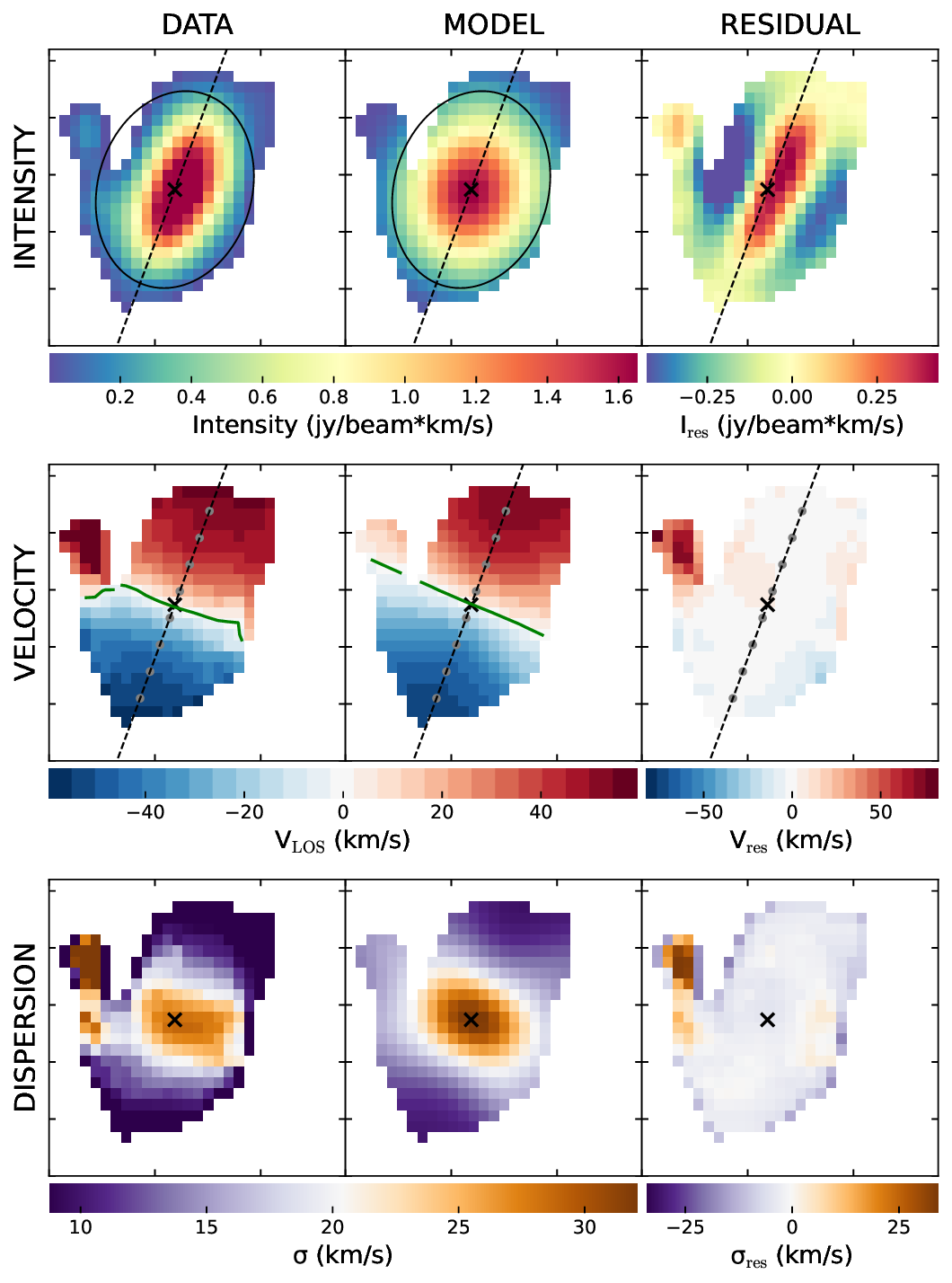}}
         \hspace{1cm}
          \subfigure[HI 0432+2926 (VLA-C)]
        { \includegraphics[width=8cm]{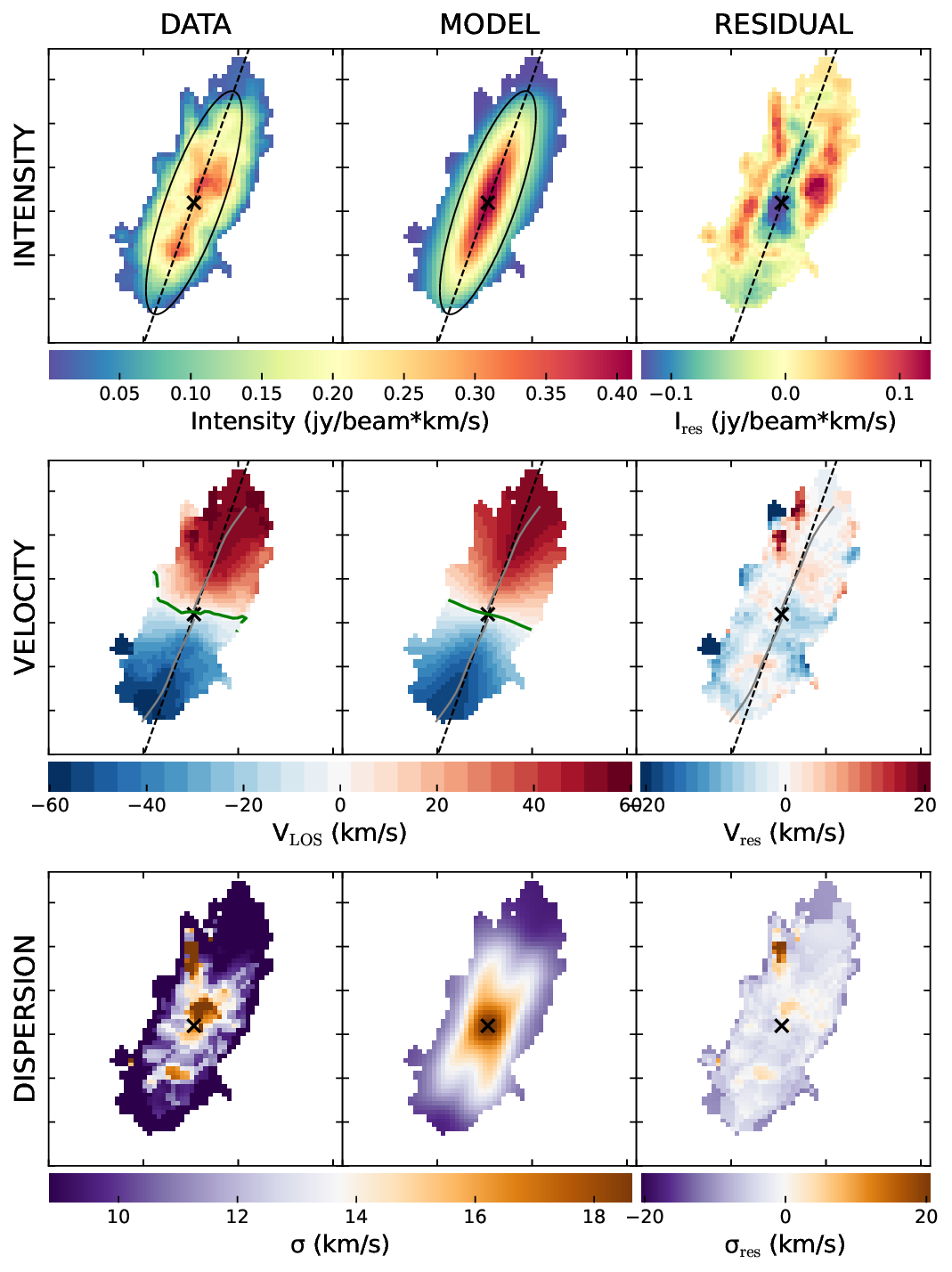}}
        
         \caption{
        \textsc{3D-Barolo} modeling results for IRAS~04296+2923 (panels (a)–(b)) and \HI~0432+2926 (panels (c)–(d)), based on the VLA D- and C-configuration \HI\ data.
For each subpanel, the columns show (from left to right) the observed data, the best-fit model, and the residuals for the integrated intensity (top row), velocity field (middle row), and velocity dispersion (bottom row).  
         }
         \label{fig.model}
     \end{figure*}

\begin{figure*}
    \centering
    \subfigure[M1 stage]
    {\includegraphics[width=9cm]{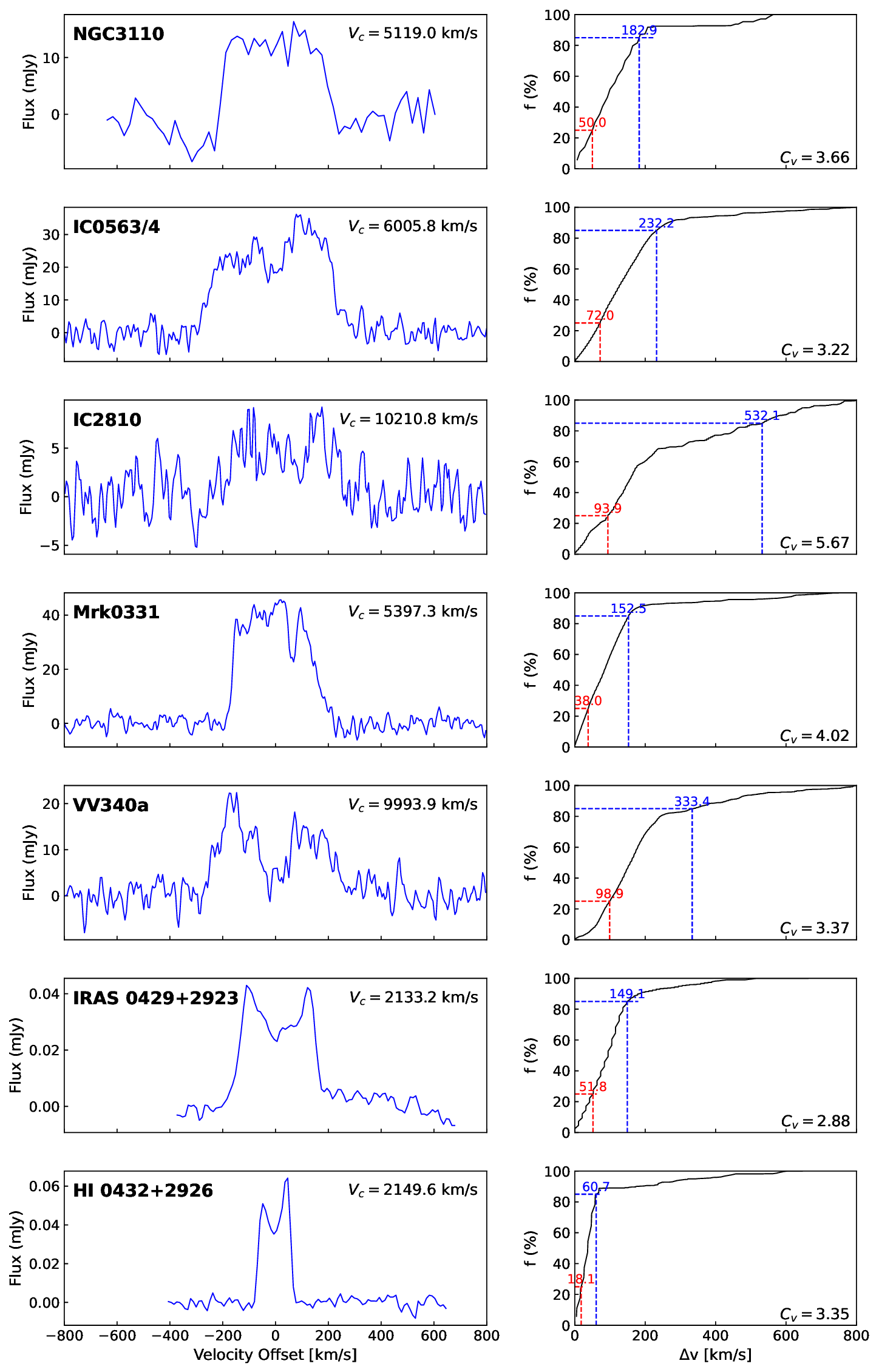}}
    \subfigure[M2 stage]
    {\includegraphics[width=9cm]{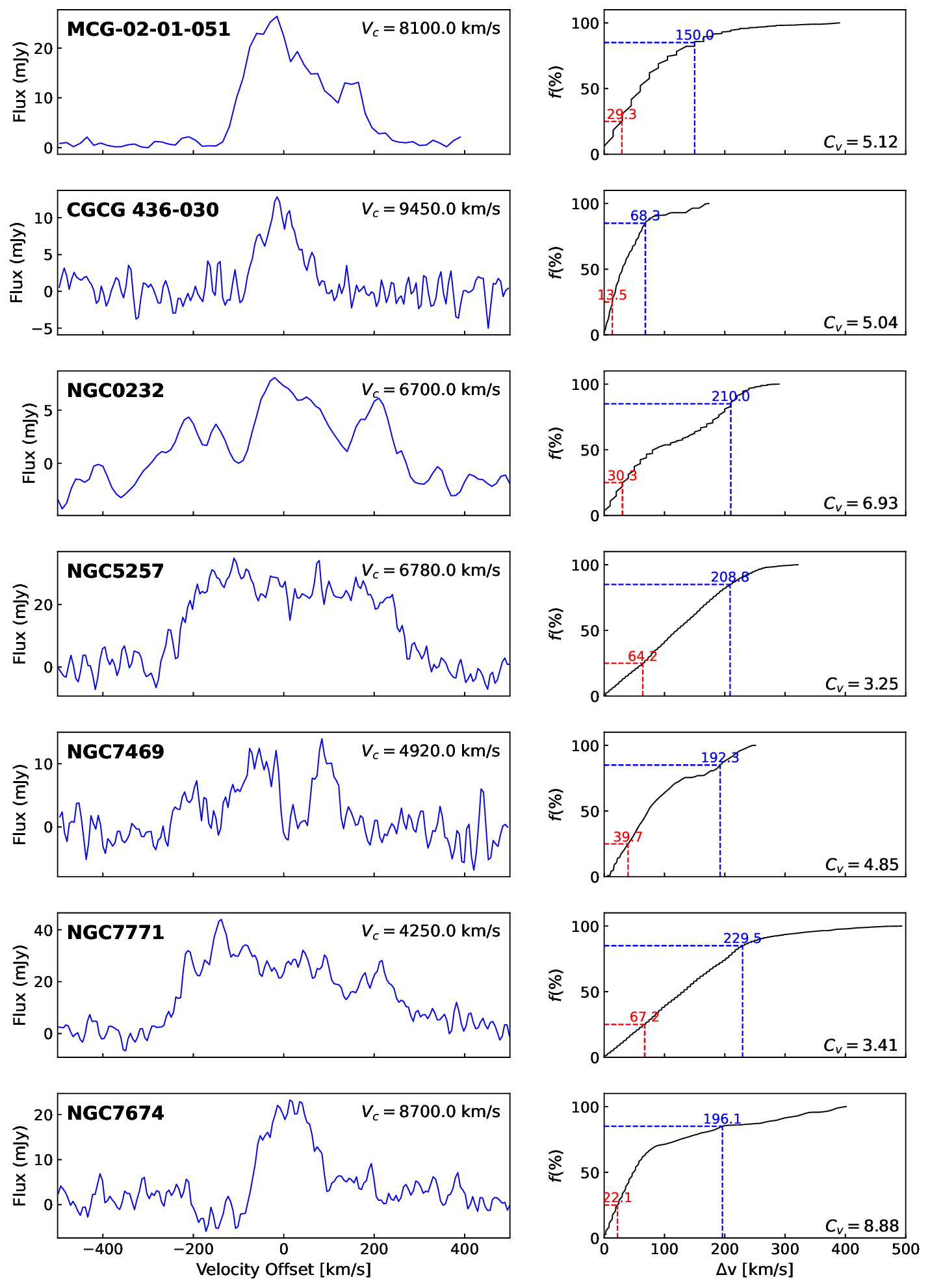}}
    \caption{\HI\ spectra and line concentration parameter ($C_v = V_{85}/V_{25}$) for early-stage merging galaxies from \cite{2016ApJ...825..128L}. Panel (a) shows the \HI\ spectra (left column) and cumulative rotation curves (right column) for five M1-stage mergers, with data retrieved from the NED database except for NGC~3110 \citep{2018ApJ...861...49H}. Panel (b) presents the same for seven M2-stage mergers, with MCG~02-01-051 from \cite{1991A&A...245..393M}; all others are from NED. Curves of growth are constructed following the method in \cite{2022ApJ...929...15Z}.}
    \label{fig:Cv}
\end{figure*}

\begin{table*}
\caption{Parameters of the HI emission line regions in IRAS 04296+2923 \label{region}}
    \centering
    \begin{tabular}{lcccccc}
        \hline
        \hline
        Region &Proj.& Center Coords  & $V_{\rm HI}$ & $W_{50}$ & $S_{21}$ &$Peak$\\
         & &(J2000)  & (km/s) & (km/s) & (Jy$\cdot$km/s) &(mJy)\\
        \hline
        1 & D & 04:32:51.4 &  2032.3$\pm$1.6 & 107.8$\pm$3.9 & 0.86$\pm$0.030& 7.5$\pm$0.4 \\  
          & C & +29:30:25.9 &  2024.4$\pm$2.1 & 75.3$\pm$5.0 & 0.87$\pm$0.05& 10.8$\pm$0.6 \\   
          
        2 & D & 04:32:48.7 &  2133.4$\pm$3.5 & 238.8$\pm$8.3 & 1.12$\pm$0.036& 4.4$\pm$0.1\\   
          & C & +29:30:25.6 &  2123.0$\pm$2.8 & 106.5$\pm$7.2 & 0.86$\pm$0.054 & 7.6$\pm$0.4\\   
          
        3 & D & 04:32:45.9 &  2222.9$\pm$3.2 & 104.4$\pm$7.6 & 1.03$\pm$0.070 &9.3$\pm$0.3\\   
          & C & +29:30:25.3 &  2214.6$\pm$2.2 & 96.5$\pm$5.4 & 1.34$\pm$0.070 &12.9$\pm$0.6\\    
          
        4 & D & 04:32:51.4 &  2016.4$\pm$1.5 & 84.3$\pm$3.5 & 0.85$\pm$0.033 &9.5$\pm$0.2\\    
          & C & +29:29:53.4 &  2008.1$\pm$2.3 & 66.4$\pm$5.4 & 0.79$\pm$0.061&11.3$\pm$0.8 \\    
          
        5 & D & 04:32:48.6 &  2230$\pm$3 & 69$\pm$7 & 0.14$\pm$0.01 &1.9$\pm$0.1\\
          &   &            &  2139$\pm$4 & 69$\pm$13& 0.10$\pm$0.02 &1.3$\pm$0.1\\
          &   &            &  2014$\pm$3 & 120$\pm$9& 0.25$\pm$0.02 &2.0$\pm$0.1\\
        
          & C & +29:29:53.2 &  2170$\pm$3  & 58$\pm$7  & -0.31$\pm$0.04  &-5.0$\pm$0.6\\  
          &   &             &  2038$\pm$9  & 86$\pm$23 & -0.18$\pm$0.04  &-2.0$\pm$0.4\\

        6 & D & 04:32:45.9  &  2235.2$\pm$2.3 & 86.6$\pm$5.4 & 0.90$\pm$0.053 &9.8$\pm$0.4\\ 
          & C & +29:29:53.4 &  2236.4$\pm$2.4 & 69.4$\pm$5.8 & 0.93$\pm$0.073 &12.6$\pm$0.9 \\ 
          
        7 & D & 04:32:51.5 &  2032.4$\pm$1.5 & 103$\pm$3.6 & 0.60$\pm$0.020 &5.4$\pm$0.1\\  
          & C & +29:29:20.2 &  2039.1$\pm$2.6 & 61.8$\pm$6.2 & 0.40$\pm$0.038&6.1$\pm$0.5 \\ 
          
        8 & D & 04:32:48.7 &  2076.1$\pm$3.0 & 192.8$\pm$7.2 & 0.66$\pm$0.023 &3.2$\pm$0.1\\
          & C & +29:29:20.3 &  2107.2$\pm$3.0 & 67.1$\pm$7.1 & 0.40$\pm$0.040&5.6$\pm$0.5 \\ 
          
        9 & D & 04:32:45.9 &  2192.7$\pm$4.0 & 192.5$\pm$9.5 & 0.66$\pm$0.030 &3.2$\pm$0.1\\
          & C & +29:29:20.4 &  2183.9$\pm$5.6 & 80.2$\pm$13.3 & 0.41$\pm$0.064 &4.8$\pm$0.6\\
          
       10 & D & 04:32:41.9 &  2438.8$\pm$11.1 & 127.4$\pm$26.3 & 0.24$\pm$0.048 &2.8$\pm$0.4 \\
          & C & +29:27:44.4 &  - & - & - \\ 
          
       11 & D & 04:32:48.0 & 2252.6$\pm$19.6 & 424$\pm$46 & 2.1$\pm$0.22 &4.8$\pm$0.4\\
          & C & +29:32:21.5 & - & -& -\\
        \hline
    \end{tabular}
       \vskip 0.1 true cm 
       \raggedright
       \noindent  Notes. Column (1) The selected regions as shown in Fig. \ref{Fig:jiugongge} and \ref{fig:regionline}. Column(2): The VLA configuration. Column (3): the coordiation of the regions. Column (4)-(6) the fitted parameters including the central velocity, FWHM, and integrated line flux  density of the \HI line profile.
   \par
\end{table*}

\begin{figure*}
    \centering
        \includegraphics[width=6cm]{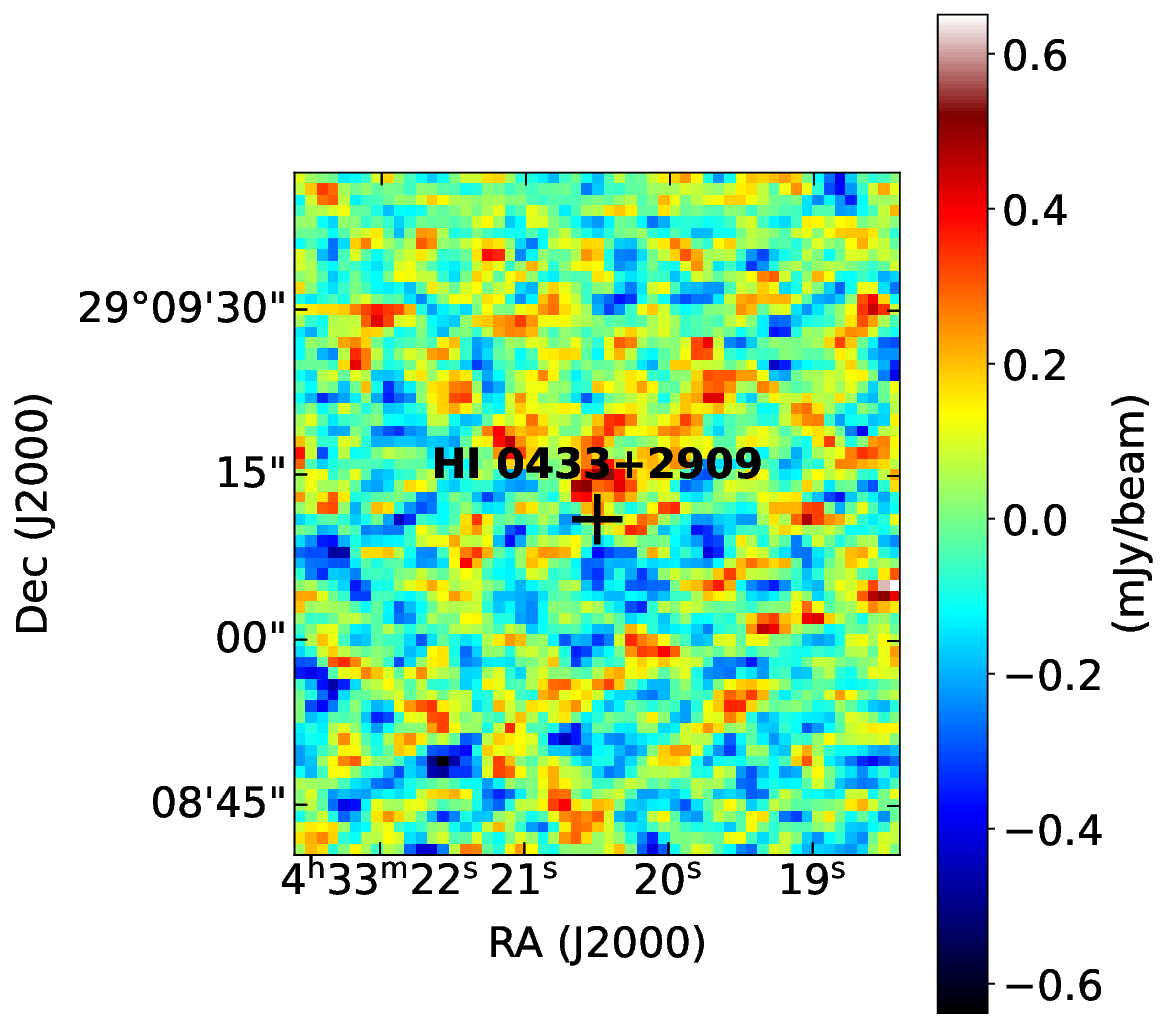}
        \includegraphics[width=6cm]{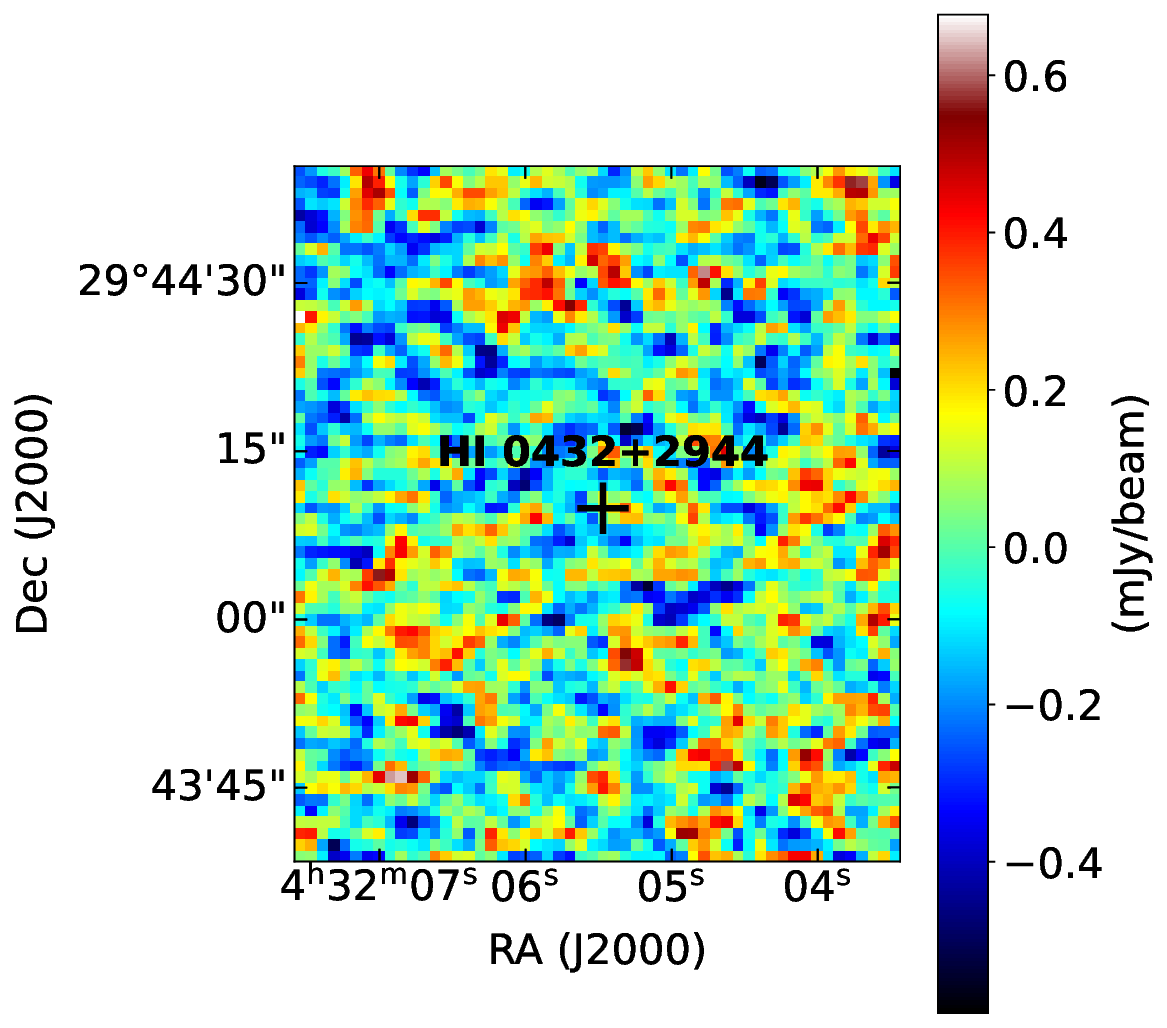}
        \includegraphics[width=6cm]{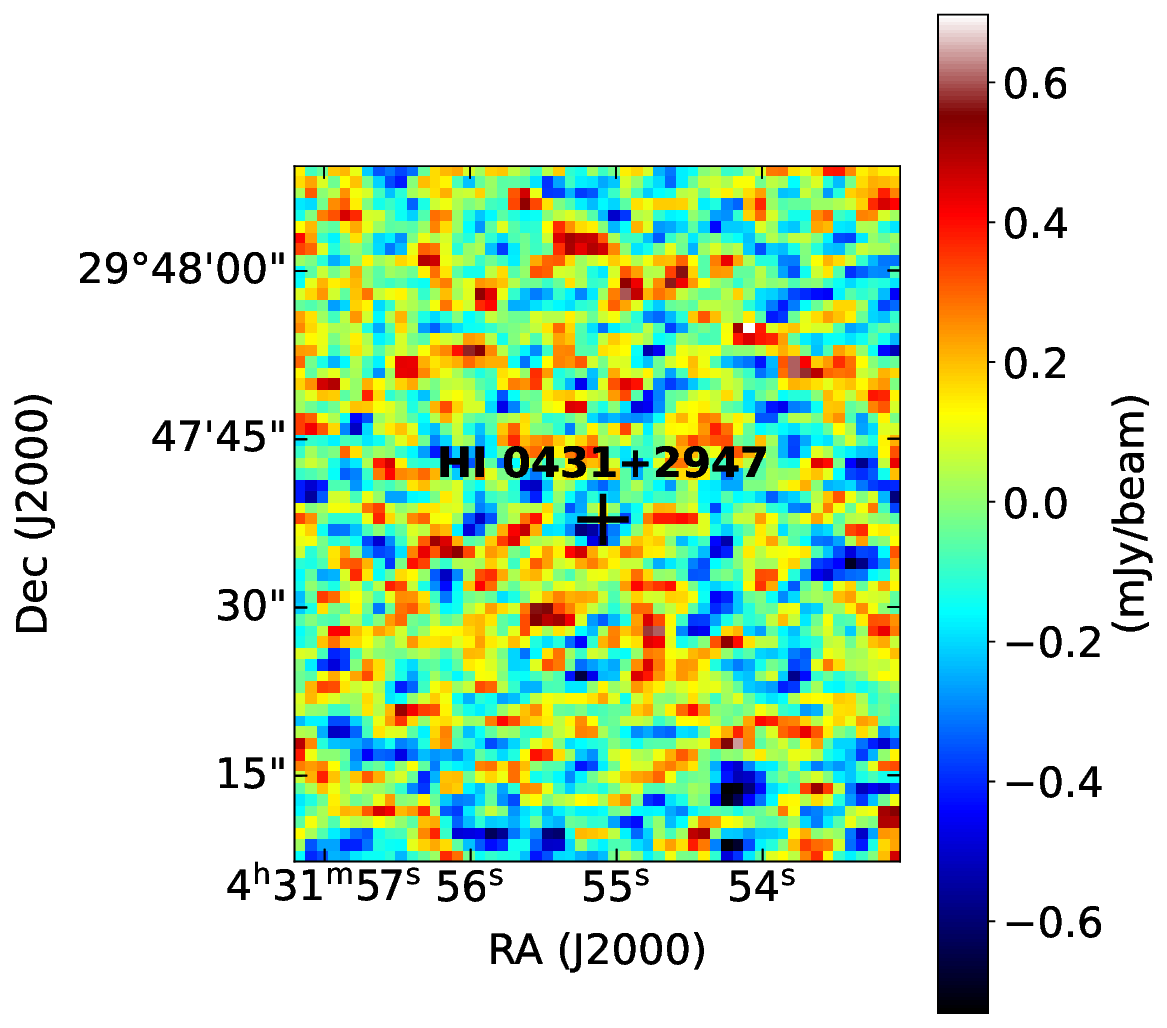}
        \caption{
VLASS 3\,GHz continuum images of the \HI galaxies \HI~0433+2909, \HI~0431+2947, and \HI~0432+2944 (see Table~\ref{sources}). The black crosses indicate the optical coordinates listed in Table~\ref{sources}. The rms noise in the images is approximately 0.2\,mJy\,beam$^{-1}$.
}
        \label{fig:vlassandvlac}
\end{figure*}

\end{appendix}

\end{document}